\documentclass[preprint,amsmath,amssymb,aps,eqsecnum]{revtex4-1}
\usepackage[dvipdfmx]{graphicx}

\usepackage{subfigure}
\usepackage{color}

\begin{document}
\title{Stability of small charged anti-de Sitter black holes 
in the Robin boundary}
\author{Takuya Katagiri}
\email{katagiri@rikkyo.ac.jp}
\author{Tomohiro Harada}
\email{harada@rikkyo.ac.jp}
\affiliation{
Department of Physics, Rikkyo University, Toshima, Tokyo 171-8501, Japan}
\date{\today}
\preprint{RUP-20-21}

\begin{abstract}
We analytically and numerically 
study quasinormal frequencies (QNFs) of neutral and charged scalar  
fields in the charged anti-de Sitter (AdS) black holes and discuss the stability of the black holes in terms of the QNFs. We focus on the range 
of the mass squared $\mu^2$ of the scalar fields for which the Robin boundary condition parametrised by $\zeta$
applies at the conformal  
infinity. We find that if the black hole of radius $r_{+}$ and charge $Q$ 
is much smaller than the AdS length $\ell$, 
the instability of the charged scalar field can be understood in terms of superradiance 
in the reflective boundary condition. Noting that the s-wave normal frequency in the AdS spacetime is a decreasing function of $\zeta$, we find that if $|eQ|\ell/r_{+}$ is greater than $(3+\sqrt{9+4\mu^2\ell^2})/2$, 
where $e$ is the charge of the scalar field, 
the black hole is superradiantly unstable irrespectively of $\zeta$.
On the other hand, if $|eQ|\ell/r_{+}$ is equal to or smaller than this critical value, 
the stability crucially depends on $\zeta$ and there appears 
a purely oscillating mode at the onset of the instability.
We argue that as a result of the superradiant instability,
the scalar field gains charge from the black hole and energy from its ambient electric field, 
while the black hole gives charge to the scalar field and gains energy from the scalar field 
but decreases its asymptotic mass parameter.
\end{abstract}

\maketitle

\tableofcontents

\newpage

\section{Introduction}
The anti-de Sitter (AdS) spacetime is an exact solution of Einstein's
field equations with a negative cosmological constant. The AdS spacetime
is maximally symmetric and geodesically complete, while it fails to be
globally hyperbolic. This means that we cannot uniquely predict the
complete time development from the initial data on a spacelike
hypersurface. 
Wald~\cite{Wald:1988} gives a method to define the initial value problem
for a massless scalar field in static and non-globally hyperbolic
spacetimes in terms of the self-adjoint extension. Ishibashi and Wald~\cite{Ishibashi:2004wx} discuss the same problem for scalar-vector-tensor linear perturbations in the AdS spacetime.

The stability of a spacetime is a fundamental and important problem. 
Usually, physically useful solutions are more or less stable. On the other hand, the onset of instability suggests the existence of a branch of solutions nonlinearly involving a stationary or oscillating perturbation. Therefore, the analysis of instability can lead to
finding new solutions~\cite{4,Ishii:2018oms,Ishii:2019wfs,Bizon:2020yqs}. In the context of AdS/CFT
correspondence~\cite{Maldacena:1997re}, an asymptotically AdS black hole
is dual to quantum field theory at finite temperature, and we can
calculate the relaxation time of the dual theory from the decaying time
scale of the perturbation on the AdS black hole spacetime. Also, its
instability corresponds to the phase transition of the dual theory as
seen in holographic superconductor
\cite{Gubser:2008px,Hartnoll:2008vx,Hartnoll:2008kx}. Thus, we can see
the phase structure of the (strongly coupled) quantum field theory in
terms of higher-dimensional gravitational physics.

For an asymptotically flat rotating black hole, it is known that a bosonic wave scattered against the potential barrier of the black hole can be amplified. This process is called superradiance~\cite{Z,Stro}. Superradiance may cause the instability of the black hole in a confined system. This is called superradiant instability. The confined system is realised by the mass of a scalar field~\cite{DDR,Yoshino:2012kn} or the negative cosmological constant~\cite{Hawking:1999dp,Dias:2016pma} or reflecting boundary conditions which guarantee no energy loss at a distant region from the black hole~\cite{PT,Cardoso:2004nk}. Superradiance can occur around not only rotating black holes but also the Reissner-Nordstr\"{o}m black hole~\cite{BK}. For superradiance to occur in the Reissner-Nordstr\"{o}m black hole, the scalar field needs to be charged. 

The other instability of the charged AdS black hole is known as near-horizon scalar condensation~\cite{Gubser:2008px}.
In the extremal limit, the black hole admits $AdS_2\times S^2$ near the horizon. Then, the charged scalar field can effectively violate the 2-dimensional Breitenlohner and Freedman bound~\cite{BF1,BF2} near the black hole. That leads to dynamical instability. This process also requires the scalar field to be charged.

For the study of stability, it is useful to investigate quasinormal frequencies (QNFs). Uchikata and Yoshida~\cite{Uchikata:2011zz} discuss the QNFs of neutral and charged massless scalar fields in the asymptotically AdS charged black hole. They claim that the spacetime can suffer from the superradiant instability only if $eQ>0$, while the spacetime is stable if $eQ\le 0$, where $e$ and $Q$ are the scalar field charge and the black hole charge, respectively. As we will show later, the result of the instability caused by superradinace is just a consequence of the assumption that the real part of the QNFs is positive. We will discuss that  the negative real part of the QNFs should also be discussed because there is no physical reason to rule out this possibility. In this respect, their analysis does not suffice. 

In addition, their analysis has room to extend with respect to boundary conditions at the conformal infinity. The boundary conditions play important roles in the dynamical evolution of a test field in spacetimes without global hyperbolicity~\cite{Wald:1988,Ishibashi:2004wx,Ishibashi:1999vw}. Hence, for the classification of the stability of (asymptotically) AdS spacetimes, the perturbation analysis should be done for all possible boundary conditions at the conformal infinity. For massless neutral scalar fields, in order to define well-posed and stable evolution, the suitable one is unique to the Dirichlet boundary condition. Of course, Uchikata and Yoshida~\cite{Uchikata:2011zz} follow this and obtain desirable results. In contrast, for \textit{massive} neutral fields with some range of the (effective) mass squared, e.g. in the case of a conformally coupled massless scalar field, the Robin boundary condition is allowed. This is characterised by one parameter and includes the Dirichlet and Neumann boundary conditions. In this sense, the Robin boundary condition is more general~\cite{Dappiaggi:2017pbe}. Also, even for charged fields, it can be adopted if the electromagnetic interaction between the black hole and the field is sufficiently small at the asymptotic infinity. For the above reasons, the study of QNFs of the neutral and charged massive scalar fields in the charged AdS black hole under the Robin boundary condition can shed light on a  relation between the boundary conditions and the dynamical evolution of the fields, such as the time scale of the (in)stability. Then, it is  possible to directly compare the result with that of~\cite{Uchikata:2011zz}. 

It is meaningful to discuss the stability of spacetimes against linear perturbations imposed the Robin boundary condition in many directions. In gravitational physics, there emerge new solutions through the analysis of linear perturbation on the pure AdS background~\cite{Bizon:2020yqs,Hertog:2004ns,Masachs:2019znp}. Through the analysis in the current paper, it is suggested that there are new black hole solutions 
which are asymptotically AdS and satisfy the Robin boundary condition 
at the conformal infinity. Since the background scalar field is trivial, the ``backreaction" contributions of the scalar field to the metric and Maxwell fields appear only from the second order. This is why the dynamics of a scalar field as a test field in the background metric and Maxwell fields provides a fully consistent linear perturbation solution. The appearance of a zero-mode linear perturbation solution of the scalar field suggests a branch off of a sequence of hairy-black-hole solutions there for it has been shown that this is the case in similar systems~\cite{Ishii:2019wfs,Dias:2016pma,Iizuka:2015vsa,Maeda:2010hf}. In the context of the AdS/CFT, the Robin boundary condition corresponds to a double-trace or multi-trace deformation~\cite{Aharony:2001pa} and the parameter of the boundary condition determines its coupling constant~\cite{Witten:2001ua}.

In this paper, we analytically and numerically investigate
the QNFs of neutral and charged massive scalar fields in the AdS spacetime and in the Reissner-Nordstr\"{o}m-AdS black holes with the Robin boundary condition at the conformal infinity. In the latter case, we assume that black holes are much smaller than the AdS length scale. We take this assumption for the following reasons. First, the most profound growing and decaying modes usually appear in the small-horizon limit~\cite{Uchikata:2011zz}. Second, two kinds of instability caused by charged fields in Reissner-Nordstr\"{o}m-AdS spacetimes, the superradiant instability and the near-horizon scalar condensation, are mutually entangled in general but should be of different natures in appropriate limits: in the small (large)-horizon limit, the superradiant instability (the near-horizon scalar condensation) is singled out~\cite{Dias:2016pma}. 
Third, in the current paper, we are interested in the qualitative change of the QNFs and the stability of black holes caused by the boundary conditions which are generalised due to the presence of even an infinitesimally small negative cosmological constant, rather than in the quantitative change caused by the large negative one.
Finally, as long as the authors know, no one has ever fully numerically evaluated the QNFs because of potential difficulty in handling the Robin boundary condition. In this context, it is useful to obtain the QNFs in the small-black-hole limit in the Robin boundary for future fully numerical studies for more general situations. For these reasons, we shall obtain the QNFs under the small-black-hole assumption as a first step. We then discuss the dynamical properties of the scalar field in terms of the QNFs. 

This paper is organised as follows. In Section II, we introduce the system and the basic equations. In Section III, we briefly review Ref.~\cite{Ishibashi:2004wx} for a massive scalar field in the AdS spacetime and present a numerical result consistent with their result. In Section IV, 
we introduce symmetries in the QNFs and justify the matched asymptotic expansion method for the neutral and charged scalar fields in the charged small AdS black holes. 
In Section V, we analytically show that the small charged AdS black hole suffers from superradiant instability if the QNFs in the pure AdS satisfy the superradaint condition.
In Section VI, we show numerical results for QNFs for the neutral and charged scalar fields in the charged small AdS black hole. 
In Section VII, we give physical interpretation of the superradiant instability in the AdS black hole. In the final section, we summarise this paper.
In the Appendix, we present mathematical notions, show the validity of the matching procedure in the matched asymptotic expansion, derive the asymptotic behaviours of the solutions, and explicitly confirm the symmetry in the QNFs. We adopt the sign convention and 
abstract index notation in Wald~\cite{Waldtextbook} and 
the unit in which $c=\hbar=1$. 

\section{System and basic equations}
We consider the Einstein-Maxwell-scalar system with a negative cosmological constant. The action is given by 
 \begin{equation}
 \label{action}
 S=\int d^{d+2}x\sqrt{-g}\left[\frac{1}{16\pi}(R-2\Lambda)+L_{m}+L_{em}\right]
 \end{equation} 
 with
 \begin{eqnarray}
 L_{m}&=&-\left[\left(D_{a} \Psi\right)\left(D^{a} \Psi\right)^*+\mu^2|\Psi|^2\right], \\
 L_{em}&=&-\frac{1}{16\pi}F_{ab}F^{ab},
 \end{eqnarray}
where $R$ is the scalar curvature and $\Lambda(<0)$ is the cosmological constant, 
$F_{ab}=\nabla_{a}A_{b}-\nabla_{b}A_{a}=\partial_{a}A_{b}-\partial_{b}A_{a}$ is the field strength,  $D_{a}:= \nabla_{a}-ieA_{a}$,
$\nabla_{a}$ is the Levi-Civita covariant derivative, $A_{a}$ is a gauge field, $e$ and $\mu^2$ are the charge and mass of the complex scalar field $\Psi$, respectively. We assume $e \ge 0$ without loss of generality and do not specify the sign of $\mu^2$.

Varying Eq.~(\ref{action}) with respect to $g_{ab}$, we obtain 
Einstein's equations 
\begin{equation}
 G_{ab}+\Lambda g_{ab}=8\pi T_{ab},
\end{equation}
where $T_{ab}=T_{(m) ab}+T_{(em) ab}$ with
\begin{eqnarray}
 T_{(m) ab}&=&\left[(D_{a}\Psi) (D_{b}\Psi)^{*}+(D_{a}\Psi)^{*}(D_{b}\Psi)\right]
-g_{ab}\left[g^{cd}(D_{c}\Psi) (D_{d}\Psi)^{*}+\mu^{2}\Psi \Psi^{*} \right], \\
 T_{(em)ab}&=&\frac{1}{4\pi}\left(F_{ac}F_{b}^{~c}-\frac{1}{4}g_{ab}F_{cd}F^{cd}\right).
\end{eqnarray}
Varying Eq.~(\ref{action}) with respect to $A_{a}$, we obtain some components of Maxwell's equations
\begin{equation}
 \nabla_{b}F^{ba}=-4\pi j_{(e)}^{a},
\end{equation}
where $j_{(e)}^{a}$ is the conserved electric current density given by 
\begin{equation}
 j_{(e)}^{a}=-ie\left[\Psi^{*}(D^{a}\Psi)-\Psi (D^{a}\Psi)^{*}\right].
\label{eq:electric_current}
\end{equation}
The rest of the components of Maxwell's equations 
\begin{equation}
 \nabla_{[a}F_{bc]}=0
\end{equation}
are automatically satisfied by construction.
Varying Eq. (\ref{action}) with respect to $\Psi$, we obtain the equation of motion for 
$\Psi$
\begin{equation}
\label{eomcs}
\left[\left(\nabla_{a}-i e A_{a}\right)\left(\nabla^{a}-i e A^{a}\right)-\mu^2\right]\Psi=0.
\end{equation}
For non-minimally coupled scalar field, $\mu^{2}$ is replaced with $\mu^2+\xi R$ in the above, 
where $\xi$ is a coupling constant to gravity.
From the above, we can find the conserved particle number current density $j^{a}$~\cite{Waldtextbook}:
\begin{equation}
j^{a}= -i  [\Psi^{*} (D^{a}\Psi)-\Psi(D^{a}\Psi)^{*}].
\label{eq:number_current}
\end{equation}
Since the electric current density $j_{(e)}^{a}$ is related to $j^{a}$ through 
$ j_{(e)}^{a}=ej^{a}$, 
we can identify the charge of the associated particle with $e$.

In the presence of the timelike Killing vector $\xi^{a}$, we can define the conserved energy current $J^{a}$ as $J^{a}:=-T_{b}^{a}\xi^{b}$, which can be decomposed as $J^{a}=J_{(m)}^{a}+J_{(em)}^{a}$, where $J_{(m)}^{a}:=-T_{(m) b}^{a}\xi^{b}$ and $J_{(em)}^{a}:=-T_{(em) b}^{a}\xi^{b}$. In spherical symmetry, the conserved charge associated with $J^{a}$ is the Misner-Sharp quasi-local energy $E_{\rm MS}$~\cite{Misner:1964je}.

We will focus on Eq.~(\ref{eomcs}) regarding $\Psi$ as a test field in the fixed 
spacetime and gauge field in Sections II--V. 
In Sections VI and VII, we will also discuss the other equations and the effect of the scalar field onto the spacetime and the gauge field.

\section{Scalar field in the AdS spacetime \label{sec:Ishibashi_Wald}}
\subsection{Brief review of exact results}
We briefly review a neutral scalar field in the AdS spacetime based on Ishibashi and 
Wald~\cite{Ishibashi:2004wx}.
From Eq. (\ref{eomcs}), the scalar field obeys the following equation:
\begin{equation}
\label{EOMS}
\left(\nabla_{\mu}\nabla^{\mu}-\mu^2\right)\Psi=0.
\end{equation} 
Since the AdS spacetime is static, Eq. (\ref{EOMS}) can be written in the form of 
 \begin{equation}
 \label{RADEQ}
-\frac{\partial^2}{\partial t^2}\Psi=A\Psi,
 \end{equation}
 where $t$ is a Killing parameter and $A$ is given by 
$A=-V D^i \left(V D_i\right)+\mu^2 V^2$, $V\equiv \left(-\xi^\mu \xi_\mu\right)^{1/2}$, 
$\xi^\mu=(\partial/\partial t)^{\mu}$ is a timelike Killing vector and $D^i$ is the 
Levi-Civita covariant derivative on a constant $t$ spacelike hypersurface. 

Now we can view $A$ as a linear operator $A: D\to K$, where $D$ and $K$ are the subspaces of the Hilbert space $\mathcal{H}=L^2$. The definitions of mathematical notions are given in Appendix \ref{self-adjoint_extension}. If $A$ in Eq. (\ref{RADEQ}) is symmetric and positive, $A$ has at least one self-adjoint extension $A_E$ which is positive~\cite{Reed:1975}. Moreover, if the boundary condition is specified, $A_E$ is uniquely determined~\cite{Ishibashi:1999vw} . If we find a positive and unique $A_E$, we can define stable and unique evolution of $\Psi$ from initial data~\cite{Wald:1988} . Therefore, the problem of how to define stable and unique evolution in the non-globally hyperbolic (static) spacetime boils down to choose the boundary condition such that $A_E$ is positive.
 
In the coordinates $(t,\chi,\theta^1,\theta^2,\dots,\theta^d)$, the metric of the unit $(d+2)$-dimensional AdS spacetime is given by
\begin{equation}
ds^2=\frac{\ell^2}{\sin^2{\chi}}\left(-dt^2+d\chi^2+\cos^2{\chi}d\Omega_{d}^2\right),
\end{equation}
where $\chi\in[0, {\pi}/{2}]$ is a radial coordinate, $(\theta^1,\theta^2,\dots,\theta^d)$ correspond to angular coordinates, 
$d\Omega^2_d$ is the metric of the $d$ dimensional sphere, and $\ell\equiv \sqrt{{d(d+1)}/({-2\Lambda})}$ is the AdS length scale.
The origin is located at $\chi={\pi}/{2}$, while $\chi=0$ corresponds to the conformal infinity. 

We assume that $\Psi$ is in the form,
\begin{equation}
\Psi(t,\chi,\theta^1,\dots,\theta^d)=\left(\ell\cot \chi\right)^{-d/2}\psi(\chi)Y_{\bm{k}}(\theta^1,\dots,\theta^d)e^{-i\omega t},
\end{equation}
where $Y_{\bm{k}}(\theta^1,\dots,\theta^d)$ are spherical harmonics and $\omega\in \mathbb{C}$ is the frequency. 
Now Eq. (\ref{RADEQ}) is rewritten as
\begin{equation}
\label{Schr}
A\psi(\chi)=\omega^2\psi(\chi),
\end{equation}
where
\begin{equation}
\label{Sym}
A=-\frac{d^2}{d\chi^2}+\frac{\nu^2-1/4}{\sin^2 \chi}+\frac{\rho^2-1/4}{\cos^2 \chi}
\end{equation}
with 
$
\nu^2={1}/{4}+{d(d+2)}/{4}+\mu^2\ell^2
$
and 
$\rho^2={1}/{4}+l(l+d-1)+{d(d-2)}/{4}$.
Note that $A$ given by Eq. (\ref{Sym}) is a symmetric operator.

We require that the general solution of Eq. (\ref{Schr}) be regular at the origin. On the other hand, the asymptotic behaviour of the solution near the conformal infinity is 
given by 
\begin{equation}
\label{GENSOL}
\psi(\chi)\sim C_1\psi_{fast}(\chi)+C_2\psi_{slow}(\chi)~,~C_1,C_2\in\mathbb{C},
\end{equation}
where $\psi_{fast}(\chi)$ and $\psi_{slow}(\chi)$ show the faster and slower cutoffs in the limit to the conformal infinity $\chi\to 0$, respectively. In analogy, we refer to the boundary condition $C_{2}=0$ and $C_{1}=0$ as the Dirichlet and Neumann boundary conditions, respectively. Besides, we call the boundary condition such that $C_{1}/C_{2}$ is a real constant the Robin boundary condition. Ref.~\cite{Ishibashi:2004wx} summarises the relation between the positivity of the self-adjoint extension of $A$ and the boundary conditions at the conformal infinity as the following theorem.
~\\
~\\
\textbf{Theorem 1}:~~\textit{Let $A$ and $A_{E}$ be the
symmetric operator given by Eq. (\ref{Sym}) and its self-adjoint extension, respectively. For the solution (\ref{GENSOL}) which satisfies the regularity condition at the origin, the relation between the positivity of $A_E$ and the boundary condition imposed at the conformal infinity is as follows: \\
\textbf{(1)}~~If $-{(d+3)(d-1)^2}/{4}\leq\mu^2\ell^2$, there exists unique $A_E$, and it is positive.  \\
\textbf{(2)}~~If $-{(d+1)^2}/{4}<\mu^2\ell^2<-{(d+3)(d-1)^2}/{4}$, $A_E$ is positive if and only if
\begin{equation}
\label{Robin}
\kappa\geq \kappa_c= -\left|\frac{\Gamma(-{\nu})}{\Gamma({\nu})}\right|\frac{{\Gamma(\zeta^0_{\nu,\rho})}^2}{{\Gamma(\zeta^0_{-\nu,\rho})}^2},
\end{equation}
where
$C_1=\kappa C_2$ ($\kappa\in\mathbb{R}$) and 
$\zeta^{0}_{\nu,\rho}\equiv ({\nu+\rho+1})/{2}$.
\\
\textbf{(3)}~~If $\mu^2\ell^2=-{(d+1)^2}/{4}$, $A_E$ is positive if and only if $
\kappa_{BF}\leq \kappa_{BF,c}= 2\gamma +2P(\zeta^0_{0,\rho}),
$
where $C_1=C_2/\kappa_{BF}$ ($\kappa_{BF}\in\mathbb{R}$), 
$P(\zeta^0_{0,\rho})\equiv d\log\Gamma(\chi)/d\chi|_{\chi=\zeta^0_{0,\rho}}$
and $\gamma$ is the Euler number.\\
\textbf{(4)}~~If $\mu^2\ell^2<-{(d+1)^2}/{4}$, $A$ is unbounded below. Then, any $A_E$ are unbounded below.}

\subsection{Numerical investigation}
Let us investigate the scalar field for the case $\textbf{(2)}$ of this theorem. For this purpose, we shall solve Eq. (\ref{Schr}) in the 4-dimensional AdS spacetime. Now we introduce a new variable
\begin{equation}
\label{yz}
y=1+\frac{1}{\tan^2\chi},
\end{equation} 
and a function $g(y)$ such that
\begin{equation}
\label{anza}
\psi(y)=y^{\frac{\omega\ell}{2}}(y-1)^{\frac{l}{2}}g(y).
\end{equation}
Then Eq. (\ref{Schr}) is reduced to an equation for $g(y)$,
\begin{equation}
\label{Hyfa}
y(1-y)\frac{d^2}{dy^2}g(y)+\left\{\gamma-(\alpha+\beta+1)y\right\}\frac{d}{dy}g(y)-\alpha\beta g(y)=0
\end{equation}
with
\begin{equation}
\begin{split}
\label{abca}
\alpha&=\frac{\omega\ell}{2}+\frac{l}{2}+\frac{3}{4}+\frac{1}{4}\sqrt{9+4\mu^2\ell^2},\\
\beta&=\frac{\omega\ell}{2}+\frac{l}{2}+\frac{3}{4}-\frac{1}{4}\sqrt{9+4\mu^2\ell^2},\\
\gamma&=\omega\ell+1.
\end{split}
\end{equation}
Eq. (\ref{Hyfa}) has two independent solutions
\begin{equation}
\begin{split}
\label{gys}
g(y)=&y^{-\alpha}F\left(\alpha,\alpha-\gamma+1;\alpha-\beta+1;\frac{1}{y}\right)~,~y^{-\beta}F\left(\beta,\beta-\gamma+1;\beta-\alpha+1;\frac{1}{y}\right),
\end{split}
\end{equation}
where $F(~,~;~;\frac{1}{y})$ is the Gaussian hypergeometric function. We thus obtain the general solution of Eq. (\ref{Schr}), 
\begin{equation}
\begin{split}
\label{Solfa}
\psi(y)=&Cy^{-\frac{l}{2}-\frac{3}{4}-\frac{1}{4}\sqrt{9+4\mu^2\ell^2}}(y-1)^{\frac{l}{2}}F\left(\alpha,\alpha-\gamma+1;\alpha-\beta+1;\frac{1}{y}\right)\\
&+Dy^{-\frac{l}{2}-\frac{3}{4}+\frac{1}{4}\sqrt{9+4\mu^2\ell^2}}(y-1)^{\frac{l}{2}}F\left(\beta,\beta-\gamma+1;\beta-\alpha+1;\frac{1}{y}\right),
\end{split}
\end{equation}
where $C$ and $D$ are arbitrary constants in $\mathbb{C}$.

Near the conformal infinity, Eq. (\ref{Solfa}) behaves as
\begin{equation}
\label{asympa}
\psi(\chi)\sim C\chi^{\frac{3}{2}+\frac{1}{2}\sqrt{9+4\mu^2\ell^2}}+D\chi^{\frac{3}{2}-\frac{1}{2}\sqrt{9+4\mu^2\ell^2}}.
\end{equation}
This is an explicit form of Eq. (\ref{GENSOL}). According to Theorem 1, we have a degree
of freedom in choosing the boundary condition depending on the value of $\mu^2\ell^2$. We here focus on the case $-9/4< \mu^2\ell^2<-5/4$ and 
impose the Robin boundary condition
\begin{equation}
\label{bc}
C=\kappa D,~~\kappa\in\mathbb{R}
\end{equation}
on Eq. (\ref{asympa}). 
Then, the AdS spacetime behaves as a confined system because Eq. (\ref{bc}) gives a reflecting boundary condition~\cite{Dappiaggi:2017pbe}. 
In the context of AdS/CFT correspondence, this boundary condition corresponds to the double-trace deformation and the one parameter of the boundary condition determines its coupling constant~\cite{Witten:2001ua}. 
We thus obtain the solution of Eq. (\ref{Schr}) satisfying the boundary condition at the conformal infinity as follows:  
\begin{equation}
\begin{split}
\label{solfa}
\psi(y)=D&\left[\kappa y^{-\frac{l}{2}-\frac{3}{4}-\frac{1}{4}\sqrt{9+4\mu^2\ell^2}}(y-1)^{\frac{l}{2}}F\left(\alpha,\alpha-\gamma+1;\alpha-\beta+1;\frac{1}{y}\right)\right.\\
&\left.+y^{-\frac{l}{2}-\frac{3}{4}+\frac{1}{4}\sqrt{9+4\mu^2\ell^2}}(y-1)^{\frac{l}{2}}F\left(\beta,\beta-\gamma+1;\beta-\alpha+1;\frac{1}{y}\right)\right].
\end{split}
\end{equation}
As shown in Appendix~\ref{sec:Gaussian_hypergeometric_functions}, in the limit of $y\to 1$, or $\chi\to \pi/2$, Eq. (\ref{solfa}) 
behaves as
\begin{equation}
\label{solf1a}
\psi(y)\sim D\Gamma(\alpha+\beta-\gamma)\left[D_1(\tilde{\omega},\kappa) \left(\frac
{\pi}{2}-\chi\right)^{-l-1}+D_2(\tilde{\omega},\kappa)\left(\frac
{\pi}{2}-\chi\right)^l\right],
\end{equation}
where
\begin{equation}
\begin{split}
\label{D12a}
D_1(\omega,\kappa)&=\frac{\kappa\Gamma(\alpha-\beta+1)}{\Gamma(\alpha)\Gamma(\alpha-\gamma+1)}+\frac{\Gamma(\beta-\alpha+1)}{\Gamma(\beta)\Gamma(\beta-\gamma+1)},\\
D_2(\omega,\kappa)&=\frac{\Gamma\left(\gamma-\alpha-\beta\right)}{\Gamma\left(\alpha+\beta-\gamma\right)}\left(\frac{\kappa\Gamma(\alpha-\beta+1)}{\Gamma(1-\beta)\Gamma(\gamma-\beta)}+\frac{\Gamma(\beta-\alpha+1)}{\Gamma(1-\alpha)\Gamma(\gamma-\alpha)}\right).
\end{split}
\end{equation}

We can see that the first term in the square brackets
on the right-hand side of Eq.~(\ref{solf1a}) diverges at $\chi={\pi}/{2}$ if $D_{1}(\omega,\kappa)\neq0$. 
Therefore, to guarantee the regularity at the origin, we require 
\begin{equation}
\label{rega}
D_1(\omega,\kappa)=0.
\end{equation}
This equation determines the eigenfrequency or the QNF of the mode which satisfies the boundary condition. It also gives us the relation between the evolution and the boundary condition at the conformal infinity. Here we can see that the functions $D_1(\omega,\kappa)$ and $D_2(\omega,\kappa)$ have a symmetry under the transformation $\omega\to-\omega$, and moreover satisfy the relation $D_1(\omega,\kappa)={D_1}^*(-\omega^*,\kappa)$ and $D_2(\omega,\kappa)={D_2}^*(-\omega^*,\kappa)$. 
Thus, if $\omega$ is a QNF, then, both of $-\omega$ and $-\omega^{*}$ and therefore $\omega^{*}$ are.
\begin{figure}[htbp]
\centering
\subfigure[The relation between $\zeta$ and $\omega$.]{
\includegraphics[scale=0.74]{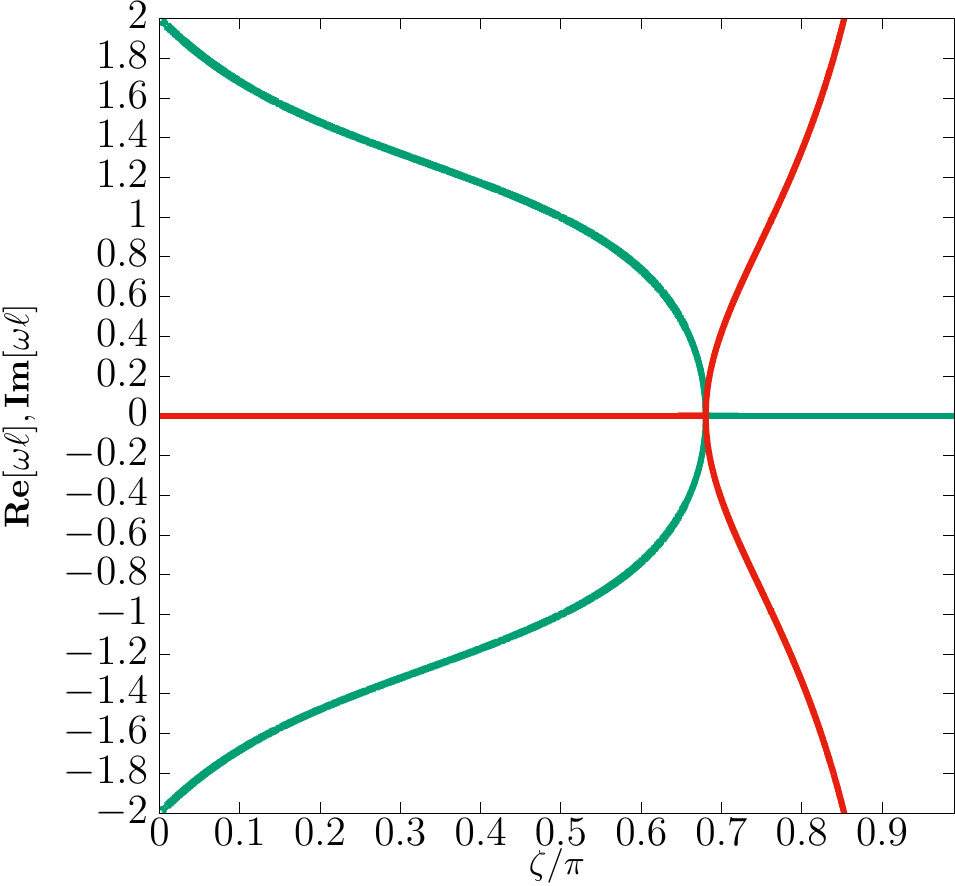}
\label{rea}}
\hspace{0.3cm}
\subfigure[The enlarged figure of Figure \ref{rea}.]{
\includegraphics[scale=0.74]{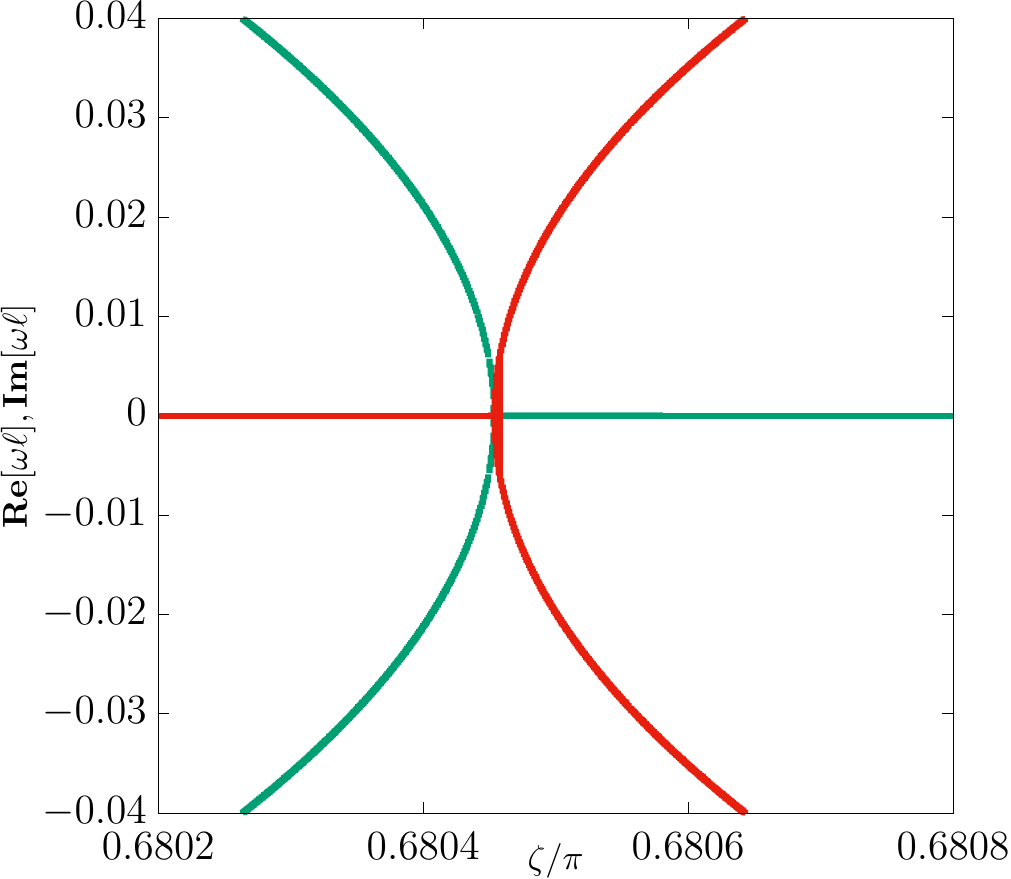}
\label{ef}}
\caption{
  The relation between $\zeta$ and $\omega$ of QNFs 
     with  $d=2,~l=0$, and $\mu^2\ell^2=-2$. The real and imaginary parts of $\omega$ are denoted by green and red lines, respectively.}
\end{figure}

Now we redefine the parameter of the boundary condition as 
$\zeta\equiv\textrm{ArcTan}\left[1/\kappa\right]\in (0,\pi)$, 
where $\zeta$ is a monotonically decreasing function of $\kappa \in\mathbb{R}$ with $\zeta=0$ and $\zeta = \pi/2$ corresponding to the Dirichlet and Neumann conditions, respectively. Then the stability criterion $\kappa\geq\kappa_c$ corresponds to $\zeta\leq\zeta_c$ in terms of $\zeta$, where we have defined 
\begin{equation}
\zeta_c\equiv \textrm{ArcTan}\left[-\left|\frac{\Gamma({\nu})}{\Gamma(-{\nu})}\right|\frac{{\Gamma(\zeta^0_{-\nu,\rho})}^2}{{\Gamma(\zeta^0_{\nu,\rho})}^2}\right].
\end{equation}
For s-wave ($l=0$) in the 4-dimensional AdS spacetime, substituting $d=2$, $\mu^2\ell^2=-2$, and $l=0$, we find
\begin{equation}
\label{condition}
\zeta_c\simeq 0.68045 \pi.
\end{equation}
We numerically solve Eq. (\ref{rega}) by the Newton-Raphson method. 
Figures \ref{rea} and \ref{ef} present the relation between $\zeta$ and $\omega$. Here we choose $d=2,~l=0$, and $\mu^2\ell^2=-2$. The vertical axis denotes $\textrm{Re}[\omega\ell]$ and $\textrm{Im}[\omega\ell]$. The horizontal axis denotes the normalised parameter $\zeta/\pi$. The green and red lines denote $\textrm{Re}[\omega\ell]$ and $\textrm{Im}[\omega\ell]$, respectively. Figure \ref{ef} is the enlarged figure of Figure \ref{rea} in the region given by $\textrm{Re}[\omega\ell],\textrm{Im}[\omega\ell]\in[-0.04,0.04]$ and $\zeta/\pi\in[0.6802,0.6808]$. These figures show that $\textrm{Re}[\omega\ell]$ decreases as $\zeta$ increases for $\zeta\leq\zeta_c\simeq 0.68\pi$ and $\textrm{Re}[\omega\ell]=0$ for $\zeta_c<\zeta$. We can also see that $\textrm{Im}[\omega\ell]$ is zero for $\zeta\leq\zeta_c$, while it takes two values, the one is positive and the other negative with the same absolute value for $\zeta_c<\zeta$. Hence, the operator $A$ of Eq. (\ref{Schr}) fails to be positive
  for $\zeta_c<\zeta$ because there is a mode of which the imaginary part of frequency is positive. This means that the mode is unstable for $\zeta_c<\zeta$ or $\kappa<\kappa_c$. 

\begin{figure}[htbp]
\centering
\includegraphics[scale=0.4]{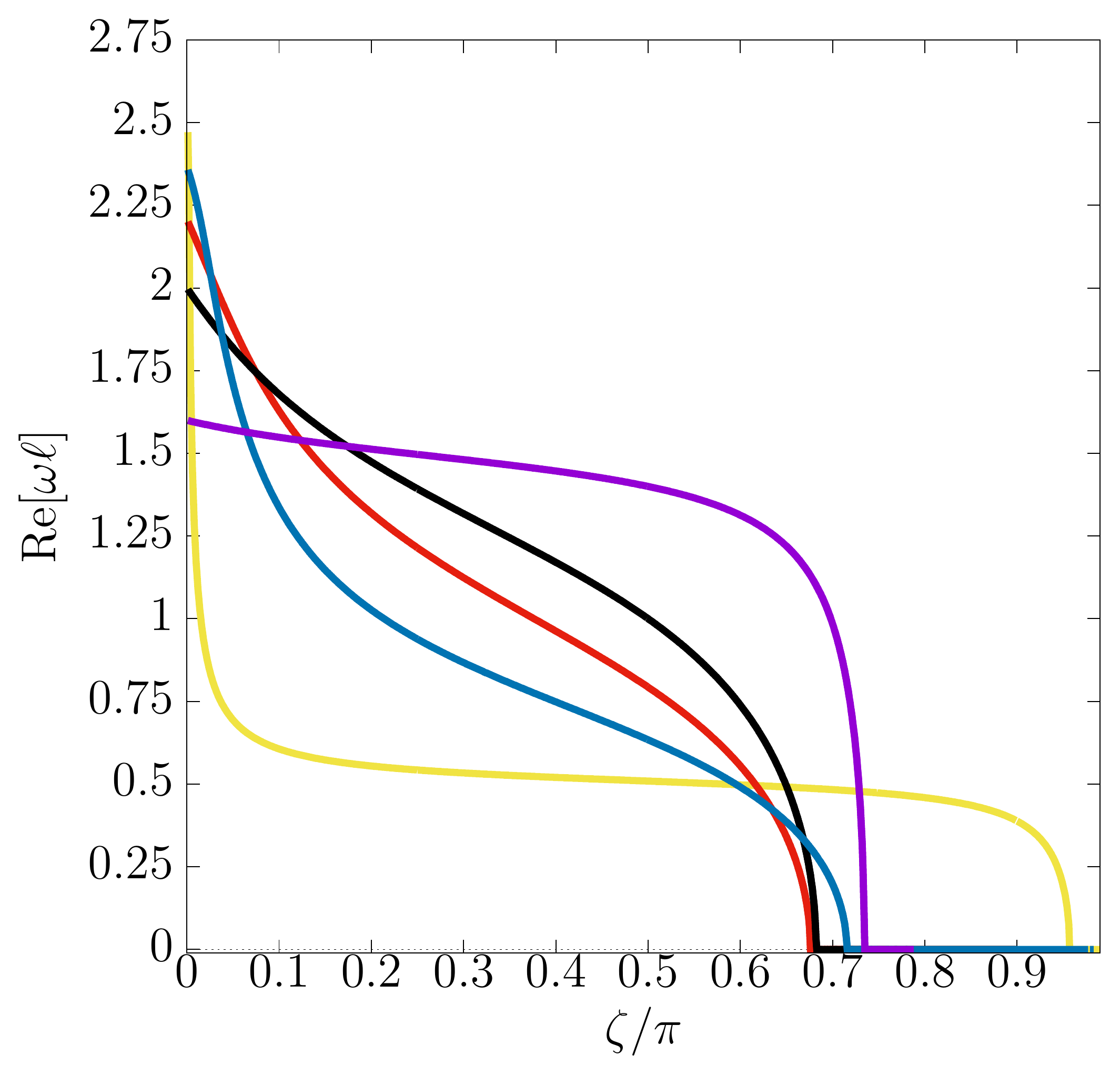}

\caption{The relation between $\zeta$ and ${\rm Re}[\omega\ell]$ of QNFs with $d=2,~l=0$, and $\mu^2\ell^2=-1.26,-1.50,-1.75,-2.00$, and $-2.24$, which are denoted by the yellow, blue, red, black, and purple lines, respectively.}
\label{OtherMassSquared}
\end{figure}

Figure~\ref{OtherMassSquared} gives the relation between $\zeta$ and ${\rm Re}[\omega\ell]$ with $d=2$, $l=0$, and $\mu^2\ell^2=-1.26,-1.50,-1.75,-2.00,-2.24$, which are denoted by yellow, blue, red, black, and purple lines, respectively. The vertical and horizontal axes denote ${\rm Re}[\omega\ell]$ and $\zeta$, respectively.  At $\zeta=0$, ${\rm Re}[{\omega}]=(3+\sqrt{9+4\mu^2\ell^2})/2$ is larger as $\mu^2\ell^2$ is increased. We have checked that for all the mass squared cases, the behaviour of $\omega$ is the same as that in Figures~\ref{rea} and \ref{ef} qualitatively. Namely, as $\zeta$ is increased for $0<\zeta<\zeta_c$, $\omega$ is real and is monotonically decreasing, and for $\zeta_c<\zeta$ it vanishes and then $\omega$ becomes pure imaginary. The critical value $\zeta_c$ becomes larger in the order $\mu^2\ell^2=-2.24,-2.00,-1.75$, while it takes smaller values in the order $\mu^2\ell^2=-1.75,-1.50,-1.26$. We have also checked that as $\mu^2\ell^2$ is further increased for $-1.26<\mu^2\ell^2<-5/4$, $\zeta_c$ monotonically increases and approaches $\pi$.

\section{Neutral and charged scalar fields in AdS black holes \label{sec:formulation}}

\subsection{Field equations, boundary conditions, and symmetries \label{sec:field_equations}}
Using the Schwarzschild-like coordinates $(t,r,\theta,\phi)$, the line element in the Reissner-Nordstr\"{o}m-AdS spacetime in 4 dimensions is written in the form
\begin{equation}
\label{BH}
\begin{split}
ds^2=-\frac{\triangle}{r^2}dt^2+\frac{r^2}{\triangle}dr^2+r^2\left(d\theta^2+\sin^2{\theta}d\phi^2\right)
\end{split}
\end{equation}
with
\begin{equation}
\triangle := r^2-2Mr+Q^2+\frac{r^4}{\ell^2}\\
=(r-r_+)(r-r_-)R(r),
\end{equation}
\begin{equation}
R(r):=1+\left(\frac{r}{\ell}\right)^2+\left(\frac{r_+}{\ell}+\frac{r_-}{\ell}\right)\left(\frac{r}{\ell}\right)+\left(\frac{r_+}{\ell}\right)^2+\left(\frac{r_-}{\ell}\right)^2+\left(\frac{r_+}{\ell}\right)\left(\frac{r_-}{\ell}\right),
\end{equation}
where $M$ is the mass and $Q$ is the charge of the black hole. There are two Killing horizons located at $r=r_+$ and $r=r_-~(0<r_-<r_+)$. 
We call $r=r_+$ and $r=r_-$ outer and inner horizons, respectively. 
The black hole is the region given by $r<r_+$. 
Hereafter, we will consider the evolution of the scalar field in the region $r_+< r$.

The equation of motion for the scalar field is given by Eq. (\ref{eomcs}) with the 1-form of the gauge field given by 
\begin{equation}
A_{\mu}dx^{\mu}={-\left(\frac{Q}{r}+\phi_0\right)}dt,
\end{equation}
where $\phi_0\in\mathbb{R}$ is an integration constant.
Since the spacetime is static and spherically symmetric, we can expand $\Psi$ in terms of spherical harmonics $Y_{lm}\left(\theta,\phi\right)$ as
\begin{equation}
\label{anz}
\Psi(t,r,\theta,\phi)=\sum_{l,m}\frac{u_{lm}(r)}{r}Y_{lm}\left(\theta,\phi\right)e^{-i\omega_{lm} t},
\end{equation}
where $\omega_{lm}\in\mathbb{R}$ is the frequency.
{Note that the above sign convention is naturally consistent with that for the gauge covariant derivative.}
Hereafter, we focus on a single set of $(l,m)$ and just write $u_{lm}(r)$ and $\omega_{lm}$ as $u(r)$ and $\omega$, respectively.

Since we discuss the dynamical properties of a scalar field in Eq. (\ref{BH}) in terms of quasinormal modes,  we now assume 
$\omega\in\mathbb{C}$.  Redefining the radial part by $\psi(r)={u(r)}/{r}$,  Eq. (\ref{eomcs}) can be written in the form
\begin{equation}
\label{req}
\left[\frac{d}{dr}\left(\triangle \frac{d}{dr}\right)+\frac{r^4}{\triangle}\left(\tilde{\omega}-\frac{eQ}{r}\right)^2-l(l+1)-\mu^2r^2\right]\psi(r)=0,
\end{equation} 
where $\tilde{\omega}:=\omega-e\phi_0$.

Introducing the tortoise coordinate $x$ by 
\begin{equation}
dx=\frac{r^2}{\triangle}dr,~x\in(-\infty,+\infty),
\end{equation}
in the limit $x\to -\infty$ or to the outer horizon, Eq. (\ref{req}) becomes
\begin{equation}
\label{neaeq}
\left[\frac{d^2}{dx^2}+\left(\tilde{\omega}-\frac{eQ}{r_+}\right)^2\right]\psi (x)\simeq 0,
\end{equation}
and the general solution of Eq. (\ref{neaeq}) behaves as
\begin{equation}
\label{gesolb}
\begin{split}
\psi(x)\sim B_{\textrm{in}}e^{-i\left(\tilde{\omega}-\frac{eQ}{r_+}\right)x}+B_{\textrm{out}}e^{i\left(\tilde{\omega}-\frac{eQ}{r_+}\right)x},~
B_{\textrm{in}},B_{\textrm{out}}\in\mathbb{C}.
\end{split}
\end{equation}
Here the subscripts ``in'' and ``out'' denote the ingoing and outgoing modes at the outer horizon, respectively.
We demand that $B_{\textrm{out}}=0$ so that the ingoing-wave condition at the outer horizon
can be satisfied.

As for the boundary condition at the conformal infinity, we expect that the classification of Theorem 1 does not change even in the presence of a black hole because that classification is based on the asymptotic behaviour of the field near the conformal infinity, although the critical values 
$\kappa_c$ and $\tilde{\kappa}_{c}$ could change because the regularity condition at the origin is replaced with the ingoing-wave condition at the black hole horizon. 
If both the scalar field and the black hole are charged, there can also appear electromagnetic superradiant instability. The electromagnetic superradiance in the asymptotically flat charged black hole will be briefly reviewed in Section~\ref{RNBH_superradiance}. It should also be noted that the equation of motion for the charged scalar field cannot be written in the Schr\"{o}dinger form as in Eq. (\ref{RADEQ}). Hence, Theorem 1 does not apply for the charged scalar field. 

We simultaneously demand the ingoing-wave condition $B_{\textrm{out}}=0$ at the horizon and the Robin boundary condition (\ref{bc}) 
with the asymptotic form (\ref{asympa}) at the conformal infinity.
In general, this can be met only a discrete set of complex frequencies. These are called QNFs. Here it is useful to see what symmetry this system has.
Eq.~(\ref{req}) trivially admits symmetry $(\omega,eQ,\psi) \to (-\omega,-eQ,\psi)$, while we can find 
other symmetries $(\omega,eQ,\psi) \to (\omega^{*},eQ,\psi^{*})$ and $(\omega,eQ,\psi) \to (-\omega^{*},-eQ,\psi^{*})$. 
Since $\kappa\in \mathbb{R}$ and the asymptotic form (\ref{asympa}) does not depend on $\tilde{\omega}$, if the Robin boundary condition is
satisfied by $\psi$, it is also by $\psi^{*}$. Therefore, all these three transformations are consistent with the Robin boundary condition at the conformal infinity.   
However, only  the transformation $(\omega,eQ,\psi) \to (-\omega^{*},-eQ,\psi^{*})$ among the three can satisfy the ingoing-wave condition at the outer horizon. 
Therefore, only this transformation remains as the symmetry of the system, which necessarily involves charge 
conjugation. In other words, $\omega\to -\omega^{*}$ is no longer the symmetry of the system for the charged scalar field if we fix $eQ$.

\subsection{Matched asymptotic expansion for small AdS black holes}
To treat Eq. (\ref{req}) analytically, we apply the matched asymptotic expansion method.  The strategy is as follows. 
We first assume $ -9/4<\mu^2\ell^2<-5/4$, for which the Robin boundary condition applies.
We also assume that the black hole radius is much smaller than the AdS length scale: $r_+\ll \ell$. 
Then, we consider two spatial regions outside the black hole. We call the regions given by $r_{+}<r\ll\ell$ and  $r_+\ll r$ near and far regions, respectively.
Then, we obtain an approximate analytic solution which satisfies the imposed boundary condition in each region. 
We will show that the two regions have an overlapping region satisfying $r_+\ll r\ll \ell$,  where the both approximate solutions are valid.
Then, we match the two solutions there and the matching condition gives an eigenvalue equation for the frequency.
Thus, we obtain QNFs and the approximate analytic solution which satisfies the boundary conditions 
both at the horizon and at the conformal infinity.

\subsubsection{Near region: $r_+<r \ll \ell$}
In the near region, since 
$\triangle\simeq (r-r_+)(r-r_-)$,
the metric is approximated by the Reissner-Nordstr\"{o}m metric.
Now we introduce a new coordinate,
\begin{equation}
z=\frac{r-r_+}{r-r_-},
\end{equation}
and a function $f(z)$ such that
\begin{equation}
\psi(z)=z^{i\sigma}(1-z)^{l+1}f(z),
\end{equation}
where 
\begin{equation}
\sigma=\frac{\left(\tilde{\omega}-\frac{eQ}{r_+}\right){r_+}^2}{r_+-r_-}.
\end{equation}
Then Eq. (\ref{req}) is reduced to
\begin{equation}
\label{Hyn}
z(1-z)\frac{d^2}{dz^2}f(z)+\left\{c-(a+b+1)z\right\}\frac{d}{dz}f(z)-\left(ab+\mathcal{E}_1(r)\right)f(z)=0
\end{equation}
with
\begin{equation}
\label{abc}
a=2i\sigma+l+1,b=l+1,c=2i\sigma+1,
\end{equation}
and 
\begin{equation}
\begin{split}
\label{fue1}
\mathcal{E}_1(r):=\mu^2r^2\left(\frac{r-r_-}{r_+-r_-}\right)+\frac{1}{\left(r_+-r_-\right)\left(r-r_+\right)}\left\{{r_+}^4\left(\tilde{\omega}-\frac{eQ}{r_+}\right)^2-r^4\left(\tilde{\omega}-\frac{eQ}{r}\right)^2\right\}.
\end{split}
\end{equation}
In Appendix~\ref{sec:validity_matching}, it is shown that $\mathcal{E}_{1}(r)\ll 1$ in the near region. 

The two independent solutions of Eq. (\ref{Hyn}) in the region $1<x\leq x_0$ can be written in terms of the Gaussian hypergeometric functions $F(~,~;~;z)$ as
\begin{equation}
f(z)=F(a,b;c;z)~,~z^{1-c}F(1-c+a,1-c+b;2-c;z).
\end{equation}
We thus obtain the general solution of Eq. (\ref{req}) in the region $x\leq x_0$,
\begin{equation}
\begin{split}
\label{eq1}
\psi(z)=&Az^{i\sigma}(1-z)^{l+1}F(a,b;c;z)+Bz^{-i\sigma}(1-z)^{l+1}F(1-c+a,1-c+b;2-c;z),
\end{split}
\end{equation}
where $A$ and $B$ are arbitrary constants. The first and second terms on the right-hand side are the outgoing and ingoing modes, respectively. Now we impose the ingoing-wave condition on Eq. ($\ref{eq1}$), so that the first term should vanish. Thus, we obtain the following approximate analytic solution in the region $x\leq x_0$:
\begin{equation}
\begin{split}
\label{soln}
\psi(z)=Bz^{-i\sigma}(1-z)^{l+1}F(1-c+a,1-c+b;2-c;z).
\end{split}
\end{equation}

\subsubsection{Far region: $r_+\ll r$}
In the far region, since 
$\triangle\simeq r^2\left(1+{r^2}/{\ell^2}\right)$,
the metric is approximated by the AdS metric.
Using a new variable
\begin{equation}
y=1+\frac{r^2}{\ell^2},
\end{equation} 
which is equivalent to Eq. (\ref{yz}), and Eq. (\ref{anza}) with the replacement of $\omega$ with $\tilde{\omega}$, Eq. (\ref{req}) is reduced to the equation for $g(y)$ given by
\begin{equation}
\label{Hyf}
y(1-y)\frac{d^2}{dy^2}g(y)+\left\{\gamma-(\alpha+\beta+1)y\right\}\frac{d}{dy}g(y)-\left(\alpha\beta +\mathcal{E}_2(r)\right)g(y)=0,
\end{equation}
where $\alpha,\beta$, and $\gamma$ are defined by Eq. (\ref{abca}) with the replacement of $\omega$ with $\tilde{\omega}$ and 
\begin{equation}
\begin{split}
\mathcal{E}_2(r):=\frac{1}{4}\left(eQ\right)\left(\frac{r}{\ell}\right)^{-1}\left(1+\frac{r^2}{\ell^2}\right)^{-1}\left\{2\left(\tilde{\omega}\ell\right)-\left(eQ\right)\left(\frac{r}{\ell}\right)^{-1}\right\}.
\end{split}
\end{equation}
In Appendix~\ref{sec:validity_matching}, it is shown that $|\mathcal{E}_{2}(r)|\ll 1$ in the far region. 

Near the conformal infinity, the scalar field behaves as
\begin{equation}
\label{asymp}
\psi(r)\sim C\left(\frac{r}{\ell}\right)^{-\frac{3}{2}-\frac{1}{2}\sqrt{9+4\mu^2\ell^2}}+D\left(\frac{r}{\ell}\right)^{-\frac{3}{2}+\frac{1}{2}\sqrt{9+4\mu^2\ell^2}},
\end{equation}
where $C$ and $D$ are arbitrary constants. 
Since we have assumed $-9/4< \mu^2\ell^2<-5/4$, we impose the Robin boundary condition Eq. (\ref{bc}). 
Thus, we find Eq. (\ref{solfa}) with the replacement of $\omega$ by $\tilde{\omega}$ as an approximate solution in the far region.

\subsubsection{Matching in the overlapping region}
Next we match the near-region and far-region approximate solutions, given by Eqs.~(\ref{soln}) and (\ref{solfa}), respectively, 
in the overlapping region. 
In Appendix \ref{sec:validity_matching}, we show that there really exists an overlapping region, where both the near-region and far-region approximate solutions are valid, 
if $\tilde{\omega}{\ell}=O(1)$ and $eQ=o(\epsilon^{1/3+\delta})$ with $\epsilon=r_{+}/\ell$ and $0<\delta<2/3$.
We consider the asymptotic behaviour of them there. First, we see the asymptotic behaviour of 
the near-region solution (\ref{soln}) at $z\sim 1$.  
As shown in Appendix \ref{sec:Gaussian_hypergeometric_functions}, in the limit of $z\to 1$, Eq. (\ref{soln}) 
behaves as
\begin{equation}
\begin{split}
\label{soln2}
\psi(z)\sim B\Gamma(1-2i\sigma)\left[B_1(\tilde{\omega},\sigma)r^{-l-1}+B_2(\tilde{\omega},\sigma)r^l\right],
\end{split}
\end{equation}
where
\begin{equation}
\begin{split}
\label{B12}
B_1(\tilde{\omega},\sigma)=\frac{\Gamma(-2l-1)(r_+-r_-)^{l+1}}{\Gamma(-2i\sigma-l)\Gamma(-l)}~,~B_2(\tilde{\omega},\sigma)=\frac{\Gamma(2l+1)(r_+-r_-)^{-l}}{\Gamma(-2i\sigma+l+1)\Gamma(l+1)}.
\end{split}
\end{equation}
Note that the functions $B_1(\tilde{\omega},\sigma)$ and $B_2(\tilde{\omega},\sigma)$ satisfy the relation $B_1(\tilde{\omega},\sigma)=B_1^*(-\tilde{\omega}^*,-\sigma^{*})$ and $B_2(\tilde{\omega},\sigma)=B_2^*(-\tilde{\omega}^*,-\sigma^{*})$, respectively.

Next, as derived in Appendix \ref{sec:Gaussian_hypergeometric_functions}, in the limit of $y\to 1$, 
the asymptotic form of the far-region solution (\ref{solfa}) is given by Eq. (\ref{solf1a}). 
 Using $r=1/\tan\chi$, we find that Eq. (\ref{solf1a}) has the same form as Eq. (\ref{soln2}). 

Since both $y\sim 1$ and $z\sim 1$ are satisfied for $x_{1}<x<x_{0}$, we can match Eq. (\ref{soln2}) and Eq. (\ref{solf1a}) and obtain
\begin{equation}
\label{match}
\frac{B_1(\tilde{\omega},\sigma)}{B_2(\tilde{\omega},\sigma)}=\ell^{2l+1}\frac{D_1(\tilde{\omega},\kappa)}{D_2(\tilde{\omega},\kappa)},
\end{equation}
where $D_1(\tilde{\omega},\kappa)$ and $D_2(\tilde{\omega},\kappa)$ are defined by Eq. (\ref{D12a}). The above gives the relation between $\kappa$ and $\tilde{\omega}$. 
It is shown in Appendix \ref{sec:symmetry} that the above equation is consistent with the symmetry $(\omega,eQ)\to (-\omega^{*},-eQ)$. 

\section{Analytic results \label{sec:analytic_results}}
We analytically solve the matching condition~\eqref{match} under assumption $|{\rm Im}[\tilde{\omega}\ell]|\ll1$  
and discuss stability in terms of the obtained QNFs. We explain here our results without explicit calculations, and complement them in Appendix~\ref{appendix:Theexplicitcalculations}.

\subsection{Approximate forms of the matching condition under $|{\rm Im}[\tilde{\omega}\ell]|\ll1$}
We shall derive approximate forms of the matching condition~\eqref{match} under the assumption $|{\rm Im}[\tilde{\omega}\ell]|\ll1$. The real and imaginary parts of Eq.~\eqref{match} are
\begin{equation}
\label{Rematch}
{\rm Re}\left[D_1\right]{\rm Re}\left[D_2\right]+{\rm Im}\left[D_1\right]{\rm Im}\left[D_2\right]=\left({\rm Re}\left[D_2\right]^2+{\rm Im}\left[D_2\right]^2\right){\rm Re}\left[\frac{B_1}{B_2}\right],
\end{equation}
and
\begin{equation}
\label{Immatch}
{\rm Im}\left[D_1\right]{\rm Re}\left[D_2\right]-{\rm Re}\left[D_1\right]{\rm Im}\left[D_2\right]=\left({\rm Re}\left[D_2\right]^2+{\rm Im}\left[D_2\right]^2\right){\rm Im}\left[\frac{B_1}{B_2}\right],
\end{equation}
respectively. By solving these for ${\rm Re}[D_1]$ and ${\rm Im}[D_1]$ simultaneously, we obtain
\begin{equation}
\begin{split}
\label{ReD1}
&{\rm Re}\left[D_1\right]={\rm Re}\left[D_2\right]{\rm Re}\left[\frac{B_1}{B_2}\right]-{\rm Im}\left[D_2\right]{\rm Im}\left[\frac{B_1}{B_2}\right],
\end{split}
\end{equation}
and
\begin{equation}
\begin{split}
\label{ImD1}
&{\rm Im}\left[D_1\right]={\rm Im}\left[D_2\right]{\rm Re}\left[\frac{B_1}{B_2}\right]+{\rm Re}\left[D_2\right]{\rm Im}\left[\frac{B_1}{B_2}\right].
\end{split}
\end{equation}
As will be shown later, under $|{\rm Im}[\tilde{\omega}\ell]|\ll1$, $D_1(\tilde{\omega},\kappa)$ and $D_2(\tilde{\omega},\kappa)$ are written in the form
\begin{equation}
\begin{split}
\label{SD1}
D_1\left(\tilde{\omega},\kappa\right)=&\Sigma_{1,R}-i\frac{\Sigma_{1,I}}{2}{\rm Im}\left[\tilde{\omega}\ell\right]+\mathcal{O}\left(\left({\rm Im}\left[\tilde{\omega}\ell\right]\right)^2\right),
\end{split}
\end{equation}
and 
\begin{equation}
\label{SD2}
D_2\left(\tilde{\omega},\kappa\right)=\frac{2^{2l+1}\left(-1\right)^{l+1}}{\left(2l+1\right)!!\left(2l-1\right)!!}\left[\Sigma_{2,R}-i\frac{\Sigma_{2,I}}{2}{\rm Im}\left[\tilde{\omega}\ell\right]\right]+\mathcal{O}\left(\left({\rm Im}\left[\tilde{\omega}\ell\right]\right)^2\right),
\end{equation} 
where $\Sigma_{1,R}$, $\Sigma_{1,I}$, $\Sigma_{2,R}$, and $\Sigma_{2,I}$ are real functions, and depend on $l,\mu^2\ell^2$, and ${\rm Re}[\tilde{\omega}\ell]$ but not $r_+$, $r_-$, $eQ$, and ${\rm Im}[\tilde{\omega}\ell]$. The explicit forms of them are given in Eqs.~\eqref{Sigma1R},~\eqref{Sigma1I},~\eqref{Sigma2R}, and~\eqref{Sigma2I}, respectively. Using Eqs.~\eqref{SD1} and~\eqref{SD2}, Eqs.~\eqref{ReD1} and~\eqref{ImD1} are rewritten as
\begin{equation}
\begin{split}
\label{matchingReal}
\Sigma_{1,R}=\frac{2^{2l+1}\left(-1\right)^{l+1}}{(2l+1)!!(2l-1)!!}\left(\Sigma_{2,R}{\rm Re}\left[\frac{B_1}{B_2}\right]+\frac{{\rm Im}\left[\tilde{\omega}\ell\right]}{2}\Sigma_{2,I}{\rm Im}\left[\frac{B_1}{B_2}\right]\right)\ell^{-2l-1}+\mathcal{O}\left(\left({\rm Im}\left[\tilde{\omega}\ell\right]\right)^2\right),
\end{split}
\end{equation}
and
\begin{equation}
\begin{split}
\label{matchingImaginary1}
-\frac{{\rm Im}\left[\tilde{\omega}\ell\right]}{2}\Sigma_{1,I}=&\frac{2^{2l+1}\left(-1\right)^{l+1}}{(2l+1)!!(2l-1)!!}\left(\Sigma_{2,R}{\rm Im}\left[\frac{B_1}{B_2}\right]-\frac{{\rm Im}\left[\tilde{\omega}\ell\right]}{2}\Sigma_{2,I}{\rm Re}\left[\frac{B_1}{B_2}\right]\right)\ell^{-2l-1}\\
&+\mathcal{O}\left(\left({\rm Im}\left[\tilde{\omega}\ell\right]\right)^2\right).
\end{split}
\end{equation}
We note here that ${\rm Re}[{B_1}/{B_2}]=\mathcal{O}({\rm Im}[\tilde{\omega}\ell])$ and ${\rm Im}[{B_1}/{B_2}]=\mathcal{O}(({\rm Im}[\tilde{\omega}\ell])^0)$ as seen in Eqs.~\eqref{RB1B2} and~\eqref{IB1B2}. Hence, the second term in the bracket of the right-hand side in Eq.~\eqref{matchingImaginary1} is $\mathcal{O}(({\rm Im}[\tilde{\omega}\ell])^2)$. Thus, Eq.~\eqref{matchingImaginary1} is
\begin{equation}
\begin{split}
\label{matchingImaginary}
-\frac{{\rm Im}\left[\tilde{\omega}\ell\right]}{2}\Sigma_{1,I}=&\frac{2^{2l+1}\left(-1\right)^{l+1}}{(2l+1)!!(2l-1)!!}\Sigma_{2,R}{\rm Im}\left[\frac{B_1}{B_2}\right]\ell^{-2l-1}+\mathcal{O}\left(\left({\rm Im}\left[\tilde{\omega}\ell\right]\right)^2\right).
\end{split}
\end{equation}
We thus obtain the approximate forms of the matching condition~\eqref{match} for $|{\rm Im}[\tilde{\omega}\ell]|\ll1$, i.e., Eqs.~\eqref{matchingReal} and~\eqref{matchingImaginary}.

\subsection{Real part of QNFs}
We first discuss ${\rm Re}[\tilde{\omega}\ell]$. We note that ${\rm Re}[{B_1}/{B_2}]\ell^{-2l-1}=\mathcal{O}((r_+/\ell)^{2(l+1)})$ and ${\rm Im}[{B_1}/{B_2}]\ell^{-2l-1}=\mathcal{O}((r_+/\ell)^{2(l+1)})$ as seen in Eqs.~\eqref{RB1B2} and~\eqref{IB1B2}. Then, it can be seen from Eqs.~\eqref{matchingReal} and~\eqref{matchingImaginary} that
\begin{equation}
\label{weakregularitycondition1}
\Sigma_{1,R}=\mathcal{O}\left(\left(\frac{r_+}{\ell}\right)^{2(l+1)}\right),
\end{equation}
and
\begin{equation}
\label{weakregularitycondition2}
\Sigma_{1,I}=\mathcal{O}\left(\left(\frac{r_+}{\ell}\right)^{2(l+1)}\right),
\end{equation}
because $\Sigma_{2,R}$ and $\Sigma_{2,I}$ are independent of $r_+$. In the limit of $r_+\to 0$, these equations are reduced to the equation determining the QNF of the mode, which satisfies the regularity condition at the origin, in the pure AdS spacetime, i.e., Eq.~\eqref{rega}. Hence, Eqs.~\eqref{weakregularitycondition1} and~\eqref{weakregularitycondition2} imply
\begin{equation}
\label{realpartofQNF}
{\rm Re}\left[\tilde{\omega}\right]={\rm Re}\left[\omega_{AdS}\right]+\mathcal{O}\left(\left(\frac{r_+}{\ell}\right)^{2(l+1)}\right),
\end{equation}
where ${\rm Re}\left[\omega_{AdS}\right]$ is the real part of the QNF in the AdS spacetime. The relation between the QNFs in the AdS spacetime and the boundary condition parameter $\zeta={\rm ArcTan}[1/\kappa]$ is given in Figures~\ref{rea} and~\ref{ef}. Thus, the real part of the QNFs in the present spacetime coincides with that in the pure AdS spacetime within $\mathcal{O}(({r_+}/{\ell})^{2(l+1)})$.

\subsection{Imaginary part of QNFs}
\begin{figure}[htbp]
\centering
\subfigure[$\mu^2\ell^2=-2$]{\includegraphics[scale=0.60] {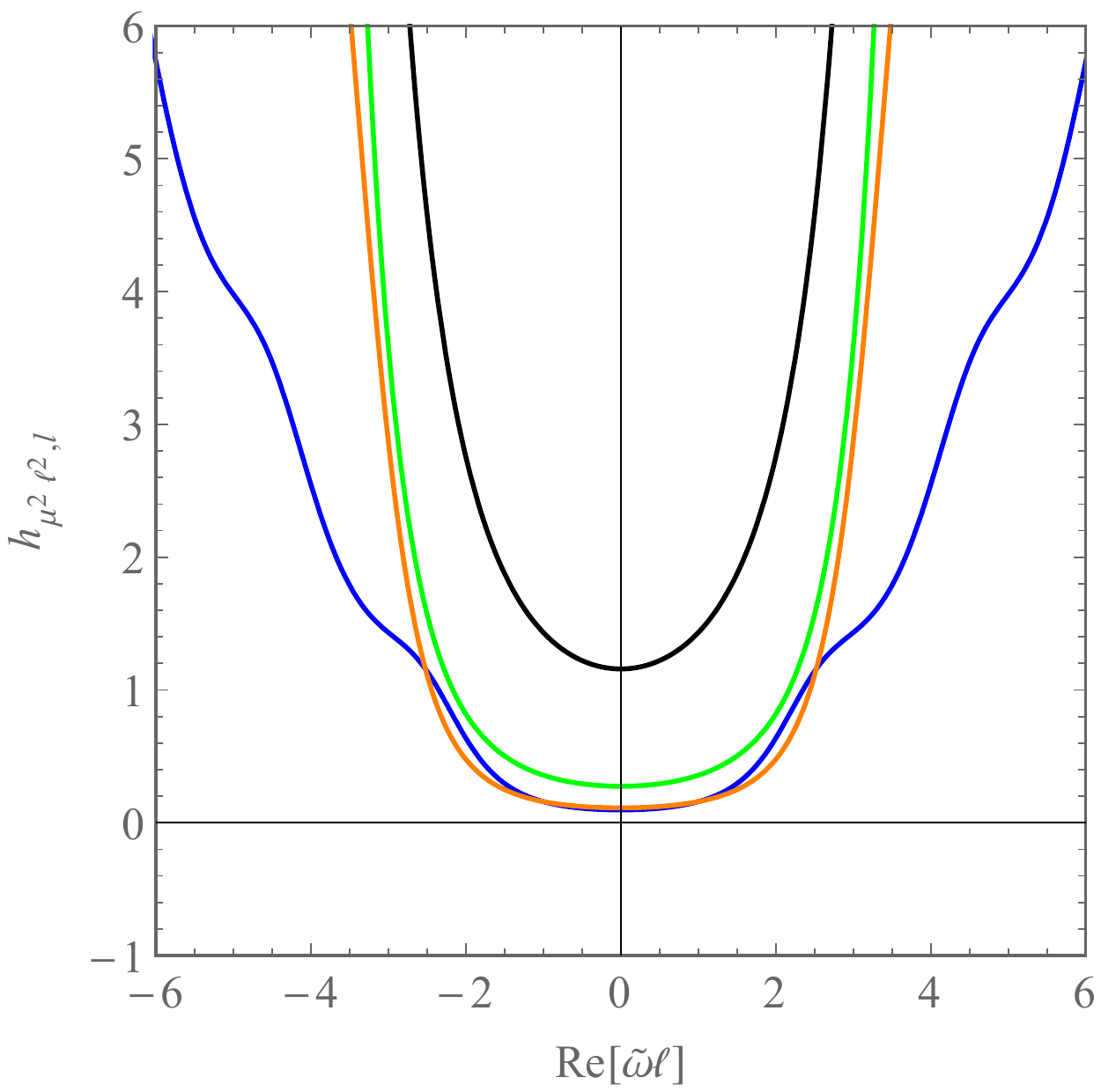}\label{SS1}}
\subfigure[$l=0$]{\includegraphics[scale=0.61] {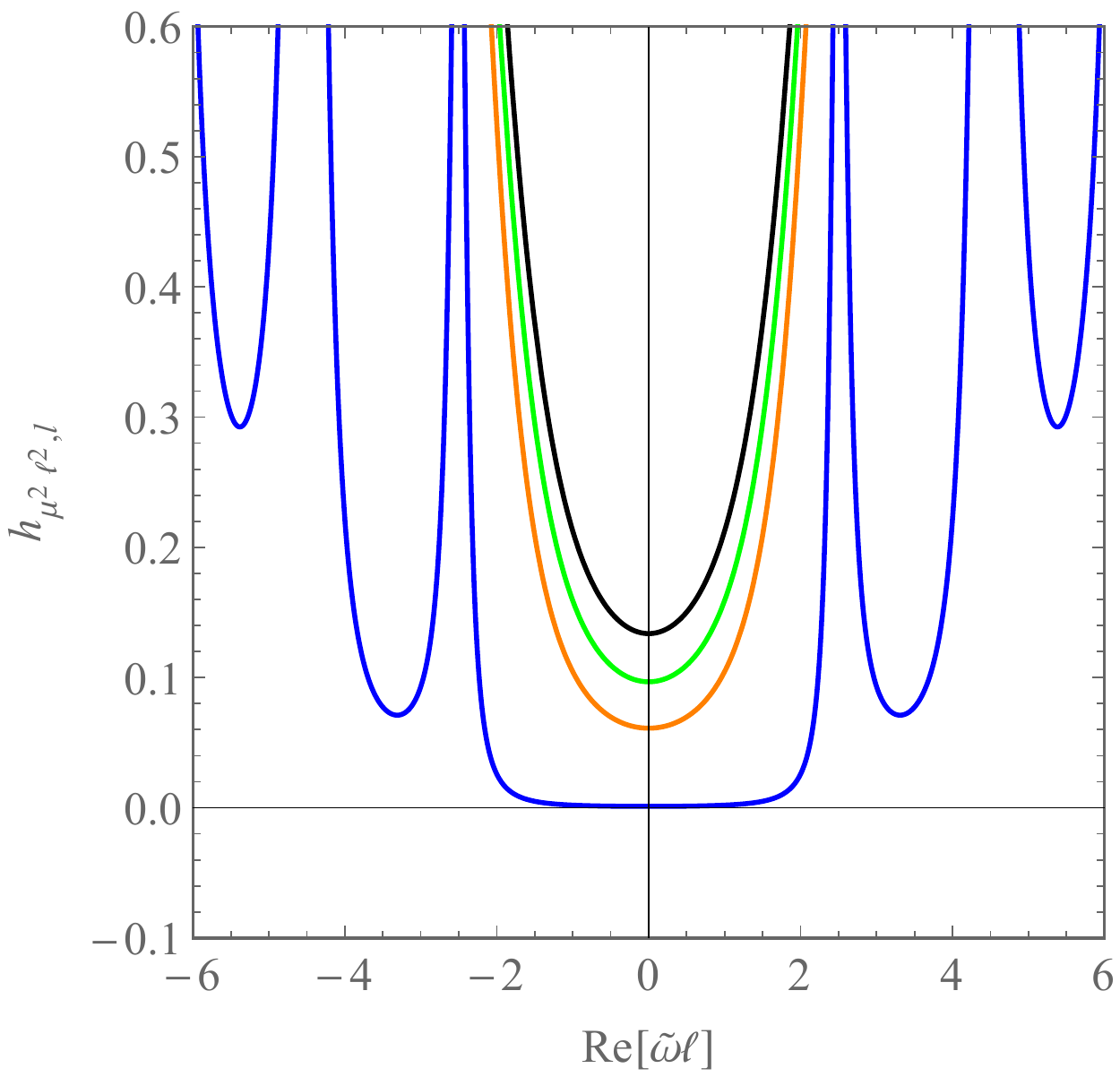}\label{SS2}}
\caption{{\it left panel}: the relation between ${\rm Re}[\tilde{\omega}\ell]$ and $h_{\mu^2\ell^2,l}$ for $\mu^2\ell^2=-2$ and $l=0,1,2,3$, which are denoted by the blue, orange, green, and black lines, respectively. {\it right panel}: the relation between ${\rm Re}[\tilde{\omega}\ell]$ and $h_{\mu^2\ell^2,l}$ for $l=0$ and $\mu^2\ell^2=-1.26,-1.75,-2.00,-2.24$, which are denoted by the blue, orange, green, and red lines, respectively. Note that although the curves get out of the plotted region for some range of ${\rm Re}[\tilde{\omega}\ell]$, we confirm that the values of $h_{\mu^2\ell^2,l}$ are positive and finite.} 
\end{figure}

We next discuss ${\rm Im}[\tilde{\omega}\ell]$. Using the explicit form of ${\rm Im}[B_1/B_2]$, which is given by Eq.~\eqref{IB1B2}, Eq.~\eqref{matchingImaginary} is 
\begin{equation}
\begin{split}
\label{fulldelta}
{\rm Im}\left[\tilde{\omega}\right]
=&-\frac{\mathcal{A}\left({\rm Re}\left[\tilde{\omega}\right]-\frac{eQ}{r_+}\right)}{{\rm Re}\left[\tilde{\omega}\ell\right]}h_{\mu^2\ell^2,l}+\mathcal{O}\left(\left({\rm Im}\left[\tilde{\omega}\ell\right]\right)^2\right),
\end{split}
\end{equation}
where
\begin{equation}
\begin{split}
\label{mathcalA}
\mathcal{A}=&2^{2l+3}\left(\frac{{r_+}}{\ell}\right)^2\left(\frac{r_+}{\ell}-\frac{r_-}{\ell}\right)^{2l}\frac{\left(l!\right)^2}{\left(2l\right)!\left(2l+1\right)!(2l+1)!!(2l-1)!!}\prod_{k=1}^{l}\left(k^2+4\sigma^2 \right),
\end{split}
\end{equation}
and
\begin{equation}
\begin{split}
\label{functionh}
h_{\mu^2\ell^2,l}=&\left(-1\right)^{l+1}{\rm Re}\left[\tilde{\omega}\ell\right]\frac{\Sigma_{2,R}}{\Sigma_{1,I}}.
\end{split}
\end{equation}
It follows from Eq.~\eqref{fulldelta} that ${\rm Im}[\tilde{\omega}\ell]$ vanishes if ${\rm Re}[\tilde{\omega}]=eQ/r_+$. Hence, there exists a purely oscillating mode if ${\rm Re}[\tilde{\omega}]=eQ/r_+$ for the charged scalar field. For the neutral field, ${\rm Re}[{\omega}]=0$ does not necessarily imply the existence of a static mode because $h_{\mu^2\ell^2,l}$ in Eq.~\eqref{functionh} is a finite positive value at ${\rm Re}[{\omega}]=0$ as will be seen in Figures~\ref{SS1} and~\ref{SS2}.
 
In order to determine the sign of ${\rm Im}[\tilde{\omega}\ell]$ in Eq.~\eqref{fulldelta}, we have to discuss that of $h_{\mu^2\ell^2,l}$ in Eq.~\eqref{functionh}. With Eq.~\eqref{weakregularitycondition1}, the boundary condition parameter $\kappa$ is written in the form of a function of ${\rm Re}[\tilde{\omega}\ell]$, $\mu^2\ell^2$, and $l$. Using that, the leading term of $h_{\mu^2\ell^2,l}$ are expressed by ${\rm Re}[\tilde{\omega}\ell],~\mu^2\ell^2$, and $l$. In Figures~\ref{SS1} and~\ref{SS2}, we numerically show that $h_{\mu^2\ell^2,l}$ is positive. The vertical and horizontal axes denote the value of $h_{\mu^2\ell^2,l}$, respectively. In Figure~\ref{SS1}, we fix $\mu^2\ell^2=-2$ but $l=0,1,2,3$ which are denoted by the blue, orange, green, and black lines, respectively, while in Figure~\ref{SS2}, we fix $l=0$ but $\mu^2\ell^2=-1.26,-1.75,-2.00,-2.24$ which are denoted by the blue, orange, green, and black lines, respectively. We note that the blue line in Figure~\ref{SS2} is in $h_{\mu^2\ell^2,l}>0$. Outside the plotted region as well, the qualitative properties do not change, i.e., $h_{\mu^2\ell^2,l}$ increases with oscillation as $|{\rm Re}[\tilde{\omega}\ell]|$ is increased. We can also see $\left(-1\right)^{l+1}\Sigma_{2,R}/\Sigma_{1,I}\to-\left(-1\right)^{l+1}\Sigma_{2,R}/\Sigma_{1,I}$ under ${\rm Re}[\tilde{\omega}]\to-{\rm Re}[\tilde{\omega}]$. We will explicitly show it in Appendix~\ref{proof}.

Thus, Eq.~\eqref{fulldelta} shows that for the charged scalar field, the growing modes appear if for $eQ>0$,
\begin{equation}
\label{superradiacecondition}
0<{\rm Re}[\tilde{\omega}]<\frac{eQ}{r_+},
\end{equation}
or for $eQ<0$,
\begin{equation}
\label{superradiacecondition2}
\frac{eQ}{r_+}<{\rm Re}[\tilde{\omega}]<0,
\end{equation}
are satisfied, while other modes are stable. These are conditions for electromagnetic supperadiace to occur. Hence, we interpret that this is superradiant instability in the present system. We will straightforwardly derive Eqs.~\eqref{superradiacecondition} and~\eqref{superradiacecondition2} from Eq.~\eqref{req} and discuss the physical interpretation in Section~\ref{sec:interpretation}. As for the neutral field, all of the modes decay. We further can see the symmetry $(\tilde{\omega},eQ)\to(-\tilde{\omega}^*,-eQ)$ in Eq.~\eqref{fulldelta}. 

\section{Numerical results \label{sec:numerical_results}}
We numerically solve Eq. (\ref{match}) by the Newton-Raphson method. We present the result in this section. Hereafter, we fix the mass parameter $\mu^{2}\ell ^{2}=-2$ except for subsection~\ref{Other mass squared cases} and focus the s-wave, i.e., $l=0$. Our numerical results are consistent with the analytic results in the previous section.

\subsection{Real part of QNFs}
\begin{figure}[tbp]
\centering
\subfigure[AdS black hole]{
\includegraphics[scale=0.78] 
{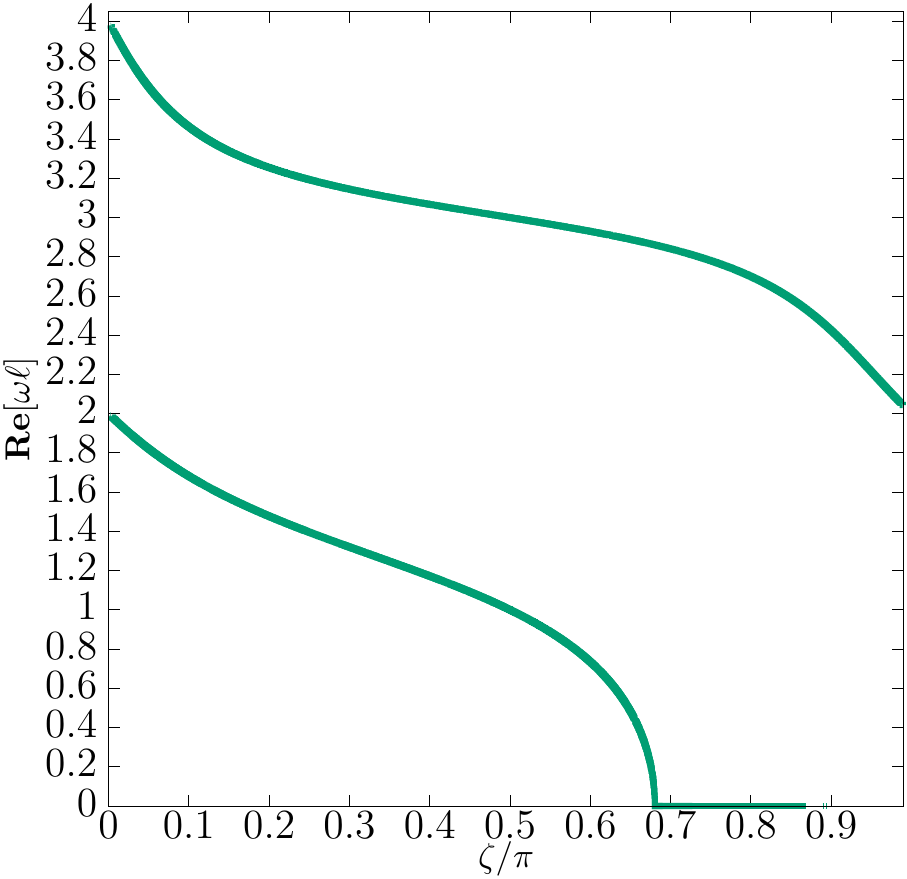}
\label{res}}
\hspace{0.3cm}
\subfigure[AdS spacetime]{
\includegraphics[scale=0.78] 
{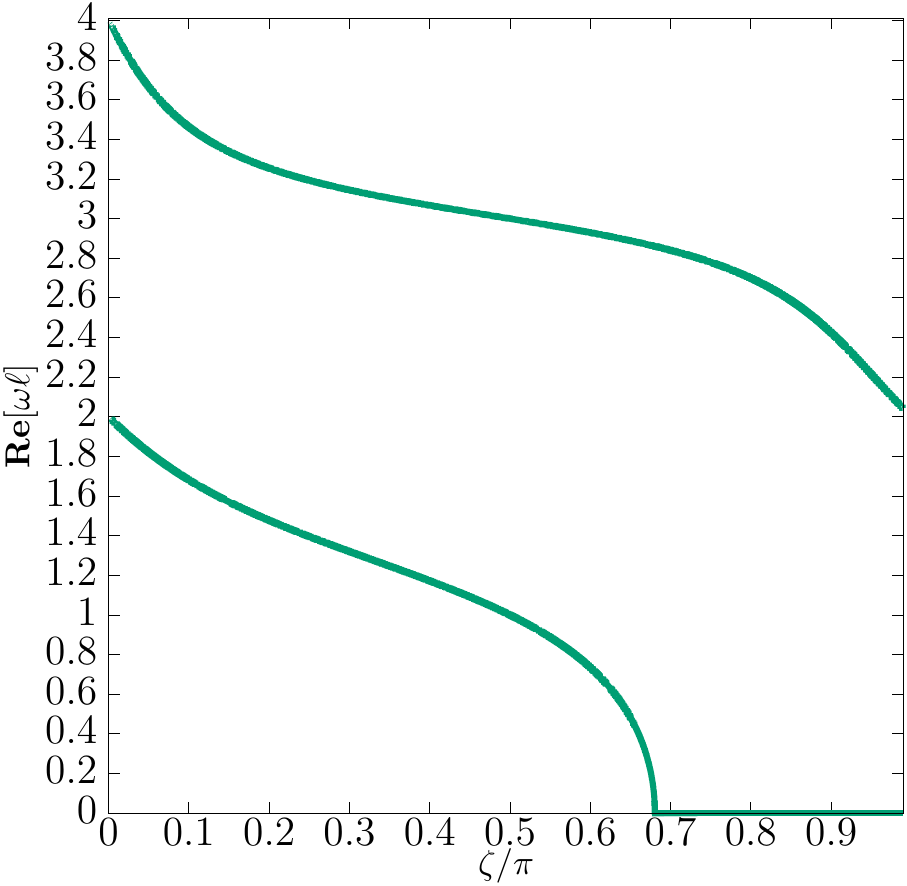}
\label{resads}}
 \caption{The relation between $\zeta$ and $\textrm{Re}[\omega\ell]\geq0$ for the neutral field in the AdS black hole 
with $\left(r_+,r_-\right)=\left(0.01\ell, 0.001\ell\right)$ and the AdS spacetime.}
\end{figure}

Figures \ref{res} and \ref{resads} give the relation between $\zeta$ and
$\textrm{Re}[\omega]\geq0$ for the neutral field in the AdS black hole with $\left(r_+,r_-\right)=\left(0.01\ell,0.001\ell\right)$ and the AdS spacetime, respectively. 
The vertical and horizontal axes denote $\textrm{Re}[\omega\ell]$ and the normalised parameter $\zeta/\pi$, respectively.
The symmetry $\omega\to -\omega$ implies that the whole graph has reflectional symmetry with respect to the line $\textrm{Re}[\omega\ell]=0$.

\begin{figure}[htbp]
\centering
\subfigure[$\textrm{Re}(\omega) \ge 0$]{\includegraphics[scale=0.74] {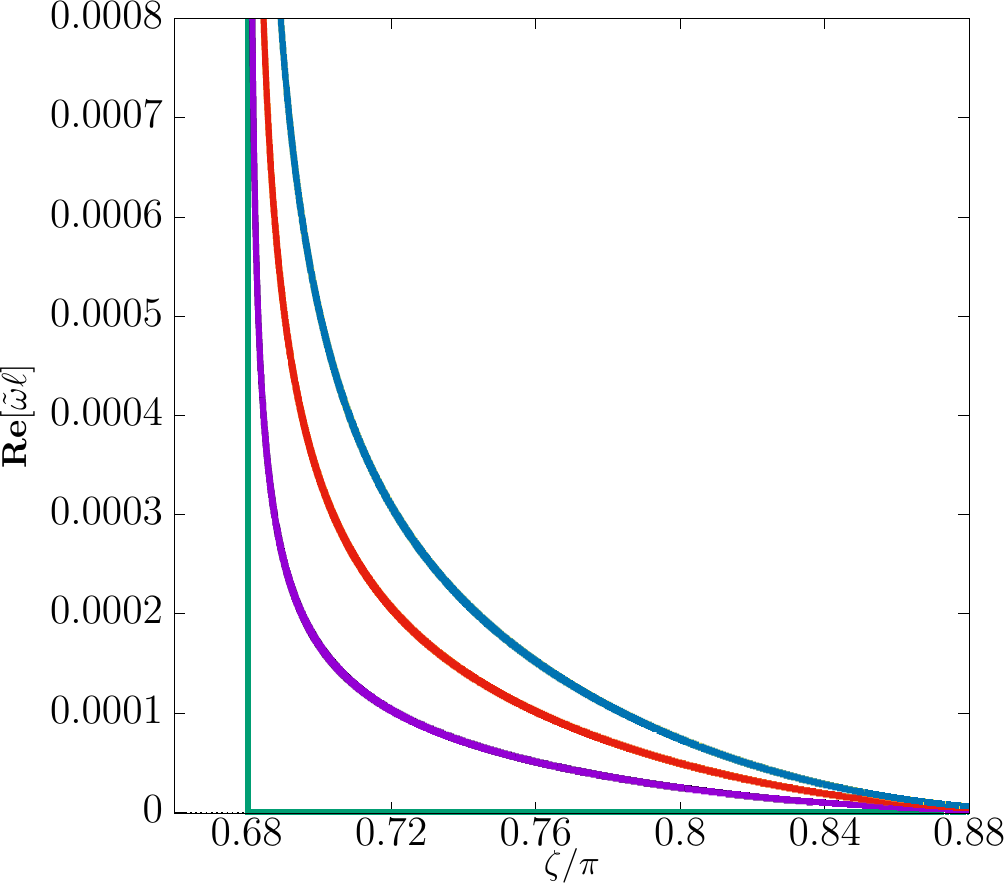}\label{resf}}
\hspace{0.3cm}
\subfigure[$\textrm{Re}(\omega) \le 0$]{
\includegraphics[scale=0.74] 
{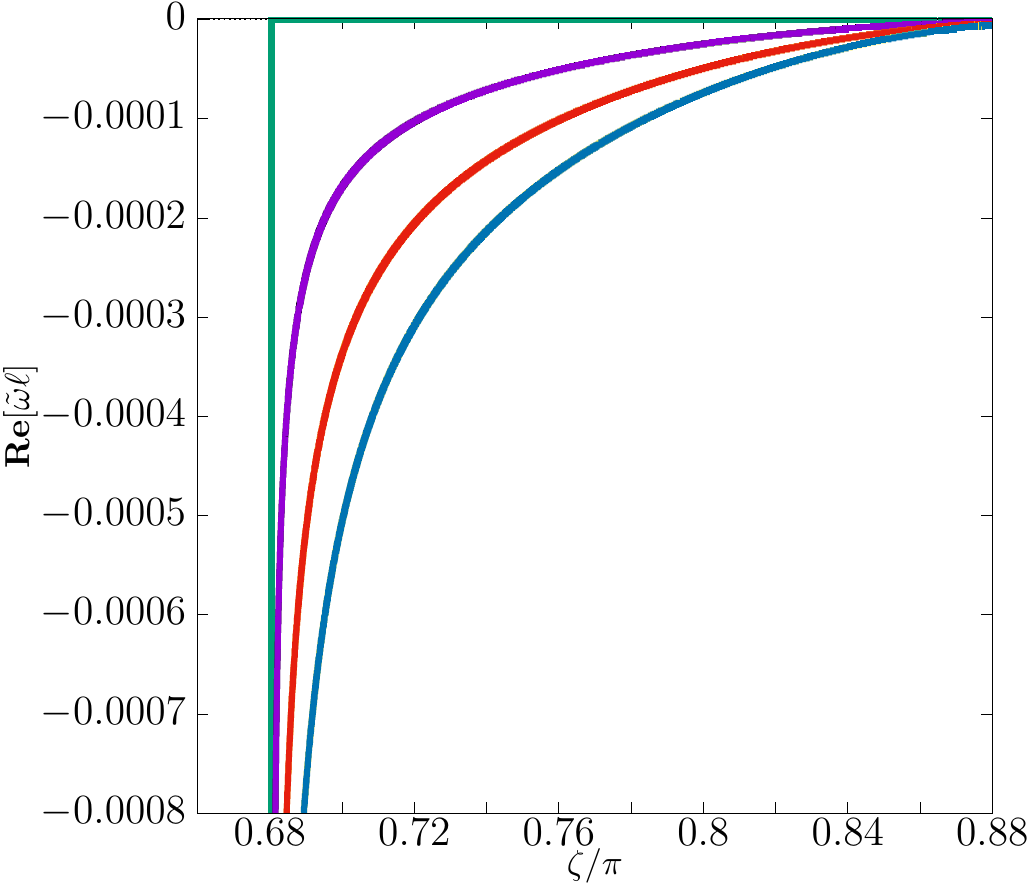}
\label{resf2}}
\caption{Same as Figure~\ref{res} but magnified near $\zeta= 0.68\pi$ for the neutral and charged fields in the AdS black hole 
with $\left(r_+,r_-\right)=\left(0.01\ell, 0.001\ell\right)$. The green, purple, red, and blue lines correspond to the result with 
{$eQ=0,0.01,0.02$, and $0.03$}
, respectively.} 
\end{figure}

We note that only the result for the neutral field is plotted since the results for the neutral and charged fields are indistinguishable in this plot with this parameter set. We call the modes with the smallest and second smallest real parts of frequency first and second fundamental modes, respectively. On the first fundamental mode of the AdS black hole, we stop the numerical calculation at $\zeta=0.88\pi$, above which the numerical error becomes large. The results for the second fundamental mode are plotted in the whole region $0\le \zeta\le \pi$ for both cases. 
Figures \ref{res} and \ref{resads} show that the real part of the QNF of the neutral field in the AdS black hole 
has the same behaviour as in the AdS spacetime qualitatively. Namely, $|\textrm{Re}[\omega]|$ of the second fundamental mode decreases as $\zeta$ increases, and moreover $|\textrm{Re}[\omega]|$ of the first fundamental mode decreases as $\zeta$ increases for $\zeta \le \zeta_{0}\simeq 0.68\pi$ but becomes zero for $\zeta_{c}\le \zeta$. The detailed result of the first fundamental mode near $\zeta= 0.68\pi$ is shown in Figures \ref{resf} and \ref{resf2}. 
Hereafter we focus on the first fundamental mode because it is most relevant to the stability of the system.
Figures \ref{resf} and \ref{resf2} show the relation between $\zeta$ and $\textrm{Re}[\tilde{\omega}\ell]$ for the charged fields in the AdS black hole with $\left(r_+,r_-\right)=\left(0.01\ell, 0.001\ell\right)$. The green, purple, red, and blue lines denote the result with 
{$eQ=0, 0.01, 0.02$, and $0.03$,} 
respectively. We stress that they are not the reflection of each other with respect to the line $\textrm{Re}[\omega]=0$ because the symmetry of the system is deformed to $\left(\tilde{\omega}, eQ\right)\to\left(-\tilde{\omega}^{*}, -eQ\right)$ as stated in Section~\ref{sec:field_equations}. 
However, since the result for the negative $eQ$ is not distinguishable from that  for the positive $eQ$ of the same 
absolute value in this plot, we plot the result only for the positive $eQ$. As we will see later, they can be distinguishable for a larger black hole.
We can also see that the neutral field has a critical value $\zeta_0\simeq 0.68 \pi$ at which $\textrm{Re}[\omega]$ becomes zero within the numerical error, while
the charged field does not. For this reason, we define the critical value $\zeta_0$ only for the neutral field, while the charged field does not have such a critical value. 
 
\subsection{Stability of the scalar field}
We judge stability of the scalar field by the sign of the imaginary part of the QNF. If the imaginary part is positive, the mode is unstable.

\subsubsection{Neutral field}
\begin{figure}[tbp]
\centering
\subfigure[The relation between $\zeta$ and $\textrm{Im}(\omega\ell)$.]{
\includegraphics[scale=0.78] 
{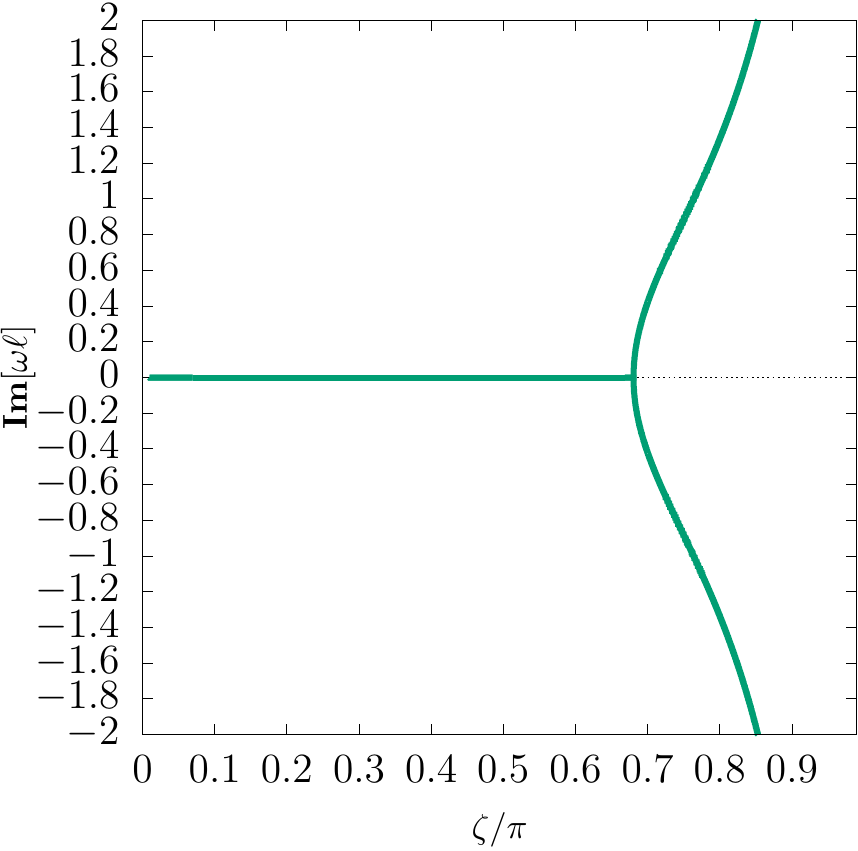}
\label{imns}}
\hspace{0.3cm}
\subfigure[The enlarged figure of Figure \ref{imns}.]{
\includegraphics[scale=0.78] 
{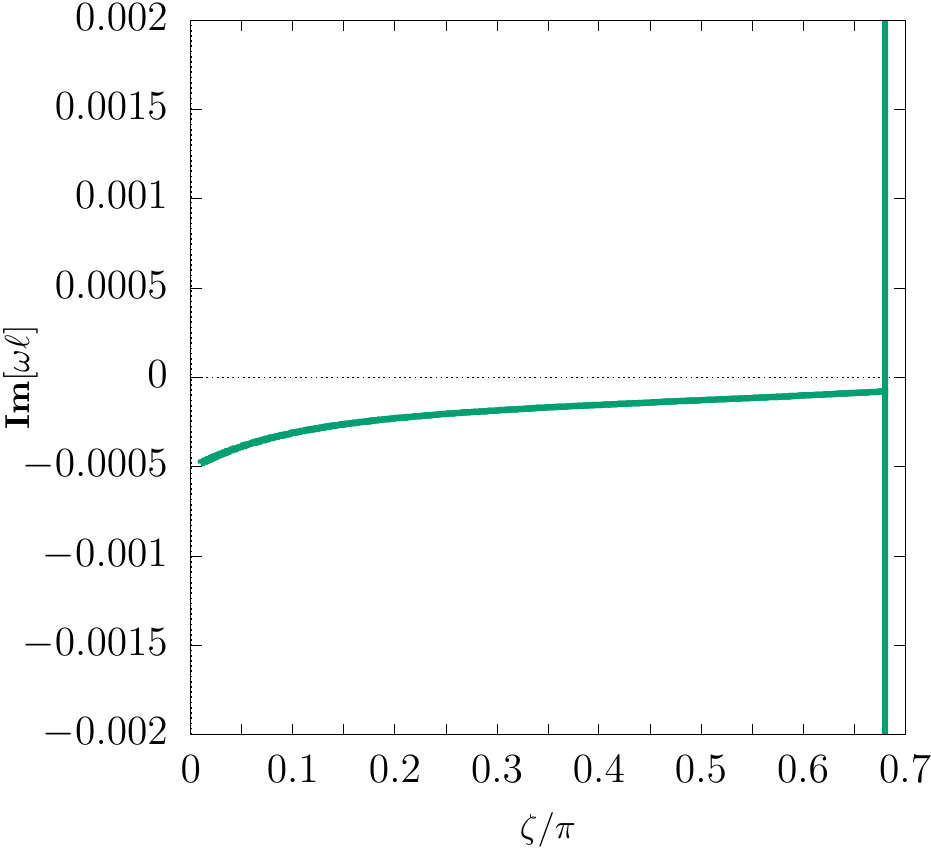}
\label{imnsf}}
\caption{The relation between $\zeta$ and $\textrm{Im}[\omega\ell]$ for 
the neutral field in the AdS black hole with $\left(r_+,r_-\right)=\left(0.01\ell,0.001\ell\right)$.}
\end{figure}

Figures \ref{imns} and \ref{imnsf} give the relation between $\zeta$ and $\textrm{Im}[\omega]$ for the neutral field in the AdS black hole with 
$\left(r_+,r_-\right)=\left(0.01\ell,0.001\ell\right)$. 
The imaginary 
parts of the both modes with positive and negative real parts give
 the same value because of the symmetry $\omega\to -\omega^{*}$ as stated 
in Section~\ref{sec:field_equations}.
Figure \ref{imns} shows that the positive $\textrm{Im}[\omega]$ appears for $\zeta_c<\zeta$, where $\zeta_c$ is defined as the critical value at which 
there appears a mode with $\textrm{Im}[\omega]>0$. 
It can be seen that $\textrm{Im}[\omega]<0$ of the mode for $\zeta<\zeta_{c}$ and $\textrm{Im}[\omega]>0$ for $\zeta>\zeta_{c}$. Furthermore, as we will see that $\zeta_0<\zeta_c$ is satisfied. 
Figure \ref{imns} demonstrates that $\textrm{Im}[\omega]$ of the unstable mode increases as $\zeta$ increases. This instability is not due to superradiance but purely of boundary origin. Figure \ref{imnsf} is the enlarged figure of \ref{imns} in the range of $\zeta/\pi\in[0,0.7]$, and $\textrm{Im}[\omega\ell]\in[-0.002,0.002]$. This figure shows that the mode splits into two at $\zeta\simeq\zeta_0$.  We have checked that this numerical result for $\zeta\lesssim\zeta_0$ is in good agreement with the analytic result in Eq.~\eqref{fulldelta}.

\subsubsection{Charged field}
\begin{figure}[htbp]
\centering
\subfigure[$eQ=0.01$]{
\includegraphics[scale=0.74] 
{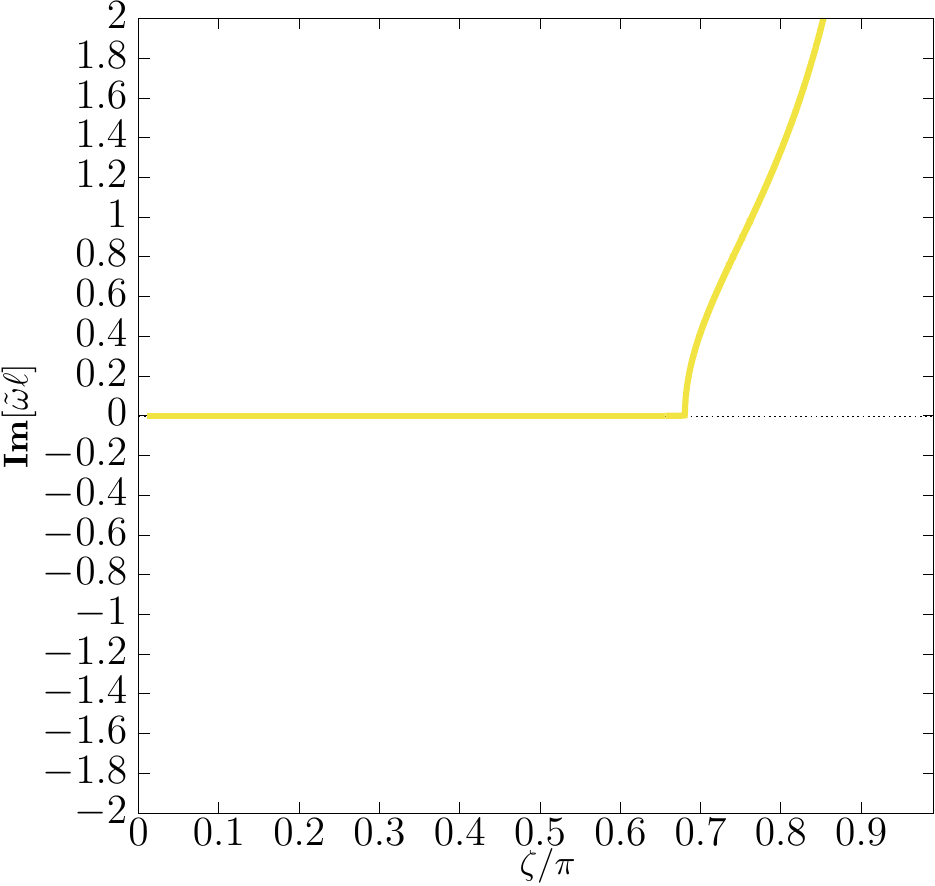}
\label{imcs}}
\hspace{0.3cm}
\subfigure[$eQ=-0.01$]{
\includegraphics[scale=0.74] 
{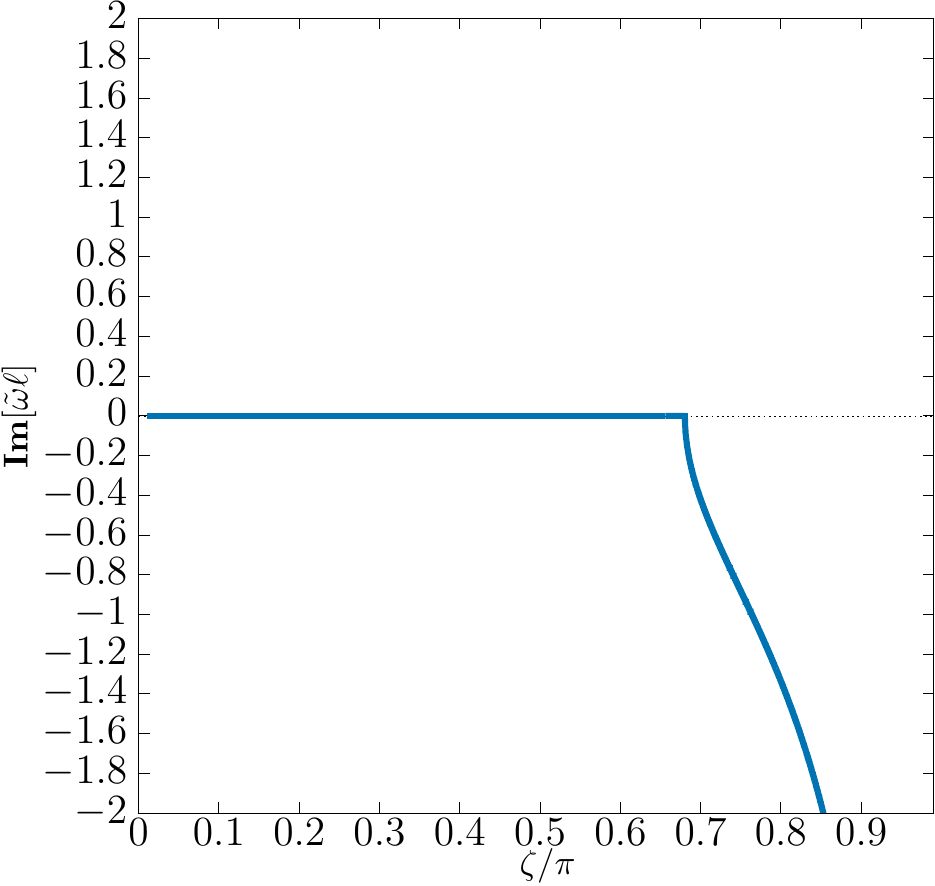}
\label{imcs1}}
  \caption{Same as Figure~\ref{imns} for the mode with $\textrm{Re}[\tilde{\omega}]>0$ for the charged field in the AdS black hole with $\left(r_+,r_-\right)=\left(0.01\ell,0.001\ell\right)$. The yellow and blue lines denote $eQ=0.01$ and  $-0.01$, respectively.}
\end{figure}
\begin{figure}[htbp]
\centering
\subfigure[$\textrm{Re}(\tilde{\omega})>0$]{
\includegraphics[scale=0.74] 
{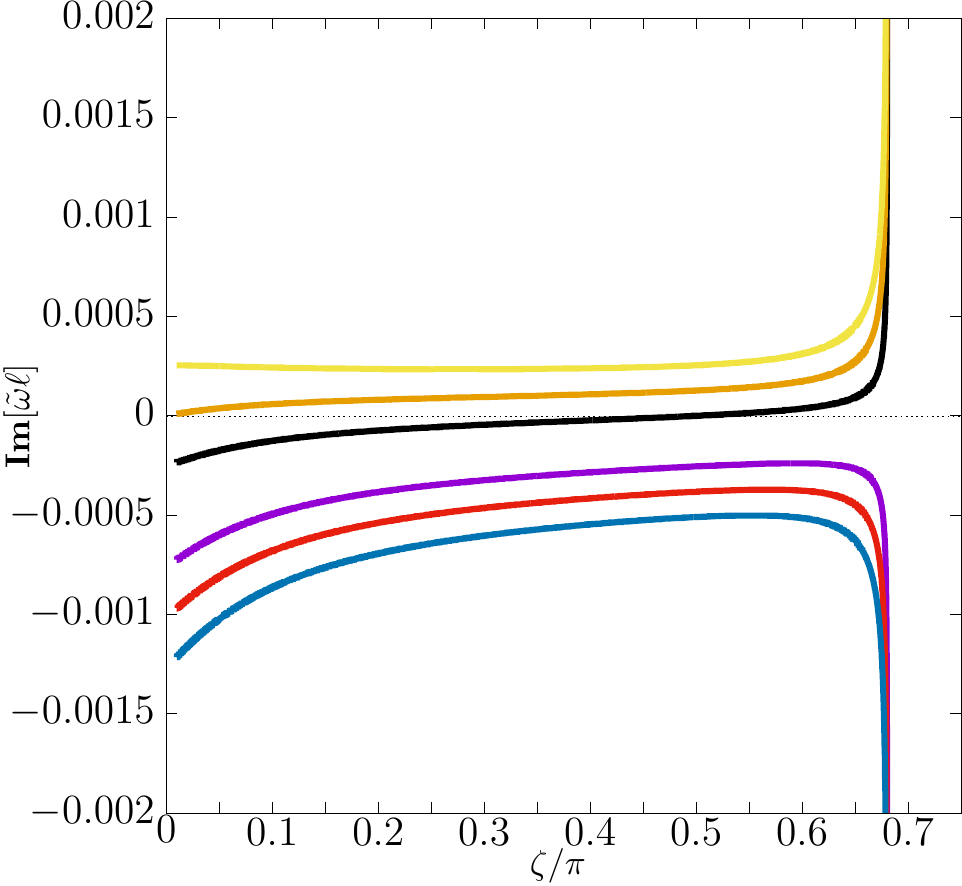}
\label{imcsf}}
\hspace{0.3cm}
\subfigure[$\textrm{Re}(\tilde{\omega})< 0$]{
\includegraphics[scale=0.74] 
{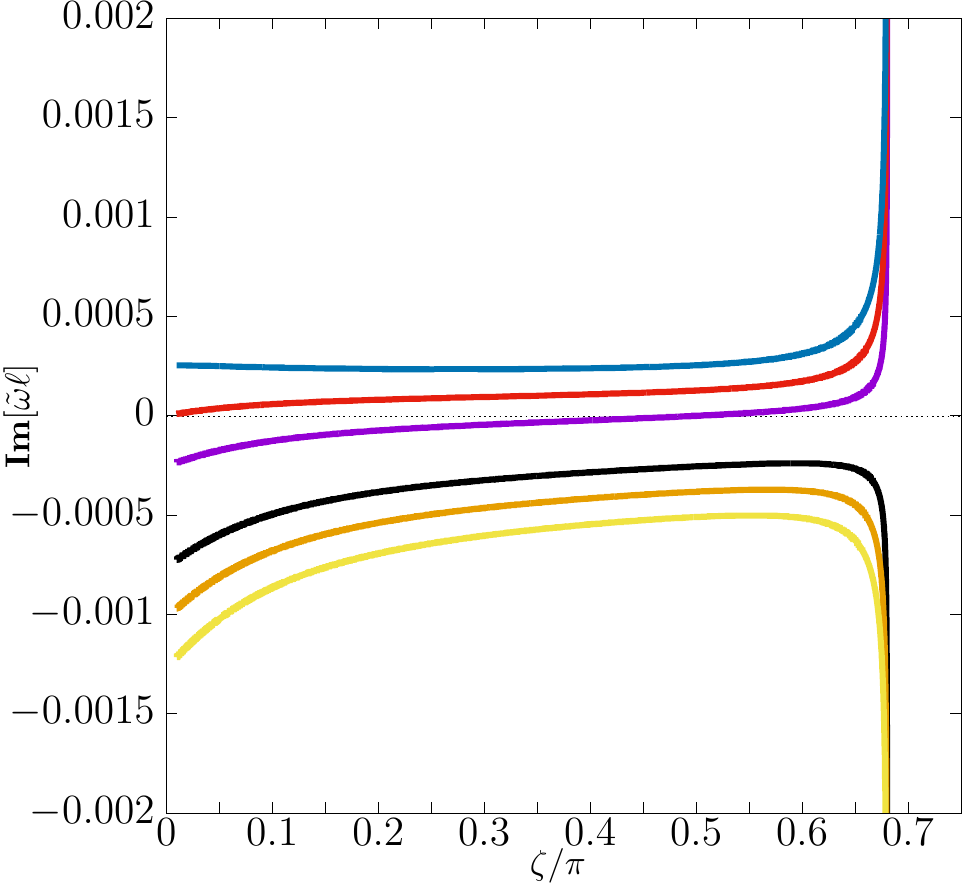}
\label{imcsf1}}  \caption{Same as Figure~\ref{imnsf} but for the charged field in the AdS black hole with $\left(r_+,r_-\right)=\left(0.01\ell,0.001\ell\right)$. The purple, black, red, orange, blue and yellow lines denote $eQ=-0.01, 0.01,-0.02,0.02,-0.03$, and $0.03$,
respectively.}
\end{figure}
Figures \ref{imcs} and \ref{imcs1} give the relation between $\zeta$ and $\textrm{Im}[\tilde{\omega}]$ for the mode with $\textrm{Re}[\tilde{\omega}]>0$ for the charged field in the AdS black hole with $\left(r_+,r_-\right)=\left(0.01\ell,0.001\ell\right)$. The yellow and blue lines denote the results with $eQ=0.01$ and $-0.01$, respectively. If we also plot curves for other positive values of $eQ$, they cannot be distinguished from that for $eQ=0.01$ in this scale, and this is also the case for the negative values of $eQ$. Therefore, we plot only for the mode with $eQ=-0.01$ and $0.01$.  Figures \ref{imcs} and \ref{imcs1} show that if $eQ>0$, there appears an unstable mode with $\textrm{Re}[\tilde{\omega}]>0$ for $\zeta_c<\zeta$, where $\zeta_c$ is defined as the onset of instability also for the charged field. 
On the other hand, if $eQ<0$, there appears an unstable mode with $\textrm{Re}[\tilde{\omega}]<0$ for $\zeta_c<\zeta$. 
Note that the symmetry of the system is given by $(\tilde{\omega},eQ)\to (-\tilde{\omega}^{*}, -eQ)$. We have checked that all of the unstable modes satisfy the superradiance condition Eq.~\eqref{superradiacecondition} or~\eqref{superradiacecondition2}, while none of the stable modes does. Hence, this instability comes from superradiance. 

The detailed feature of the charged field for $\zeta\simeq \zeta_c$ is shown in Figures \ref{imcsf} and \ref{imcsf1}.
They present the relation between $\zeta$ and $\textrm{Im}[\tilde{\omega}]$ 
for the charged field in the AdS black hole with $\left(r_+,r_-\right)=\left(0.01\ell,0.001\ell\right)$. 
The results with $eQ=-0.01, 0.01,-0.02,0.02,-0.03$, and $0.03$ 
are denoted by the purple, black, red, orange, blue, and yellow lines, respectively. We can see that 
the mode with $\textrm{Re}[\tilde{\omega}\ell]>0$ can be unstable for $eQ>0$ even for $\zeta\leq\zeta_{0}$, while the mode with $\textrm{Re}[\tilde{\omega}\ell]<0$ can be unstable for $eQ<0$. As stated above, we have checked for any $\zeta$ we investigated that all of the unstable modes satisfy the superradiance condition Eq.~\eqref{superradiacecondition} or~\eqref{superradiacecondition2}, while none of the stable modes does. For this reason, we interpret that this instability arises from superradiance. Also, since the time scale of the instability for $\zeta_{0}<\zeta$ can be much shorter than that for $\zeta\leq \zeta_{0}$, superradiant instability can be enhanced by the boundary condition. We have checked that this numerical result for $\zeta\lesssim\zeta_0$ is in good agreement with the analytic result in Eq.~\eqref{fulldelta}.

\subsection{Other mass squared cases}
\label{Other mass squared cases}
\begin{figure}[htbp]
\centering
\subfigure[The real part]{
\includegraphics[scale=0.33] 
{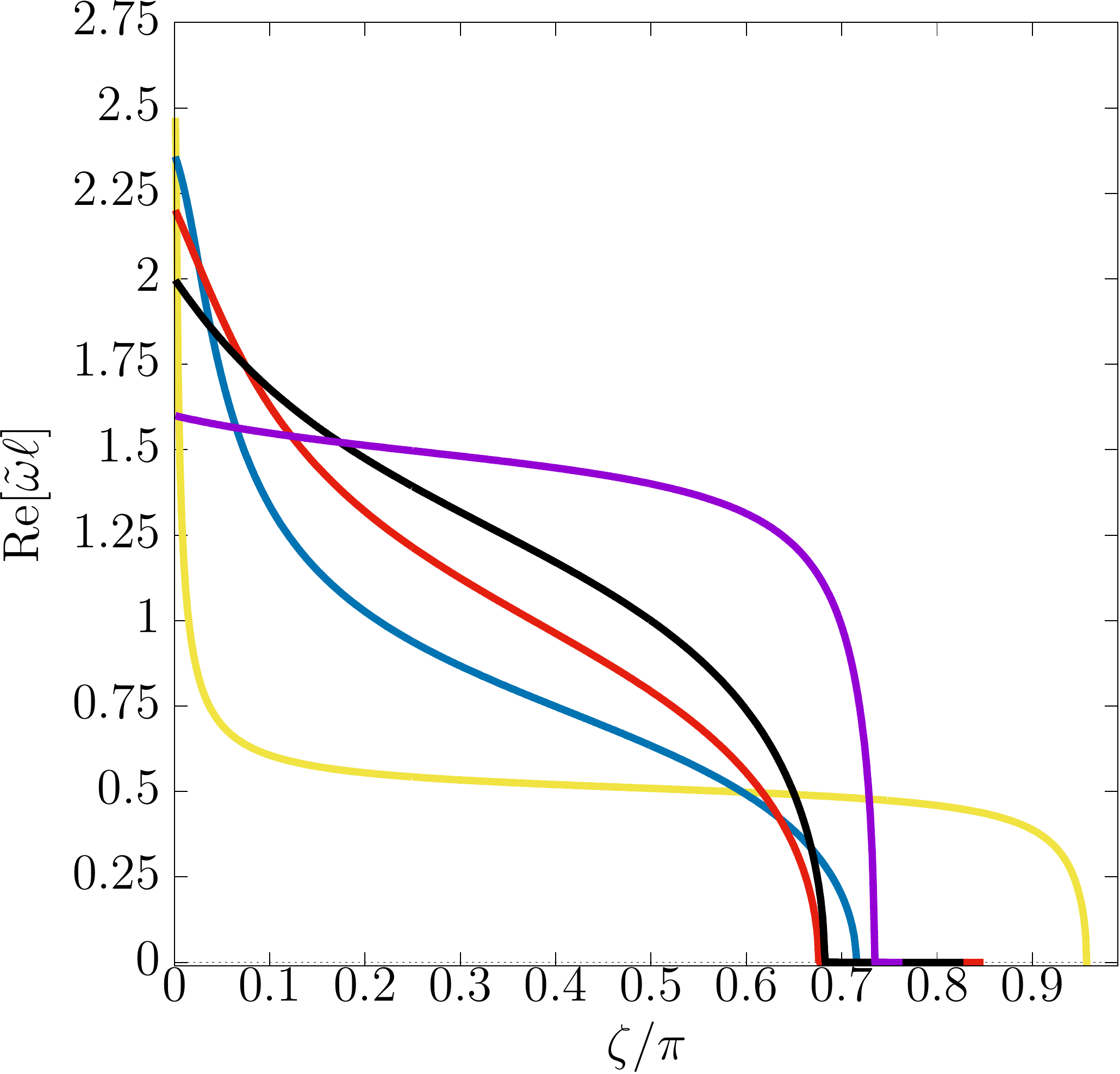}
\label{RealPart}}
\hspace{0.3cm}
\subfigure[The imaginary part]{
\includegraphics[scale=0.35] 
{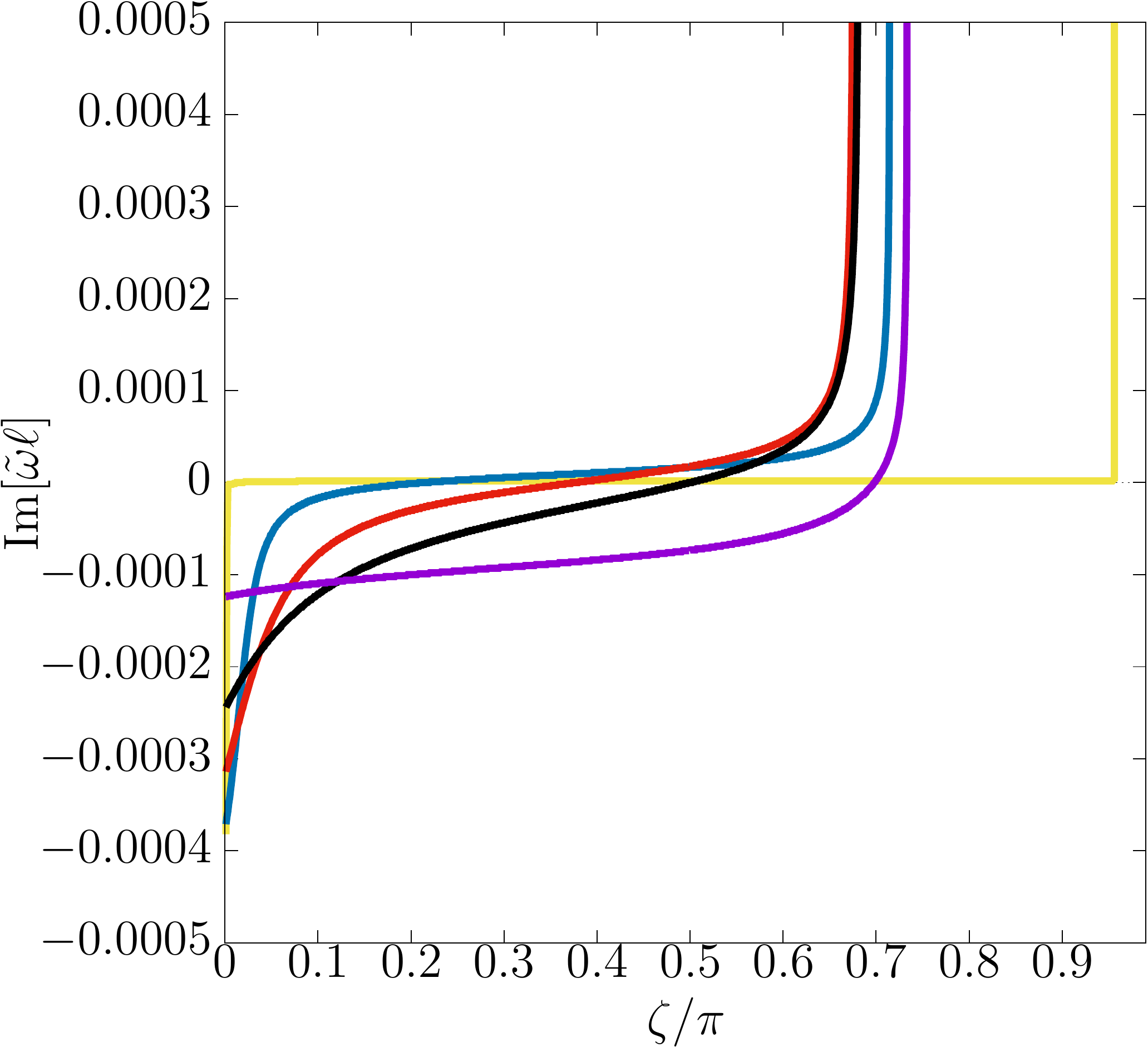}
\label{ImaginaryPart}}
 \caption{The relation between $\zeta$ and the QNFs with $\left(r_+,r_-,eQ\right)=\left(0.01\ell, 0.001\ell,0.01\right)$ and $\mu^2\ell^2=-1.26,-1.50,-1.75,-2.00,-2.24$, which are denoted by the yellow, blue, red, black, and purple lines, respectively.} 
\end{figure}
We investigate QNFs for other mass squared cases in the range of $-9/4<\mu^2\ell^2<-5/4$. Figure~\ref{RealPart} gives the relation between $\zeta$ and ${\rm Re}[\tilde{\omega}\ell]$ with $(r_+,r_-,eQ)=(0.01\ell, 0.001\ell,0.01)$ and $\mu^2\ell^2=-1.26,-1.50,-1.75,-2.00,-2.24$, which are denoted by the yellow, blue, red, black, and purple lines, respectively. The vertical and horizontal axes denote ${\rm Re}[\tilde{\omega}\ell]$ and $\zeta$, respectively. We note that Figure~\ref{RealPart} is indistinguishable from Figure~\ref{OtherMassSquared} in this parameter set. Namely, ${\rm Re}[\tilde{\omega}\ell]$ in the present system coincides with that of the neutral field in the AdS spacetime in the limit of $r_+/\ell\to0$. At $\zeta=0$, ${\rm Re}[\tilde{\omega}\ell]$ is larger as $\mu^2\ell^2$ is increased. As $\zeta$ is increased from $0$, ${\rm Re}[\tilde{\omega}\ell]$ decreases for all $\mu^2\ell^2$ we have investigated. We have checked that for the neutral field, as in the pure AdS spacetime case, ${\rm Re}[{\omega}\ell]$ becomes zero at a certain value of $\zeta$, which is denoted by $\zeta_0$. In particular, $\zeta_0$ becomes larger in the order $\mu^2\ell^2=-2.24,-2.00,-1.75$, while it takes smaller values in the order $\mu^2\ell^2=-1.75,-1.50,-1.26$. We have also checked that as $\mu^2\ell^2$ is further increased for $-1.26<\mu^2\ell^2<-5/4$, $\zeta_0$ approaches $\pi$. For the charged field, on the other hand, ${\rm Re}[\tilde{\omega}\ell]$ rapidly decreases and approaches zero but does not vanish near $\zeta=\zeta_0$ for the neutral field with the same $\mu^2\ell^2$.

Figure~\ref{ImaginaryPart} gives the relation between $\zeta$ and ${\rm Im}[\tilde{\omega}\ell]$ with the same parameters as that in Figure~\ref{RealPart}. The vertical and horizontal axes denote ${\rm Im}[\tilde{\omega}\ell]$ and $\zeta$, respectively. We note that the yellow line is positive in $\zeta/\pi\gtrsim0.012$. At $\zeta=0$, ${\rm Im}[\tilde{\omega}\ell]$ is negative for all $\mu^2\ell^2$ we have investigated and the absolute value is larger as $\mu^2\ell^2$ is increased. For all $\mu^2\ell^2$, as $\zeta$ is increased from $0$, ${\rm Im}[\tilde{\omega}\ell]$ increases, vanishes at $\zeta=\zeta_c$, and for $\zeta>\zeta_c$ it becomes a positive value and further increases. The value of $\zeta_c$ differs depending on $\mu^2\ell^2$, and in particular, it takes smaller values as $\mu^2\ell^2$ is increased for this parameter set $(eQ\ell/r_+=1.00)$. However, this will not necessarily also be the case for other $eQ$, e.g, for $eQ\ell/r_+=0.25$, $\zeta_c$ is larger in the order $\mu^2\ell^2=-1.26, -2.24,-1.50, -2.00, -1.75$. This can be understood from the fact that $\zeta_c$ is determined by the intersection of each curved line of ${\rm Re}[\tilde{\omega}\ell](\zeta)$ and the horizontal line ${\rm Re}[\tilde{\omega}\ell]=eQ\ell/r_+={\rm const.}$ in Figure~\ref{OtherMassSquared}. The growing modes satisfy the superradiace conditions~\eqref{superradiacecondition} or~\eqref{superradiacecondition2}, while the decaying modes do not. Hence, this is superradiant instability in the present system.

\subsection{Larger black hole}
\begin{figure}[htbp]
\centering
\subfigure[$\textrm{Re}(\tilde{\omega})\ge 0$]{
\includegraphics[scale=0.72] 
{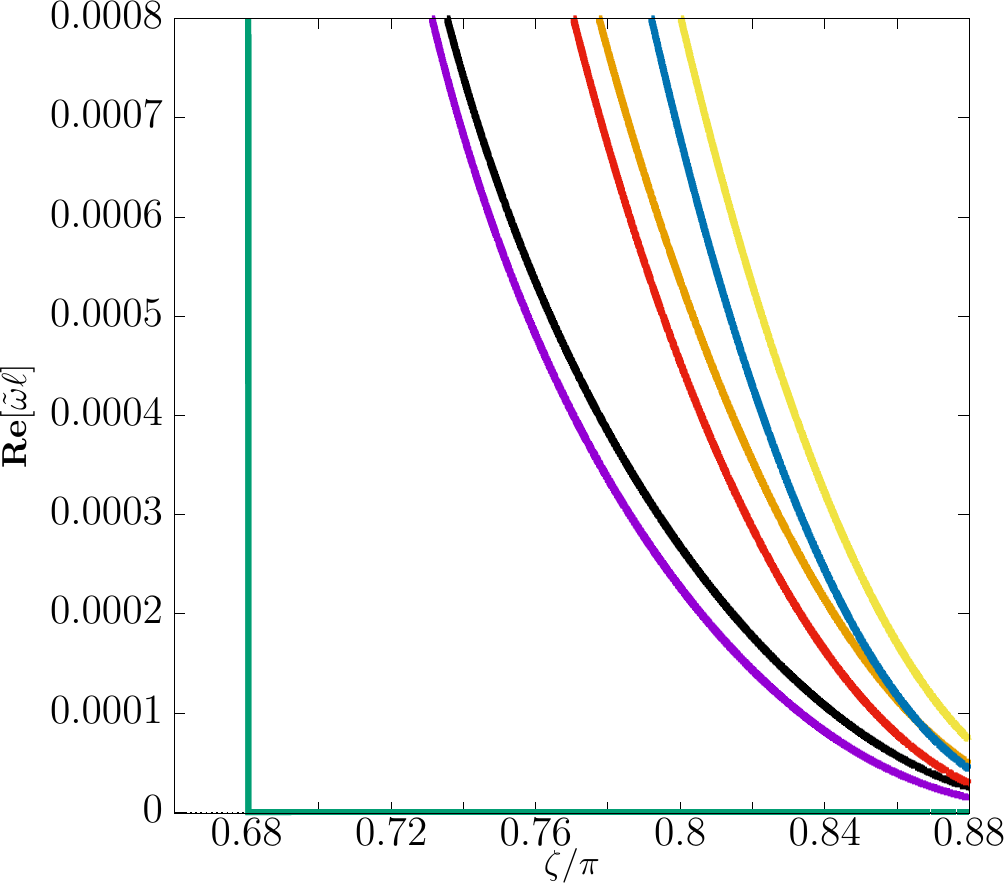}
\label{rlpf}}
\hspace{0.3cm}
\subfigure[$\textrm{Re}(\tilde{\omega})\le 0$]{
\includegraphics[scale=0.72] 
{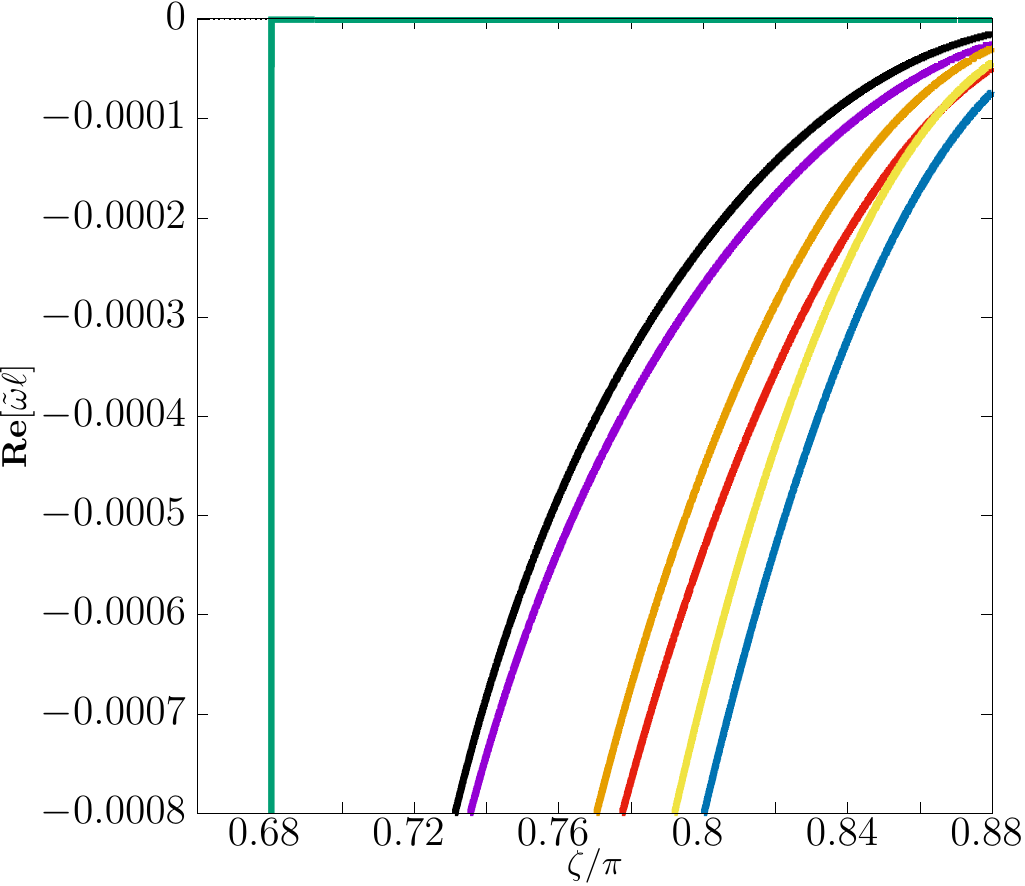}
\label{rlnf}}
 \caption{Same as Figures~\ref{resf} and \ref{resf2} but in the larger AdS black hole with $\left(r_+,r_-\right)=\left(0.1\ell, 0.001\ell\right)$. 
The green, purple, black, red, orange, blue, and yellow lines denote the results with $eQ=0,-0.01,0.01,-0.02,0.02,-0.03$, and $0.03$, respectively.} 
\end{figure}
\begin{figure}[htbp]
\centering
\subfigure[$\textrm{Re}(\tilde{\omega})\ge 0$]{
\includegraphics[scale=0.72] 
{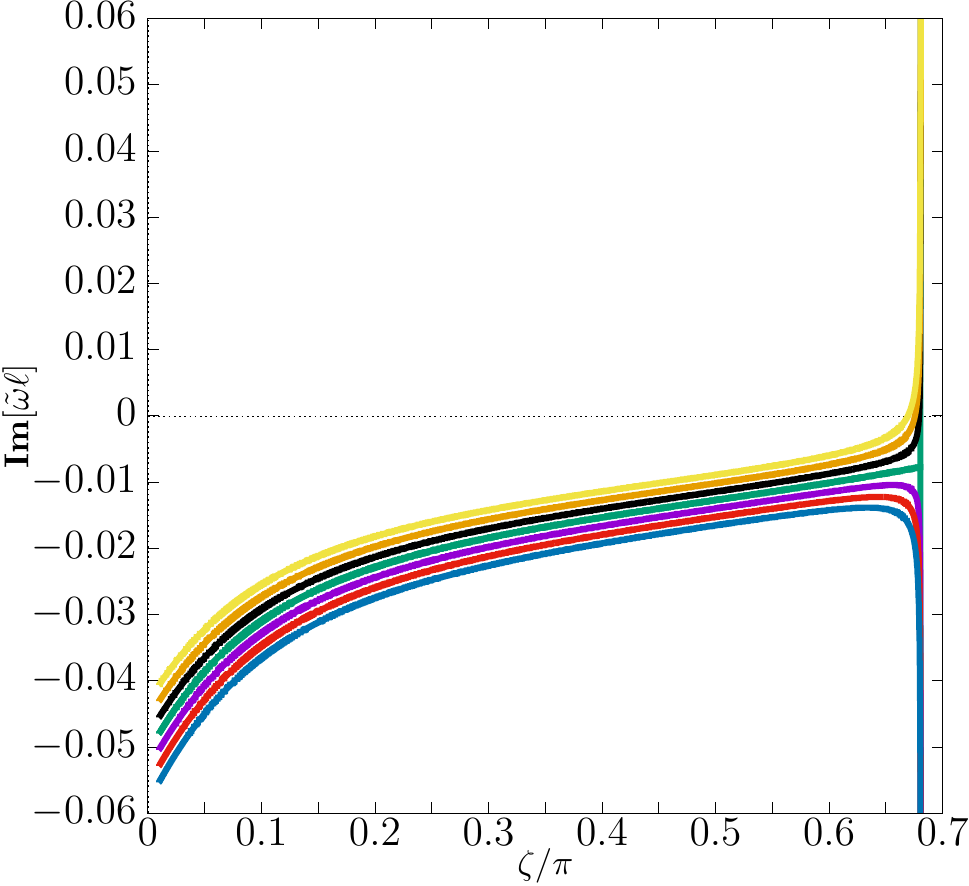}
\label{imlpf}}
\hspace{0.3cm}
\subfigure[$\textrm{Re}(\tilde{\omega})\le 0$]{
\includegraphics[scale=0.72] 
{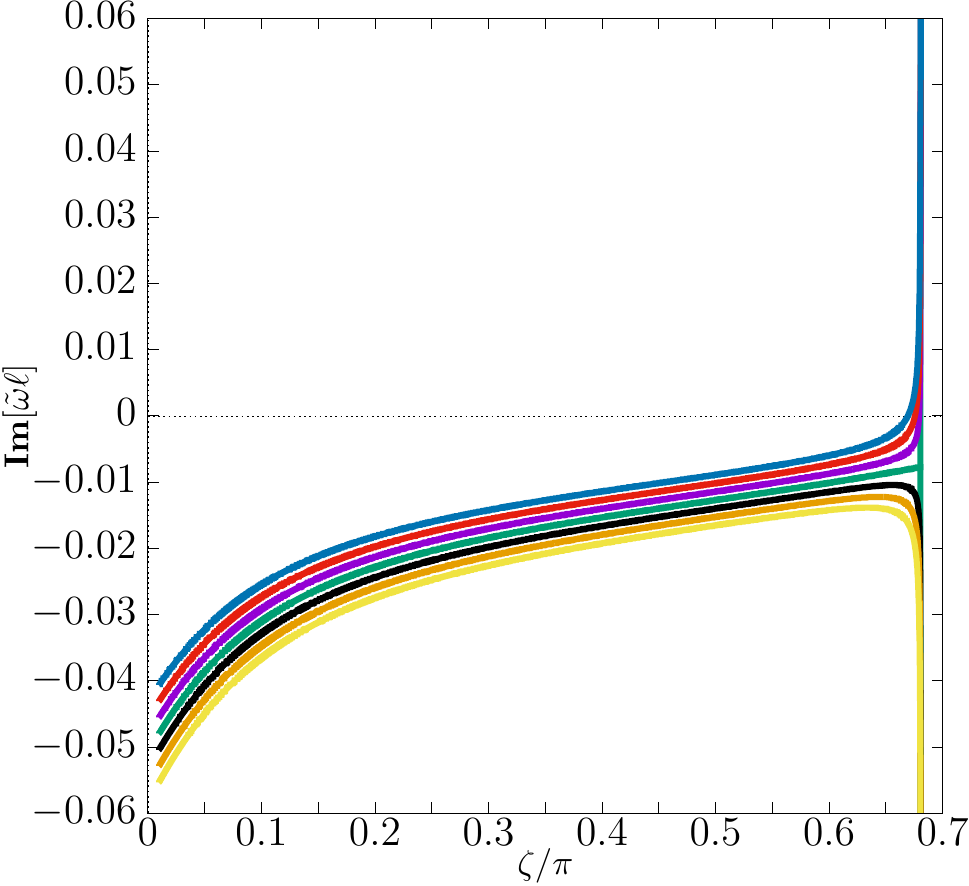}
\label{imlnf}}
 \caption{Same as Figure~\ref{imnsf} but for the charged field in the larger AdS black hole 
with $\left(r_+,r_-\right)=\left(0.1\ell, 0.001\ell\right)$. The green, purple, black, red, orange, blue, and yellow lines denote the results with 
$eQ=0,-0.01,0.01,-0.02,0.02,-0.03$ and $0.03$, respectively.}
\end{figure}
We investigate a black hole larger than the previous one. We fix $\mu^2\ell^2=-2$. For this purpose, let us consider an AdS black hole with $\left(r_+,r_-\right)=\left(0.1\ell, 0.001\ell\right)$. Although it is not clear how much the condition Eq. (\ref{e1}) is met, we would still continue to apply the matched asymptotic expansion method. 

The relation between $\zeta$ and $\textrm{Re}[\tilde{\omega}]$ for the neutral and charged fields in the AdS black hole with $\left(r_+,r_-\right)=\left(0.1\ell,0.001\ell\right)$ looks similar to Figures \ref{resf} and \ref{resf2}. However, there is a remarkable difference if it is magnified. 
Comparing Figures \ref{rlpf} and \ref{rlnf} with Figures \ref{resf} and \ref{resf2}, we can clearly see that the system does not have 
symmetry $(\omega,eQ) \to (-\omega^{*},eQ)$ but $(\omega,eQ)\to (-\omega^{*},-eQ)$. 

Figures \ref{imlpf} and \ref{imlnf} give the relation between $\zeta$ and $\textrm{Im}[\tilde{\omega}]$ for the charged field 
in the AdS black hole with $\left(r_+,r_-\right)=\left(0.1\ell,0.001\ell\right)$. 
The vertical and horizontal axes denote $\textrm{Im}[\tilde{\omega}\ell]$ and $\zeta/\pi\in[0,0.7]$, respectively. Figures \ref{imlpf} and \ref{imlnf} show the similar behaviour to that seen in the smaller black hole with $(r_{+},r_{-})=(0.01\ell, 0.001\ell)$.
Comparing these figures with Figures \ref{imnsf}, \ref{imcsf}, and \ref{imcsf1}, we can see that $|\textrm{Im}[\tilde{\omega}\ell]|$ of the stable mode for $\zeta\leq\zeta_0$ for the AdS black hole with $\left(r_+,r_-\right)=\left(0.1\ell, 0.001\ell\right)$ are generally larger than those for the smaller 
black hole with $\left(r_+,r_-\right)=\left(0.01\ell, 0.001\ell\right)$ if we fix the values of $\zeta$ and $eQ$.

\subsection{Fine structure}

\subsubsection{Neutral field}
\begin{figure}[htbp]
\centering
\subfigure[The relation between $\zeta$ and $\omega$]{
\includegraphics[scale=0.76] 
{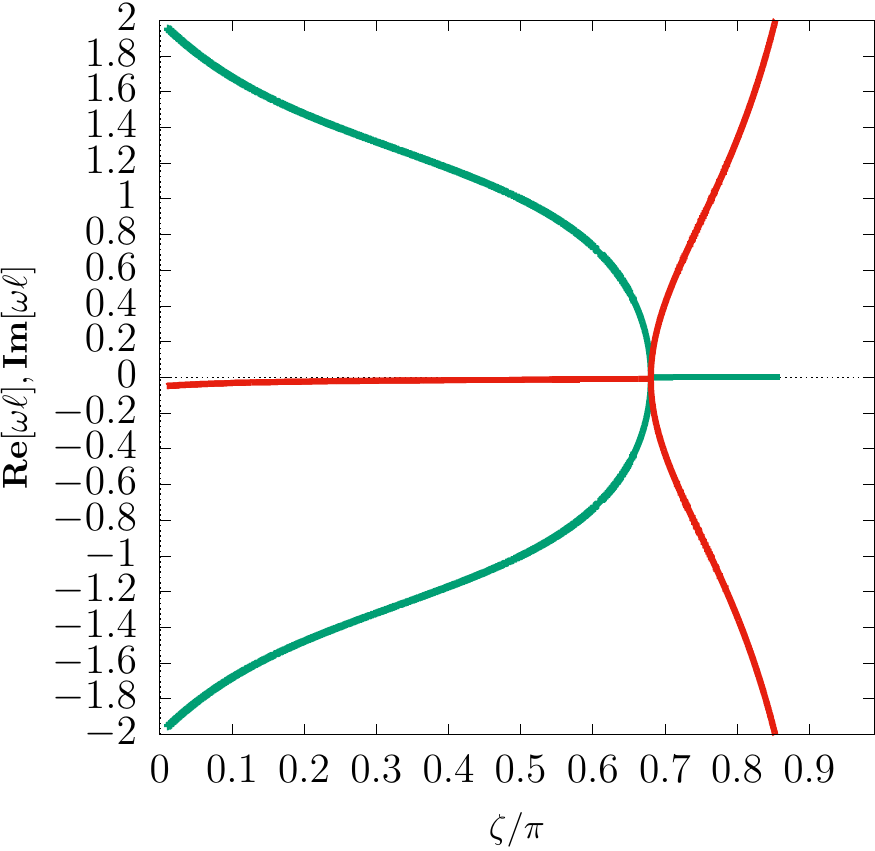}
\label{ws}}
\hspace{0.3cm}
\subfigure[The enlarged figure of Figure \ref{ws}.]{
\includegraphics[scale=0.76] 
{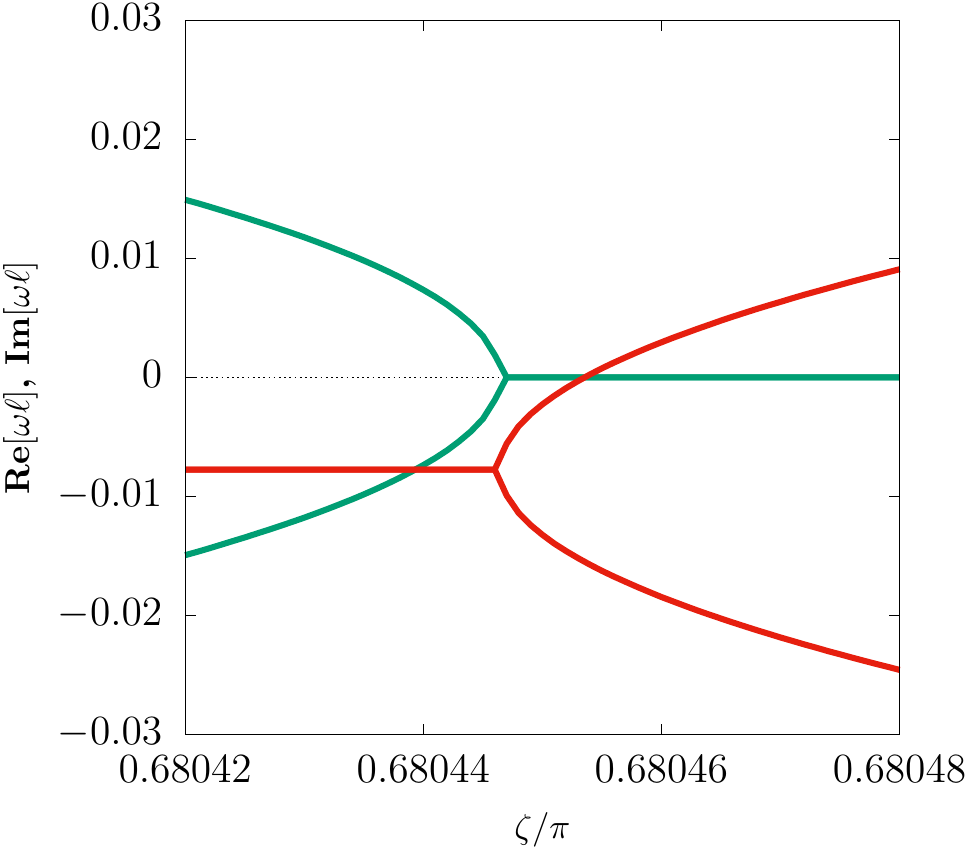}\label{wsf}}
 \caption{Same as Figures~\ref{rea} and \ref{ef} but for the neutral field in the 
AdS black hole with $\left(r_+,r_-\right)=\left(0.1\ell,0.001\ell\right)$.}
 \end{figure}

Here we shall discuss how the results for the neutral field in the AdS black hole are different from that in the AdS spacetime. Figure \ref{ws} gives the relation between $\zeta$ and $\omega\ell$ 
for the neutral field 
in the AdS black hole with 
$\left(r_+,r_-\right)=\left(0.1\ell,0.001\ell\right)$
as in Figure \ref{rea}.
Figure \ref{wsf} is the enlarged figure of Figure \ref{ws} in the range of $\zeta/\pi\in[0.68042,0.68048]$ and $\textrm{Im}[\omega\ell]\in[-0.03,0.03]$. 
Comparing these figures with Figures \ref{rea} and \ref{ef}, we can see that the symmetries $\omega\to -\omega$ and $\omega\to \omega^{*}$ for the AdS spacetime are broken for the neutral field in the AdS black hole spacetime. Here we point out that the symmetry for the real part $\omega\to -\omega^{*}$ remains. Thus, a black hole breaks the symmetry. 
Physically, the negative imaginary part of the QNF for $\zeta\leq\zeta_c$ comes from the dissipation of the scalar field into the black hole.
 
Figure \ref{wsf} shows that $\zeta_0<\zeta_c$ is actually satisfied. It is also shown that at $\zeta=\zeta_0$ within the numerical error, the imaginary part splits into two different values. Hence, there are two QNFs which have a vanishing real part but different negative imaginary parts for $\zeta_0<\zeta<\zeta_c$. Moreover, the two pure imaginary QNFs have one positive and one negative imaginary parts for $\zeta_c<\zeta$. Thus, we can see that there exists a static mode of the neutral field perturbation on the charged AdS black hole for $\zeta=\zeta_{c}$. This can be most clearly seen in Figure~\ref{fl1}.

\subsubsection{Charged field}

\begin{figure}[htbp]
\centering
\subfigure[The relation between $\zeta$ and $\tilde{\omega}$]{
\includegraphics[scale=0.76] 
{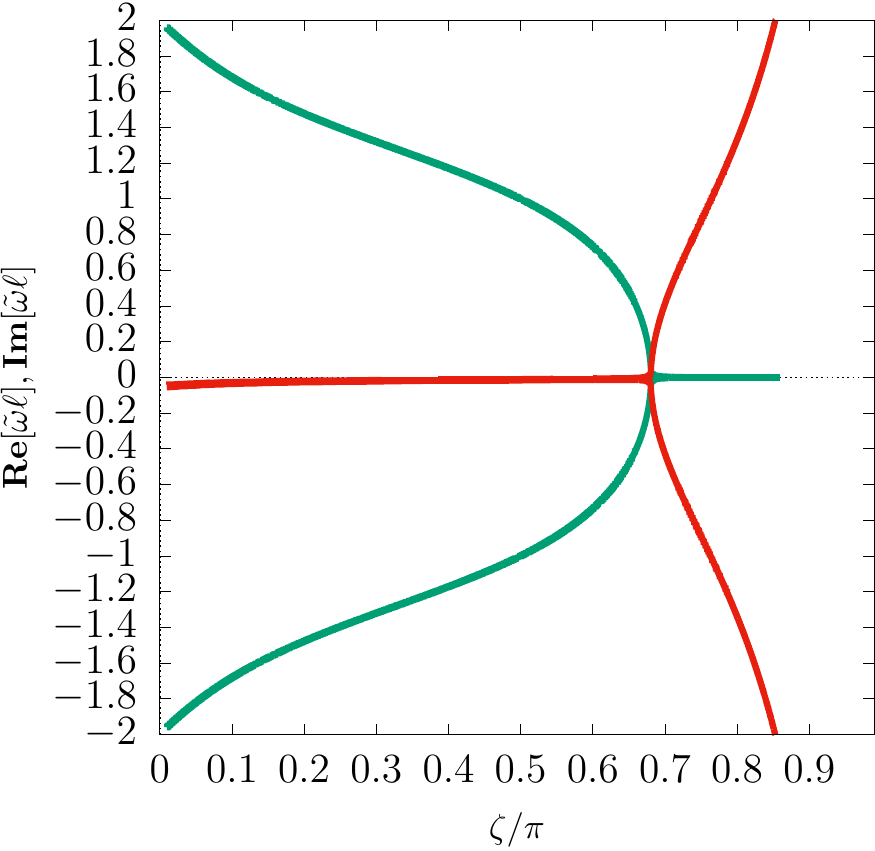}
\label{wc}}
\hspace{0.3cm}
\subfigure[The enlarged figure of Figure \ref{wc}.]{
\includegraphics[scale=0.76] 
{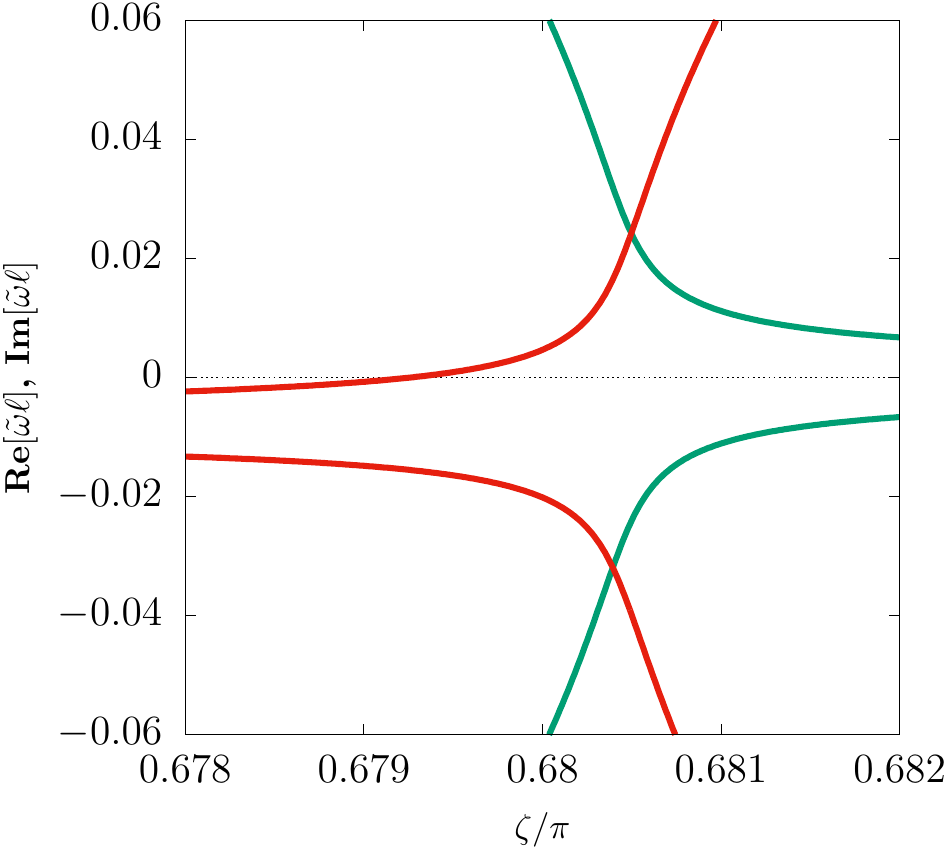}
\label{wcf}}
 \caption{Same as Figures~\ref{rea} and~\ref{ef} but for the charged field with {$eQ=0.01$} in the AdS black hole 
with $\left(r_+,r_-\right)=\left(0.1\ell,0.001\ell\right)$.}
\end{figure}

Figure \ref{wc} gives the relation between $\zeta$ and $\tilde{\omega}$ for the charged field with $eQ=0.01$ in the AdS black hole with 
$\left(r_+,r_-\right)=\left(0.1\ell,0.001\ell\right)$. 
Figure \ref{wcf} is the enlarged figure of Figure \ref{wc} in the range of $\zeta/\pi\in[0.678,0.682]$ and $\textrm{Im}[\tilde{\omega}\ell]\in[-0.06,0.06]$. We note that the result for $eQ=-0.01$ gives the same graph. 
Figure \ref{wcf} shows the existence of the boundary condition at the onset of instability $\zeta=\zeta_{c}$ such that $\textrm{Im}[\tilde{\omega}\ell]=0$ but $\textrm{Re}[\tilde{\omega}\ell]\neq0$. This mode is purely oscillative. This suggests a branch to nonlinearly oscillating solutions or charged oscillating black holes dubbed ``black resonators''~\cite{4,Ishii:2018oms,Ishii:2019wfs} in the AdS spacetime.

\subsubsection{Flow of QNFs with respect to $\zeta$}
\begin{figure}[htbp]
\centering
\subfigure[Neutral field]{
\includegraphics[scale=0.75] 
{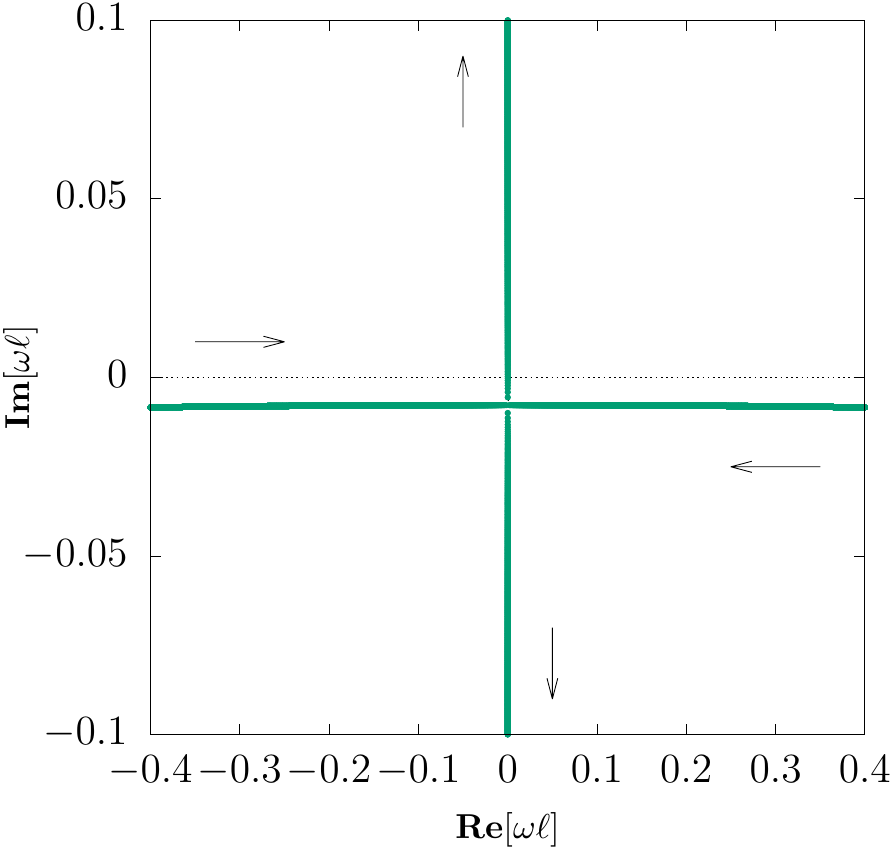}
\label{fl1}}
\hspace{0.3cm}
\subfigure[Charged field]{
\includegraphics[scale=0.75] 
{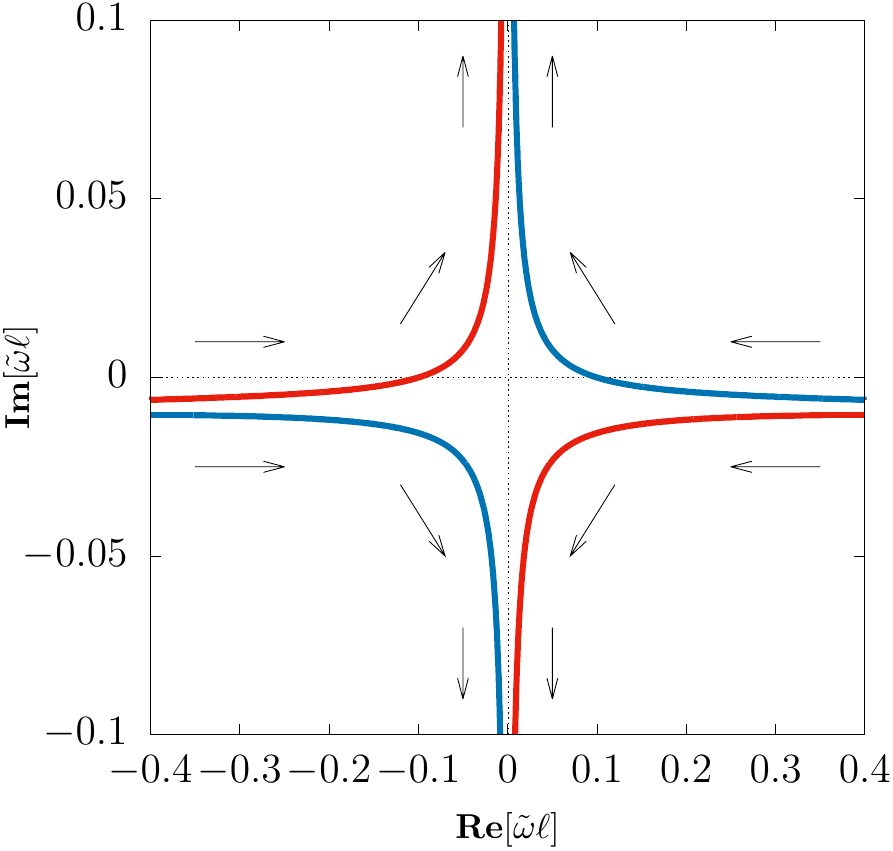}
\label{fl2}}
 \caption{The flow of QNFs in the AdS black hole with $\left(r_+,r_-\right)=\left(0.1\ell,0.001\ell\right)$. The $\tilde{\omega}$ moves as $\zeta$ is increased continuously. 
The green, red, and blue lines denote $eQ=0, -0.01$, and $0.01$, respectively.}
\end{figure}
Figures \ref{fl1} and \ref{fl2} show the flow of the QNFs in the complex plane for the neutral and charged fields, respectively, in the AdS black hole with $\left(r_+,r_-\right)=\left(0.1\ell,0.001\ell\right)$.
The green, red, and blue lines denote trajectories which the QNFs with {$eQ=0, -0.01, 0.01$} 
draw when we continuously change $\zeta$. The arrows indicate the direction in which the QNF flows as $\zeta$ increases.
In Figure \ref{fl1}, we can see that for the neutral field, the QNF for $\zeta<\zeta_{0}$ has a non-zero real part and a negative imaginary part. 
At $\zeta=\zeta_0$, the real part of these QNFs vanishes.
Then, the QNFs for $\zeta_0<\zeta\leq\zeta_c$ are pure imaginary and both move along the arrow on the imaginary axis as $\zeta$ increases,  
the one goes up and the other down in the lower half plane. 
After that, the imaginary part of the QNF which is going up becomes zero at $\zeta=\zeta_{c}$, implying the onset of instability.
The flow described above is consistent with the symmetry $\omega\to -\omega^{*}$.

Figure \ref{fl2} shows the flow for the charged field. The QNFs approach the imaginary axis as $\zeta$ increases. The one mode goes up and the other down along the axis as $\zeta$ increases. For $eQ>0$, the QNF with $\textrm{Re}[\tilde{\omega}]>0$ goes up, while that with $\textrm{Re}[\tilde{\omega}]<0$ goes down.
For $eQ<0$, the result is opposite. This is a feature quite different from the neutral field. The mode which goes up eventually causes superradiant instability at $\zeta=\zeta_{c}$. We can see that the flow is consistent with the symmetry $\left(\tilde{\omega}, eQ\right)\to \left(-\tilde{\omega}^{*}, -eQ\right)$.

\section{Physical interpretation 
\label{sec:interpretation}}
\subsection{Electromagnetic superradiance\label{RNBH_superradiance}}
From Eq. (\ref{eomcs}) and Eq. (\ref{anz}), we obtain the radial equation,
\begin{equation}
\label{rdeq}
\left[-\frac{d^2}{dx^2}+V(x)\right]u(x)=\left(\tilde{\omega}-\frac{eQ}{r}\right)^2u(x),
\end{equation}
where
\begin{equation}
V(x)=\frac{{\triangle}}{r^2}\left[\frac{1}{r}\frac{d}{dr}\left(\frac{{\triangle}}{r^2}\right)+\frac{1}{r^2}\left\{l(l+1)+\mu^2r^2\right\}\right].
\end{equation}
We assume $\omega\in \mathbb{R}$. 
The effective potential of the charged scalar field in the charged black hole can be negative near the outer horizon depending upon the charge of the field. The region, where the effective potential is negative, is called the generalised ergoregion~\cite{Denardo:1973pyo}. 

Here, we first discuss electromagnetic superradiance for a massless charged scalar field in the charged black hole spacetime with $\mu^{2}=\Lambda=0$, which gives an important background for the physical interpretation of the electromagnetic 
superradiance in the AdS system. In the limit of $x\to \infty$, Eq. (\ref{rdeq}) is reduced to
\begin{equation}
\label{infeq}
\left[\frac{d^2}{dx^2}+\tilde{\omega}^2\right]u(x)\simeq 0,
\end{equation}
and we obtain the general solution of Eq. (\ref{infeq}) as
\begin{equation}
\label{gesola}
u(x)\sim A_{\textrm{in}}e^{-i\tilde{\omega}x}+A_{\textrm{out}}e^{i\tilde{\omega}x},~A_{\textrm{in}},A_{\textrm{out}}\in\mathbb{C}.
\end{equation}
The subscripts ``in" and ``out" denote the ingoing and outgoing modes at the conformal infinity, respectively.

Noting the conservation of the number current given by Eq.~(\ref{eq:number_current}), we find
\begin{equation}
\label{wron}
\tilde{\omega}(|A_{\textrm{in}}|^2-|A_{\textrm{out}}|^2)=\left(\tilde{\omega}-\frac{eQ}{r_+}\right)|B_{\textrm{in}}|^2 r_{+}^{2},
\end{equation}
where we have used the asymptotic solutions (\ref{gesolb}) with $B_{\textrm{out}}=0$ and Eq. (\ref{gesola}). 
Thus, we can observe that the amplitude of the reflected wave is larger than
that of the incident wave, i.e.
\begin{equation}
|A_{\textrm{out}}|^2>|A_{\textrm{in}}|^2,
\label{eq:superradiance}
\end{equation}
if the mode with $eQ>0$ satisfies 
\begin{equation}
\begin{split}
\label{scon}
0<\tilde{\omega}<\frac{eQ}{r_+},\end{split}
\end{equation}
or if the mode with $eQ<0$ satisfies 
\begin{equation}
\begin{split}
\label{scon2}
\frac{eQ}{r_+}<\tilde{\omega}<0.
\end{split}
\end{equation}
There conditions for $\tilde{\omega}$ are equivalent to Eqs.~\eqref{superradiacecondition} and~\eqref{superradiacecondition2}, respectively. This amplification of the wave is called electromagnetic superradiance.

Using Eq.~(\ref{gesola}),  
the conservation of $j^{a}$ and $j_{(e)}^{a}$ are integrated to give
\begin{eqnarray}
 j^{r}(t,r,\theta,\phi)=2\tilde{\omega}(|A_{\textrm{out}}|^{2}-|A_{\textrm{in}}|^{2})\frac{|Y_{lm}(\theta,\phi)|^{2}}{r^{2}},
\end{eqnarray}
and 
\begin{equation}
 j_{(e)}^{r}(t,r,\theta,\phi)=2e\tilde{\omega}(|A_{\textrm{out}}|^{2}-|A_{\textrm{in}}|^{2})\frac{|Y_{lm}(\theta,\phi)|^{2}}{r^{2}},
\label{eq:j_e^1}
\end{equation}
respectively, for $r_{+}\le r<\infty$.
Therefore, there are a net positive (negative) 
outgoing number current and therefore net positive (negative) outgoing electric current for $\tilde{\omega}>(<)0$. We can conclude that a net positive (negative) particle number and net positive (negative) charge are extracted from the system 
for $\tilde{\omega}>(<0)$ and they must be extracted from the black hole.
We will see that this is really the case in Section \ref{sec:thermodynamical_insight}.

Here we see the energetics of the system. 
The outgoing energy current of the scalar field in the asymptotic region can be calculated to 
\begin{equation}
 J^{r}_{(m)}\simeq 2\tilde{\omega}^{2}(|A_{\textrm{out}}|^{2}-|A_{\textrm{in}}|^{2})\frac{|Y_{lm}(\theta,\phi)|^{2}}{r^{2}} \quad \mbox{as}~~r\to \infty,
\label{eq:J_m^1_infty}
\end{equation}
while the energy current by the electromagnetic field vanishes there.
Therefore, positive energy is extracted from the system in the superradiance. However, it does not necessarily mean that this positive energy must be extracted from the black hole because $J^{a}_{(m)}$ is not a conserved current as $\nabla_{a}J^{a}_{(m)}=-\nabla_{a}J_{(em)}^{a}$.
In general, since 
$\nabla^{a}T_{(em)ab}=j_{(e)}^{a}F_{ab}$,
we obtain 
$\nabla_{a}J^{a}_{(em)}=-j_{(e)}^{a}\xi^{b}F_{ab}$.
In the present system, this implies 
\begin{equation}
\nabla_{a}J_{(em)}^{a}=-j^{r}_{(e)}F_{rt}.
\end{equation}
This shows that the electromagnetic field loses energy due to the work on the 
electric current exerted by the electric field. On the other hand, this implies 
\begin{equation}
\nabla_{a}J^{a}_{(m)}=j^{r}_{(e)}F_{rt}
\label{eq:J_m^a_source}
\end{equation}
Thus, the scalar field receives energy through the work on the electric current exerted 
by the electric field. We can easily integrate Eq.~(\ref{eq:J_m^a_source}) putting
Eq.~(\ref{eq:j_e^1}) in the source term and fix the function of integration using Eq.~(\ref{eq:J_m^1_infty}). Thus, we obtain
\begin{equation}
J^{r}_{(m)}=2\tilde{\omega}\left(\tilde{\omega}-\frac{eQ}{r}\right)(|A_{\textrm{out}}|^{2}-|A_{\textrm{in}}|^{2})\frac{|Y_{lm}|^{2}}{r^{2}}
\end{equation}
for $r_{+}\le r<\infty$.
We can see that under the condition for the superradiance, $J_{(m)}^{r}$ is necessarily negative at $r=r_{+}$. Therefore, there is a net ingoing energy current at the horizon.
$J^{r}_{(m)}$ changes its sign at $r=r_{S}=eQ/\tilde{\omega}$. 
There is a net ingoing energy current for $r_{+}\le r<r_{S}$, while 
a net outgoing energy current for $r>r_{S}$. 
The radius $r_{S}$ thus characterises the energy injection from the electric field into the scalar field. In particle picture, we can interpret it as the pair creation of particles of charge $e$ and antipartcles of charge $-e$ at $r=r_{S}$. The former runs outwardly away and extracts energy and charge from the system, while the latter runs inwardly into the horizon 
of the black hole and gives energy to and discharge the horizon.

Now we argue that the superradiance decreases $M$, 
the mass of the whole system. Assuming that the superradiance proceeds in a quasi-static manner, Eqs.~(\ref{eq:j_e^1}) and 
(\ref{eq:J_m^1_infty}) imply that the ratio of $\delta  Q$ and $\delta  M$, increments of $Q$ and $M$, is given by 
\begin{equation}
 \frac{\delta  Q}{\delta  M}=\frac{e}{\tilde{\omega}}
\label{eq:DeltaQDeltaM}
\end{equation}
From $r_{+}=M+\sqrt{M^{2}-Q^{2}}$ and $ A=4\pi r_{+}^{2}$, 
$\delta  A$, increment of the area of the horizon in a quasistatic process, is calculated to 
\begin{equation}
 \delta  A=8\pi r_{+}\frac{r_{+}\delta  M-Q\delta  Q}{\sqrt{M^{2}-Q^{2}}}
\label{eq:DeltaA}
\end{equation}
where $\kappa$ is the surface gravity given by 
\begin{equation}
\kappa=\frac{\sqrt{M^{2}-Q^{2}}}{r_{+}^{2}}.
\end{equation}
Using Eq.~(\ref{eq:DeltaQDeltaM}), this gives
\begin{equation}
\delta A=\frac{8\pi}{\kappa}\left(1-\frac{eQ}{\tilde{\omega}r_{+}}\right)\delta M,
\end{equation}
Noting the superradiance condition, Eq.~\eqref{scon} or~\eqref{scon2},  
the area law $\delta  A>0$ implies $\delta  M<0$. Thus, the superradiance extracts 
energy from the whole system. Alternatively, Eq.~(\ref{eq:J_m^1_infty}) together with Eq.~(\ref{eq:superradiance}) directly implies $\delta  M<0$ and thus guarantees the area law through the superradiance condition 
 Eq.~\eqref{scon} or~\eqref{scon2}.
However, since the energy current for $r<r_{S}$ falls inwardly, the black hole horizon gains net positive energy.
The superradiance extracts energy not from the black hole horizon but from its ambient electric field.

In the AdS charged black hole, we can show that all of $\sqrt{-g}j^{r}$, $\sqrt{-g}j_{(e)}^{r}$, and $\sqrt{-g}J^{r}$ vanish at the conformal infinity by using the asymptotic behaviour~(\ref{asymp}) under the Robin boundary condition (\ref{bc}). Therefore, the outgoing and ingoing waves must cancel out each other there. This means that the Robin boundary condition makes the conformal boundary perfectly reflective. Since we can regard the overlapping region between the near and far regions as the asymptotically flat region both from the black hole and from the conformal infinity, we can interpret the superradiant instability in the present system as follows. The ingoing wave in the overlapping region is reflected and amplified by electromagnetic superradiance near the black hole. This amplified outgoing wave goes through the overlapping region and perfectly reflected by the AdS boundary under the Robin boundary. This reflected ingoing wave goes through the asymptotic region and becomes the incident wave to the black hole again. A series of these successive processes results in the superradiance instability. The scalar field takes charge from the black hole and energy from its ambient electric field and gives energy to the black hole. The black hole takes energy from and gives charge to the scalar field.

\subsection{Thermodynamical insight
\label{sec:thermodynamical_insight}}
Here we discuss a thermodynamical argument. 
We explicitly 
calculate the energy current and 
electric current ingoing into the horizon, 
$J_{a}n^{a}|_{{\cal H}_{+}}$ and $j_{(e)a}n^{a}|_{{\cal H}_{+}}$, respectively, where
$n^{a}=-\xi^{a}$ at the outer horizon for the static black hole.
To avoid the coordinate singularity at the outer horizon, we use the ingoing Eddington-Finkelstein
coordinates
\begin{equation}
 ds^{2}=-\frac{\triangle}{r^{2}}dv^{2}+2dvdr+r^{2}(d\theta^{2}+\sin^{2}\theta d\phi^{2}),
\end{equation}
where $v=t+x+\mbox{const}$ and the gauge field takes the form 
\begin{equation}
 A_{\mu}dx^{\mu}=-\left(\frac{Q}{r}+\phi_{0}\right)dv+\frac{r^{2}}{\triangle }\left(\frac{Q}{r}+\phi_{0}\right)dr.
\end{equation}
Near the outer horizon, the scalar field satisfying the ingoing-wave condition behaves as
\begin{equation}
\Psi\sim e^{-i\omega v+i \left(\frac{eQ}{r_{+}}+e\phi_{0}\right)x}
Y_{lm}(\theta,\phi).
\end{equation}
Assuming that $\tilde{\omega}=\tilde{\omega}_{R}+i \tilde{\omega}_{I}$ is complex,
we obtain the following result: 
\begin{eqnarray}
 J_{(m)a}n^{a}|_{{\cal H}_{+}}&=&2\left|\tilde{\omega}-\frac{eQ}{r_{+}}\right|^{2}|\Psi|^{2}, 
\label{eq:Jn} \\
 J_{(em)a}n^{a}|_{{\cal H}_{+}}&=&0, \\
 j_{(e)a}n^{a}|_{{\cal H}_{+}}&=&2e \left(\tilde{\omega}_{R}-\frac{eQ}{r_{+}}\right)|\Psi|^{2}.
\label{eq:jn}
\end{eqnarray}
The Misner-Sharp quasi-local energy associated with the energy current $J^{a}$ is given by 
\begin{equation}
 E_{\rm MS}:=\frac{1}{2}r\left(1-\nabla_{a}r\nabla^{a}r\right)=M-\frac{Q^{2}}{2r}
\end{equation}
in the Reissner-Nordstr\"{o}m spacetime. Eq.~(\ref{eq:Jn}) implies that no energy can be extracted from the black hole irrespective 
of the value of $\tilde{\omega}$. 

Assuming a quasistatic change, we can also find
\begin{equation}
\delta  Q=\frac{e\left(\tilde{\omega}_{R}-\frac{eQ}{r_{+}}\right)}{\left|\tilde{\omega}-\frac{eQ}{r_{+}}\right|^{2}}\delta E_{H}
\label{eq:DeltaQDeltaE}
\end{equation}
with $\delta E_{H}\ge 0$, where $\delta  E_{H}$ stands for the change of the Misner-Sharp energy $M_{\rm MS}$ at $r=r_{+}$.

From Eq.~(\ref{eq:DeltaQDeltaE}), we can see that if the superradiance condition 
is satisfied in terms of $\tilde{\omega}_{R}$, we can conclude that 
$\delta  |Q| < 0$, that is, the black hole is discharged by the scalar field. This is because if $eQ>0$, Eqs.~\eqref{scon} and (\ref{eq:jn})
imply $e \delta  Q <0$, 
whereas if $eQ<0$,  Eqs.~\eqref{scon2} and (\ref{eq:jn}) imply
$e \delta  Q >0$.

We can approximate the black hole geometry near the horizon by the Reissner-Nordtr\"{o}m spacetime.  
Using $E_{H}=M-{Q^{2}}/({2r_{+}})$ and Eq.~(\ref{eq:DeltaA}), 
we find 
\begin{equation}
 \delta  A=\frac{2A}{M}\delta  E_{H}.
\end{equation}
Therefore we conclude that the area of the black hole never decreases.

From Eqs.~(\ref{eq:DeltaA}) and (\ref{eq:DeltaQDeltaE}), 
we can derive the following relation between $\delta  M$ and $\delta  A$:
\begin{equation}
 \frac{\delta  M}{M}=\frac{-\left[\left(\frac{r_{+}}{M}-1\right)\tilde{\omega}_{R}+
\left(2-\frac{r_{+}}{M}\right)\frac{eQ}{r_{+}}\right]
\left(\frac{eQ}{r_{+}}-\tilde{\omega}_{R}\right)+\left(\frac{r_{+}}{M}-1\right)
\tilde{\omega}_{I}^{2}}{\left|\tilde{\omega}-\frac{eQ}{r_{+}}\right|^{2}}
\frac{\delta  A}{2A}.
\label{eq:DeltaM}
\end{equation}
From $M<r_{+}<2M$, the superradiance condition, and $\delta  A \ge 0$, we can conclude that $\delta  M \le 0$ if  
$\tilde{\omega}_{I}^{2}$ is sufficiently small. Therefore, the weak superradiance decreases both $M$ and $|Q|$ but increases $A$.
This can be understood as follows. Note that $M$ and $E_{H}$ stand for the total energy and the quasi-local energy within the horizon, respectively, while 
the difference $Q^{2}/(2r_{+})$ between the two
corresponds to the energy of the electromagnetic field outside the horizon. Since 
\begin{equation}
 M=E_{H}+\frac{Q^{2}}{2r_{+}},
\end{equation}
even if $E_{H}$ increases, the decrease in ${Q^{2}}/(2r_{+})$ can overcompensate it. This is the case for the weak superradiance and 
the total energy will decrease in the quasi-static process. In the asymptotically flat case, the difference in the total energy will be radiated away to infinity, while in the asymptotically AdS case in the reflective boundary, it will be converted to the energy of the ambient scalar field.

Since the null energy condition is satisfied in the present system and the boundary condition at the conformal infinity 
uniquely determines the time development of the scalar field, we can expect that the area law of the AdS black hole holds
in an appropriate formulation~\cite{Ishibashi:2020}. We can see that the superradiance conditions given by Eqs.~\eqref{scon} 
and~\eqref{scon2} in terms of $\tilde{\omega}_{R}$ imply $\delta  A\ge 0$ 
for $eQ>0$ and $eQ<0$, respectively.
Therefore,
from a thermodynamical point of view,
we conclude that 
the black hole may remain the (approximate) Reissner-Nordstr\"om solution with 
decreasing its mass and the absolute value of its charge, 
while increasing its area 
in a quasistatic manner during the superradiance instability,
as long as the black hole is much smaller than the AdS length.

\section{Conclusion}
In this paper, we have extended the result of~\cite{Ishibashi:2004wx} for a neutral massive scalar field in the AdS spacetime to charged and neutral massive scalar fields in a charged AdS black hole. In Section \ref{sec:formulation}, we have clarified the parameter range for which the matched asymptotic expansion method applies for charged massive scalar fields and derived an equation which determines QNFs.  
In Section \ref{sec:analytic_results}, we have analytically shown that the 
quasinormal mode of the small charged AdS black hole becomes superradiantly 
unstable if its frequency in the pure AdS spacetime satisfies the superradiance condition.
In Section \ref{sec:numerical_results}, we have numerically obtained the relation between the parameter of the boundary condition at the conformal infinity, $\zeta\in[0,\pi]$, and the QNFs of the scalar field in the AdS black holes with $\left(r_+,r_-\right)=\left(0.01\ell,0.001\ell\right)$ and $\left(r_+,r_-\right)=\left(0.1\ell,0.001\ell\right)$. 

We have shown that the first fundamental mode of the neutral scalar field in the AdS black hole has a critical value $\zeta_0$ beyond which the real part of the QNF vanishes and the imaginary part splits into two different values. We note that such a critical value does not exist for the second fundamental mode. 
We have also shown that there exists a critical value $\zeta_c$, corresponding to the onset of instability. Furthermore, we find $\zeta_0<\zeta_c$ and the imaginary parts are negative for $\zeta_0<\zeta<\zeta_c$. For $\zeta_c<\zeta$, there are two QNFs that have vanishing real part but positive and negative values for the imaginary part. This result indicates instability for $\zeta_c<\zeta$. 
Since the field is neutral, there is no superradiance. Hence, this instability arises from the boundary condition. The time scale of the instability becomes shorter as $\zeta$ increases. At $\zeta=\zeta_c$, there exists a static neutral perturbation in the charged AdS black hole. For $\zeta< \zeta_c$, the evolution is stable and the time scale of the decay becomes longer as $\zeta$ increases. Moreover, we analytically and numerically establish that although the system of the neutral field in the pure AdS spacetime has both symmetries 
$\omega\to -\omega$ and $\omega\to \omega^{*}$, the existence of the black hole horizon breaks them. Instead, a symmetry $\omega\to -\omega^{*}$ remains for the neutral field on the AdS black hole spacetime.

The charged scalar field has no critical value at which the real part of the QNF becomes zero. The results we have shown imply that the evolution of the charged field can be unstable by superradiance whether the charge is negative or positive.  
If $e|Q|\ell /r_{+}>(3+\sqrt{9+4\mu^2\ell^2})/2$, the system is superradiantly unstable irrespectively of $\zeta$, while if $e|Q|\ell /r_{+}\ge (3+\sqrt{9+4\mu^2\ell^2})/2$, the superradiant stability depends on $\zeta$. The instability can be caused by the charged scalar field even for $\zeta\leq\zeta_0$, where $\zeta_0$ is the critical value at which $\textrm{Re}[\omega]=0$ for \textit{the neutral field}. This is remarkably different from the neutral field. Furthermore, the evolution is unstable for $\zeta_c<\zeta$. The time scale of the instability for $\zeta_0<\zeta$ can be much shorter than that for $\zeta\leq \zeta_0$. In this sense, the superradiant instability can be enhanced by the boundary condition. We point out that the symmetry $\tilde{\omega}\to -\tilde{\omega}^{*}$, which exists for the neutral field, is broken. Actually, the symmetry is deformed 
 to 
$\left(\tilde{\omega}, eQ\right)\to\left(-\tilde{\omega}^{*}, -eQ\right)$. We also show that the weak superradiance in this system extracts energy not from the black hole but from its ambient electric field and that it extracts charge from the black hole and thus weakens its ambient electric field. The Robin boundary condition makes the conformal infinity perfectly reflective so that the superradiance induces instability.
This can be interpreted as a classical scalar-field counterpart of discharging a black hole by 
the Schwinger effect. The black hole increases its energy within the horizon, decreases the absolute value of its charge, increases its area, but decreases its mass parameter, as long as the black hole is much smaller than the AdS length.

With the symmetry of the QNFs, the superradiant instability in this system is independent of the sign of $eQ$ in contrast to the claim in~\cite{Uchikata:2011zz}. Furthermore, we expect that this symmetry gives some explanation on instability shown in~\cite{Maeda:2010hf}. They showed that charged scalar fields in the Reissner-Nordstr\"om-AdS black holes can cause instability irrespectively of the sign of $eQ$. They, however, claim that this has nothing to do with superradiance because the QNMs with $eQ<0$ cannot satisfy the superradiance condition. Although they do not assume $\textrm{Re}[\tilde{\omega}]>0$ in the analysis, they seem to assume it when they interpret the result. Our result implies that the instability they find comes from the superradiance because the symmetry shows that  the superradiance \textit{can} 
occur even in the case of $eQ<0$ if we admit $\textrm{Re}[\tilde{\omega}]<0$.

The results in this paper are restricted to black holes much smaller than 
the AdS length scale. 
On the other hand, the instability of large charged AdS black holes has been intensively discussed in terms of the holographic superconductor. In the context of AdS/CFT correspondence, the instabilities of the AdS black hole are expected to imply the phase transition of dual theories. Our result suggests the new dynamics of the dual quantum field theory. Also, we would like to know the final fate of the instabilities. Recently, Ref.~\cite{Bizon:2020yqs} has shown the existence of static solitons in the 4-dimensional AdS spacetime with the boundary condition corresponding to $\zeta_c<\zeta$. Their result suggests the existence of black hole solutions with a nontrivial scalar field satisfying the boundary condition $\zeta_c<\zeta$. Then, the hairy black hole solution with the nontrivial ambient scalar field could be a candidate for the final fate of the instabilities we have shown. For the charged case, both the black hole and the charged scalar field have charges of the same sign. These problems remain for our outlook.

\begin{acknowledgments}
The authors are grateful to Akihiro Ishibashi for his helpful comments. The authors would like to thank Takaaki Ishii, Masashi Kimura, Keiju Murata, and Norihiro Tanahashi for the informative comments. TK acknowledges Shunichiro Kinoshita and Shin Nakamura for the hospitality and the helpful comments on the seminar at particle theory group in Chuo university. The authors also thank the workshop ``GR in AdS''. This work was supported by Rikkyo University Special Fund for Research (TK) and JSPS KAKENHI Grant Numbers JP19K03876 and JP19H01895 (TH).
\end{acknowledgments}

\appendix
\section{Positive self-adjoint extension of symmetric operators \label{self-adjoint_extension}}
Here we provide definitions for the mathematical notions used in Section~\ref{sec:Ishibashi_Wald}. 

A subset $D$ of $\mathcal{H}$ is said to be \textit{dense} if $D$ satisfies $\mathcal{H}=\bar{D}$, where $\bar{D}$ is the closure of $D$. 
A vector $\psi\in \mathcal{H}$ is said to be \textit{normalisable} if and only if 
$\left(\psi,\psi\right)<\infty$.
The domain of a linear operator $\mathcal{O}$ is denoted by $D(\mathcal{O})$.

A linear operator $\mathcal{O}$ is said to be \textit{positive} if and only if 
$\left(\psi,\mathcal{O}\psi\right)>0$
for any nonzero vector $\psi\in D(\mathcal{O})$, where $(\cdot,\cdot)$ stands for the inner product. 
A linear operator $\mathcal{P}$ is said to be an \textit{adjoint} of $\mathcal{O}$ if and only if  $(\psi,\mathcal{O}\phi)=(\Theta,\phi)$ and $\mathcal{P}\psi=\Theta$
for any vectors $\phi\in D({\mathcal{O}})$, $\psi\in D(\mathcal{P})$ and $\Theta\in\mathcal{H}$. The adjoint of $\mathcal{O}$ is denoted by $\mathcal{O}^{*}$.
A linear operator $\mathcal{P}$ is said to be an \textit{extension} of $\mathcal{O}$
if and only if 
$
D(\mathcal{O}) \subset D(\mathcal{P}).
$

A linear operator $\mathcal{O}$ is said to be \textit{symmetric} if all of 
the following three conditions are satisfied: 
(i) $D(\mathcal{O})$ is dense, 
(ii) $(\psi,\mathcal{O}\phi)=(\mathcal{O}\psi,\phi)$ for any vectors $\psi, \phi \in D(\mathcal{O})$, 
and (iii) $D(\mathcal{O})\subset D({\mathcal{O}}^*)$.
Therefore, the adjoint of a symmetric operator is an extension of the symmetric operator.

A symmetric operator 
$\mathcal{O}$ is said to be \textit{bounded below if and only if 
there exists $\gamma \in \mathbb{R}$ such that $(\mathcal{O}\psi,\psi)\ge \gamma (\psi,\psi)$} for any vector $\psi\in D(\mathcal{O})$.  A symmetric operator is said to be \textit{unbounded below} if and only if it is not bounded below.

A linear operator $\mathcal{O}$ is said to be \textit{self-adjoint} if and only if all of the following three conditions are satisfied: 
(i) $D(\mathcal{O})$ is dense, 
(ii) $(\psi,\mathcal{O}\phi)=(\mathcal{O}\psi,\phi)$ for any vectors $\psi, \phi \in D(\mathcal{O})$, and (iii) $D(\mathcal{O})= D({\mathcal{O}}^*)$.
One can obtain a self-adjoint operator by extending a symmetric operator~\cite{Reed:1975,Ishibashi:1999vw}. Thus obtained self-adjoint operator is 
called \textit{self-adjoint extension}.

\section{Validity of the matching of the near-region and far-region solutions \label{sec:validity_matching}}

We introduce a small parameter
$
\epsilon={r_+}/{\ell}
$ and a nondimensional coordinate $x={r}/{r_+}$.
First we focus on the validity of the near-horizon solution. 
Eq. (\ref{fue1}) is rewritten in the form 
\begin{equation}
\begin{split}
\mathcal{E}_1(x)=-\frac{\left(\epsilon x\right)^2}{1-\frac{r_-}{r_+}}&\left[-\mu^2\ell^2\left(x-\frac{r_-}{r_+}\right)+\left(\tilde{\omega}\ell\right)^2(1+x)\left(1+\frac{1}{x^2}\right)\right.\\
&\left.-2\left(\tilde{\omega}\ell\right)\left(\frac{eQ}{\epsilon}\right)\left(1+\frac{1}{x}+\frac{1}{x^2}\right)+\left(\frac{eQ}{\epsilon}\right)^2\left(\frac{1}{x}+\frac{1}{x^2}\right)\right].
\end{split}
\end{equation}
Because $-\left(\tilde{\omega}\ell\right)(eQ)\leq|\tilde{\omega}\ell||eQ|$, we find an inequality
$
|\mathcal{E}_1(x)|\leq \tilde{\mathcal{E}}_1(x)$,
where we have defined
\begin{equation}
\begin{split}
\tilde{\mathcal{E}}_1(x):=\frac{\left(\epsilon x\right)^2}{1-\frac{r_-}{r_+}}&\left[-\mu^2\ell^2\left(x-\frac{r_-}{r_+}\right)+|\tilde{\omega}\ell|^2(1+x)\left(1+\frac{1}{x^2}\right)\right.\\
&\left.+2|\tilde{\omega}\ell|\left(\frac{|eQ|}{\epsilon}\right)\left(1+\frac{1}{x}+\frac{1}{x^2}\right)+\left(\frac{|eQ|}{\epsilon}\right)^2\left(\frac{1}{x}+\frac{1}{x^2}\right)\right].
\end{split}
\end{equation}
Here we note that $\tilde{\mathcal{E}}_1(x)$ is an increasing function of $x$.
In the limit of $1\ll x$, the asymptotic behaviour of $\tilde{\mathcal{E}}_1(x)$ can be written as 
\begin{equation}
\begin{split}
\tilde{\mathcal{E}}_1(x)\simeq x\left[\left\{|\tilde{\omega}\ell|\left( \epsilon x\right)+|eQ|\right\}^2-\mu^2\ell^2\left(\epsilon x\right)^2\right].
\end{split}
\end{equation}
As for $|ab|$, we have an inequality
\begin{equation}
1<|ab|=\sqrt{4\sigma^2 +(l+1)^2}.
\end{equation}
Hence, $|\mathcal{E}_1(x)|\ll |ab|$ is satisfied in the region given by $1< x\leq x_0$,  if $x_0$ satisfies $1\ll x_0\ll1/\epsilon$ and 
\begin{equation}
\begin{split}
\label{e1}
\tilde{\mathcal{E}}_1(x_0)\ll1\Leftrightarrow\epsilon^2 \left\{\left(|\tilde{\omega}\ell|+\frac{|eQ|}{\epsilon}\frac{1}{x_0}\right)^2-\mu^2\ell^2\right\}\ll\frac{1}{{x_0}^3}.
\end{split}
\end{equation}
Assuming $\tilde{\omega}\ell =O(1)$ and $eQ=o(\epsilon^{1/3+\delta})$, 
we find that if we take $x_{0}=c_{0}\epsilon^{-2/3+\delta}$ with $c_{0}=O(1)$ and $0<\delta <2/3$, the inequality (\ref{e1}) is satisfied.

Next we focus on the validity of the far-region solution. Clearly, Eq. (\ref{Hyf}) is reduced to 
Eq. (\ref{Hyfa}) for the neutral scalar field because of $\mathcal{E}_2(r)=0$. 
For the charged field,  
we can approximate Eq. (\ref{Hyf}) to Eq. (\ref{Hyfa}) if $\mathcal{E}_2(r)$ satisfies
$
h(y)|\mathcal{E}_2(y)|\ll1$,
where we have defined
\begin{equation}
h(y):=\frac{|g(y)|}{|y(1-y)\frac{d^2}{dy^2}g(y)|+|\left\{\gamma-(\alpha+\beta+1)y\right\}\frac{d}{dy}g(y)|+|-\alpha\beta g(y)|}.
\end{equation}
Substituting Eq. (\ref{gys}) into the above $g(y)$, we notice $\textrm{max}~h(y)=\mathcal{O}(1)$ for $y\lesssim200$ with $|\tilde{\omega}\ell|=\mathcal{O}(1)$, $l=\mathcal{O}(1)$. Hence, the charged scalar field in the far region can be described by the Gaussian hypergeometric functions if $|\mathcal{E}_2(y)|\ll1$ is satisfied.
We shall discuss whether $|\mathcal{E}_2(y)|\ll1$ is satisfied. In terms of $\epsilon$ and $x$, $\mathcal{E}_2(y)$ is rewritten as
\begin{equation}
\begin{split}
\mathcal{E}_2(x)=\frac{1}{4\left(\epsilon x\right)^2}\frac{1}{1+\left(\epsilon x\right)^2}\left\{2\left(\tilde{\omega}\ell\right)(eQ)\left(\epsilon x\right)-\left(eQ\right)^2\right\}.
\end{split}
\end{equation}
Noting $-\left(\tilde{\omega}\ell\right)(eQ)\leq |\tilde{\omega}\ell||eQ|$, we find
\begin{equation}
|\mathcal{E}_2(x)|\leq\tilde{\mathcal{E}}_2(x),
\end{equation}
where we have defined
\begin{equation}
\begin{split}
\label{E2}
\tilde{\mathcal{E}}_2(x):=\frac{1}{4\left(\epsilon x\right)^2}\frac{1}{1+\left(\epsilon x\right)^2}\left\{2|\tilde{\omega}\ell||eQ|\left(\epsilon x\right)+|eQ|^2\right\}.
\end{split}
\end{equation}
Note that $\tilde{\mathcal{E}}_2(x)$ is a decreasing function of $x$ and $\tilde{\mathcal{E}}_2(x)\ll1$ in the limit of $x\to\infty$. Therefore, if we can find $x_1$ such that $\tilde{\mathcal{E}}_2(x_1)\ll1$ and $1\ll x_1\ll1/\epsilon$,  $|\mathcal{E}_2(x)|\ll1$ is satisfied for $x_1\leq x$.  
Assuming $|eQ|\ll \epsilon x_{1}$ and $\tilde{\omega}\ell =O(1)$, 
$|\mathcal{E}_2(y)|\ll1$ holds for $x_{1}\le x<\infty$.

Finally, we discuss whether there is an overlapping region, where both the near-region and far-region solutions are valid.
For the neutral field, we can match them without further discussion. The reason is that the far-region approximate solution (\ref{solfa}) can describe the field everywhere in the region $1\ll x$ and the near-region approximate solution (\ref{soln}) is valid in $1\ll x\leq x_0$ such that $1\ll x_0\ll1/\epsilon$ because there exists $x_0$ such that the inequality (\ref{e1}) is satisfied for $eQ=0$ under the assumption $|\omega\ell|=\mathcal{O}\left(1\right)$.
For the charged field case, assuming $|eQ|=o(\epsilon^{1/3+\delta})$ and $|\tilde{\omega}\ell|=O(1)$,
if we choose $x_{0}=c_{0}\epsilon^{-2/3+\delta}$ and $x_{1}=c_{1}\epsilon^{-2/3+\delta}$ with
$0<c_{1}<c_{0}$ and $0<\delta<2/3$, both $\tilde{\mathcal{E}}_{1}\ll 1$ and $\tilde{\mathcal{E}}_{2} \ll 1$ are satisfied for $x_{1}\le x\le x_{0}$. 
Therefore, we can identify the overlapping region with the interval $x_{1}\le x\le x_{0}$.

\section{Asymptotic behaviours of the near-region and far-region solutions \label{sec:Gaussian_hypergeometric_functions}}

We are interested in the asymptotic behaviour of the far-region solution (\ref{solfa}) at $y\simeq 1$. Using the transformation formula of the Gaussian hypergeometric functions~\cite{textbook1}, we obtain
\begin{equation}
\begin{split}
\label{trsf1}
F\left(\alpha,\alpha-\gamma+1;\alpha-\beta+1;\frac{1}{y}\right)=&y^{\alpha-\gamma+1}(y-1)^{\gamma-\alpha-\beta}\frac{\Gamma(\alpha-\beta+1)\Gamma(\alpha+\beta-\gamma)}{\Gamma(\alpha)\Gamma(\alpha-\gamma+1)}\\
&\times F(1-\alpha,1-\beta;\gamma-\alpha-\beta;y-1)\\
&+y^{\alpha}\frac{\Gamma(\alpha-\beta+1)\Gamma(\gamma-\alpha-\beta)}{\Gamma(1-\beta)\Gamma(\gamma-\beta)}\\
&\times F(\alpha,\beta;\alpha+\beta-\gamma+1;y-1),
\end{split}
\end{equation}
and
\begin{equation}
\begin{split}
\label{trsf2}
F\left(\beta,\beta-\gamma+1;\beta-\alpha+1;\frac{1}{y}\right)=&y^{\beta-\gamma+1}(y-1)^{\gamma-\alpha-\beta}\frac{\Gamma(\beta-\alpha+1)\Gamma(\alpha+\beta-\gamma)}{\Gamma(\beta)\Gamma(\beta-\gamma+1)}\\
&\times F(1-\alpha,1-\beta;\gamma-\alpha-\beta;y-1)\\
&+y^{\beta}\frac{\Gamma(\beta-\alpha+1)\Gamma(\gamma-\alpha-\beta)}{\Gamma(1-\alpha)\Gamma(\gamma-\alpha)}\\
&\times F(\alpha,\beta;\alpha+\beta-\gamma+1;y-1).
\end{split}
\end{equation}
Therefore Eq. (\ref{solfa}) is rewritten in the form
\begin{equation}
\begin{split}
\label{sol2a}
\psi(y)=&Dy^{\frac{\tilde{\omega}\ell}{2}}(y-1)^{\frac{l}{2}}\\
&\times\left[y^{-\gamma+1}(y-1)^{\gamma-\alpha-\beta}\Gamma(\alpha+\beta-\gamma)\left(\frac{\kappa\Gamma(\alpha-\beta+1)}{\Gamma(\alpha)\Gamma(\alpha-\gamma+1)}+\frac{\Gamma(\beta-\alpha+1)}{\Gamma(\beta)\Gamma(\beta-\gamma+1)}\right)\right.\\
&\times F(1-\alpha,1-\beta;\gamma-\alpha-\beta;y-1)\\
&+\Gamma(\gamma-\alpha-\beta)\left(\frac{\kappa\Gamma(\alpha-\beta+1)}{\Gamma(1-\beta)\Gamma(\gamma-\beta)}+\frac{\Gamma(\beta-\alpha+1)}{\Gamma(1-\alpha)\Gamma(\gamma-\alpha)}\right)\\
&\left.\times F(\alpha,\beta;\alpha+\beta-\gamma+1;y-1)\right].
\end{split}
\end{equation}
The above expression gives the asymptotic behaviour (\ref{solf1a}).

Next we are interested in the asymptotic behaviour of the near-region solution (\ref{soln}) at $z\simeq 1$. 
Using the transformation formula of the Gaussian hypergeometric functions~\cite{textbook1},
\begin{equation}
\begin{split}
F(1-c+a,1-c+b;2-c;z)=&\frac{\Gamma(2-c)\Gamma(c-a-b)}{\Gamma(1-a)\Gamma(1-b)}\\
&\times F(1-c+a,1-c+b;-c+a+b+1;1-z)\\
&+(1-z)^{c-a-b}\frac{\Gamma(2-c)\Gamma(-c+a+b)}{\Gamma(1-c+a)\Gamma(1-c+b)}\\
&\times F(1-a,1-b;c-a-b+1;1-z).
\end{split}
\end{equation}
Therefore, the near-region solution (\ref{soln}) is rewritten in the form
\begin{equation}
\begin{split}
\label{soln1}
\psi(z)=Bz^{-i\sigma}\Gamma(1-2i\sigma)&\left[\frac{\Gamma(-2l-1)}{\Gamma(-2i\sigma-l)\Gamma(-l)}(1-z)^{l+1}\right.\\
&\times F(1-c+a,1-c+b;-c+a+b+1;1-z)\\
&+\frac{\Gamma(2l+1)}{\Gamma(-2i\sigma+l+1)\Gamma(l+1)}(1-z)^{-l}\\
&\left.\times F(1-a,1-b;c-a-b+1;1-z)\right].\\
\end{split}
\end{equation}
From the above expression, we obtain the asymptotic form (\ref{soln2}) with Eq.~(\ref{B12}).

\section{Symmetry $(\omega,eQ)\to (-\omega^{*},-eQ)$ in the matched asymptotic expansion \label{sec:symmetry}}

Note that $\sigma$ is transformed to $-\sigma^{*}$ by this transformation.
Then, the left-hand side of Eq. (\ref{match}) is transformed as 
\begin{equation}
\frac{B_1(\tilde{\omega}, \sigma)}{B_2(\tilde{\omega},\sigma)}\to
\frac{B_1(-\tilde{\omega}^{*}, -\sigma^{*})}{B_2(-\tilde{\omega}^{*},-\sigma^{*})}
=\left(\frac{B_1(\tilde{\omega},\sigma)}{B_2(\tilde{\omega},\sigma)}\right)^*,
\end{equation}
because $B_1(\tilde{\omega},\sigma)$ and $B_2(\tilde{\omega},\sigma)$ satisfy the relation $B_1(\tilde{\omega},\sigma)=B_1^*(-\tilde{\omega}^*,-\sigma^{*})$ 
and $B_2(\tilde{\omega},\sigma)=B_2^*(-\tilde{\omega}^*,-\sigma^{*})$. Next, the right-hand side of Eq. (\ref{match}) is transformed as
\begin{equation}
\frac{D_1(\tilde{\omega},\kappa)}{D_2(\tilde{\omega},\kappa)}\to
\frac{D_1(-\tilde{\omega}^{*},\kappa)}{D_2(-\tilde{\omega}^{*},\kappa)}=
\left(\frac{D_1(\tilde{\omega},\kappa)}{D_2(\tilde{\omega},\kappa)}\right)^*,
\end{equation}
because of $D_1(\tilde{\omega},\kappa)={D_1}^*(-\tilde{\omega}^*,\kappa)$ and $D_2(\tilde{\omega},\kappa)={D_2}^*(-\tilde{\omega}^*,\kappa)$ 
as stated below Eq. (\ref{rega}). Hence, this transformation brings Eq. (\ref{match}) to
\begin{equation}
\left(\frac{B_1(\tilde{\omega},\sigma)}{B_2(\tilde{\omega},\sigma)}\right)^*=
\ell^{2l+1}\left(\frac{D_1(\tilde{\omega},\kappa)}{D_2(\tilde{\omega},\kappa)}\right)^*.
\end{equation}
Since the above is nothing but the complex conjugate of Eq. (\ref{match}), the solution of the above is the same as the solution obtained before the transformation. Thus, the symmetry for the charged scalar field in the charged black hole is given by $\left(\tilde{\omega}, eQ\right)\to\left(-\tilde{\omega}^{*}, -eQ\right)$.

\section{The explicit calculations}
\label{appendix:Theexplicitcalculations}
We give the explicit forms of $\Sigma_{1,R}$, $\Sigma_{1,I}$, $\Sigma_{2,R}$, and $\Sigma_{2,I}$ in Eqs.~\eqref{SD1} and~\eqref{SD2}, and ${\rm Re}[B_1/B_2]$, ${\rm Im}[B_1/B_2]$ in Eqs.~\eqref{matchingReal} and~\eqref{matchingImaginary}.

\subsection{$\Sigma_{1,R}$ and $\Sigma_{1,I}$}
$D_1\left(\tilde{\omega},\kappa\right)$ in Eq.~\eqref{match} is explicitly
\begin{equation}
\begin{split}
\label{D1}
D_1\left(\tilde{\omega},\kappa\right)=&\frac{\kappa \Gamma\left(1+\frac{1}{2}\sqrt{9+4\mu^2\ell^2}\right)}{\Gamma\left(\frac{\tilde{\omega}\ell}{2}+\frac{l}{2}+\frac{3}{4}+\frac{1}{4}\sqrt{9+4\mu^2\ell^2}\right)\Gamma\left(-\frac{\tilde{\omega}\ell}{2}+\frac{l}{2}+\frac{3}{4}+\frac{1}{4}\sqrt{9+4\mu^2\ell^2}\right)}\\
&+\frac{ \Gamma\left(1-\frac{1}{2}\sqrt{9+4\mu^2\ell^2}\right)}{\Gamma\left(\frac{\tilde{\omega}\ell}{2}+\frac{l}{2}+\frac{3}{4}-\frac{1}{4}\sqrt{9+4\mu^2\ell^2}\right)\Gamma\left(-\frac{\tilde{\omega}\ell}{2}+\frac{l}{2}+\frac{3}{4}-\frac{1}{4}\sqrt{9+4\mu^2\ell^2}\right)}.
\end{split}
\end{equation}
Assuming $|{\rm Im}\left[\tilde{\omega}\ell\right]|\ll1$, $D_1\left(\tilde{\omega},\kappa\right)$ takes a form
\begin{equation}
\begin{split}
\label{AD1}
D_1\left(\tilde{\omega},\kappa\right)=&\Sigma_{1,R}-i\frac{\Sigma_{1,I}}{2}{\rm Im}\left[\tilde{\omega}\ell\right]+\mathcal{O}\left(\left({\rm Im}\left[\tilde{\omega}\ell\right]\right)^2\right),
\end{split}
\end{equation}
where
\begin{equation}
\begin{split}
\label{Sigma1R}
\Sigma_{1,R}=&\frac{\kappa \Gamma\left(1+\frac{1}{2}\sqrt{9+4\mu^2\ell^2}\right)}{\Gamma\left(\frac{{\rm Re}\left[\tilde{\omega}\ell\right]}{2}+\frac{l}{2}+\frac{3}{4}+\frac{1}{4}\sqrt{9+4\mu^2\ell^2}\right)\Gamma\left(-\frac{{\rm Re}\left[\tilde{\omega}\ell\right]}{2}+\frac{l}{2}+\frac{3}{4}+\frac{1}{4}\sqrt{9+4\mu^2\ell^2}\right)}\\
&+\frac{ \Gamma\left(1-\frac{1}{2}\sqrt{9+4\mu^2\ell^2}\right)}{\Gamma\left(\frac{{\rm Re}\left[\tilde{\omega}\ell\right]}{2}+\frac{l}{2}+\frac{3}{4}-\frac{1}{4}\sqrt{9+4\mu^2\ell^2}\right)\Gamma\left(-\frac{{\rm Re}\left[\tilde{\omega}\ell\right]}{2}+\frac{l}{2}+\frac{3}{4}-\frac{1}{4}\sqrt{9+4\mu^2\ell^2}\right)}.
\end{split}
\end{equation}
and
\begin{equation}
\begin{split}
\label{Sigma1I}
\Sigma_{1,I}
=&\frac{\kappa \Gamma\left(1+\frac{1}{2}\sqrt{9+4\mu^2\ell^2}\right)}{\Gamma\left(\frac{{\rm Re}\left[\tilde{\omega}\ell\right]}{2}+\frac{l}{2}+\frac{3}{4}+\frac{1}{4}\sqrt{9+4\mu^2\ell^2}\right)\Gamma\left(-\frac{{\rm Re}\left[\tilde{\omega}\ell\right]}{2}+\frac{l}{2}+\frac{3}{4}+\frac{1}{4}\sqrt{9+4\mu^2\ell^2}\right)}\\
&\times\left(P\left(\frac{{\rm Re}\left[\tilde{\omega}\ell\right]}{2}+\frac{l}{2}+\frac{3}{4}+\frac{1}{4}\sqrt{9+4\mu^2\ell^2}\right)-P\left(-\frac{{\rm Re}\left[\tilde{\omega}\ell\right]}{2}+\frac{l}{2}+\frac{3}{4}+\frac{1}{4}\sqrt{9+4\mu^2\ell^2}\right)\right)\\
&+\frac{ \Gamma\left(1-\frac{1}{2}\sqrt{9+4\mu^2\ell^2}\right)}{\Gamma\left(\frac{{\rm Re}\left[\tilde{\omega}\ell\right]}{2}+\frac{l}{2}+\frac{3}{4}-\frac{1}{4}\sqrt{9+4\mu^2\ell^2}\right)\Gamma\left(-\frac{{\rm Re}\left[\tilde{\omega}\ell\right]}{2}+\frac{l}{2}+\frac{3}{4}-\frac{1}{4}\sqrt{9+4\mu^2\ell^2}\right)}\\
&\times\left(P\left(\frac{{\rm Re}\left[\tilde{\omega}\ell\right]}{2}+\frac{l}{2}+\frac{3}{4}-\frac{1}{4}\sqrt{9+4\mu^2\ell^2}\right)-P\left(-\frac{{\rm Re}\left[\tilde{\omega}\ell\right]}{2}+\frac{l}{2}+\frac{3}{4}-\frac{1}{4}\sqrt{9+4\mu^2\ell^2}\right)\right).
\end{split}
\end{equation}
Here, $P(z)$ is the digamma function. Eq.~\eqref{AD1} is the same as Eq.~\eqref{SD1}. 

\subsection{$\Sigma_{2,R}$ and $\Sigma_{2,I}$}
$D_2(\tilde{\omega},\kappa)$ in Eq.~\eqref{match} is explicitly
\begin{equation}
\begin{split}
\label{D2}
D_2\left(\tilde{\omega},\kappa\right)=&\frac{\Gamma\left(-l-\frac{1}{2}\right)}{\Gamma\left(l+\frac{1}{2}\right)}\\
&\times\left(\frac{\kappa \Gamma\left(1+\frac{1}{2}\sqrt{9+4\mu^2\ell^2}\right)}{\Gamma\left(\frac{\tilde{\omega}\ell}{2}-\frac{l}{2}+\frac{1}{4}+\frac{1}{4}\sqrt{9+4\mu^2\ell^2}\right)\Gamma\left(-\frac{\tilde{\omega}\ell}{2}-\frac{l}{2}+\frac{1}{4}+\frac{1}{4}\sqrt{9+4\mu^2\ell^2}\right)}\right.\\
&\left.~~~~~~+\frac{ \Gamma\left(1-\frac{1}{2}\sqrt{9+4\mu^2\ell^2}\right)}{\Gamma\left(\frac{\tilde{\omega}\ell}{2}-\frac{l}{2}+\frac{1}{4}-\frac{1}{4}\sqrt{9+4\mu^2\ell^2}\right)\Gamma\left(-\frac{\tilde{\omega}\ell}{2}-\frac{l}{2}+\frac{1}{4}-\frac{1}{4}\sqrt{9+4\mu^2\ell^2}\right)}\right).
\end{split}
\end{equation}
We have
\begin{equation}
\label{D2GG}
\frac{\Gamma\left(-l-\frac{1}{2}\right)}{\Gamma\left(l+\frac{1}{2}\right)}=\frac{2^{2l+1}\left(-1\right)^{l+1}}{\left(2l+1\right)!!\left(2l-1\right)!!},
\end{equation}
by using properties of the gamma functions for a non-negative integer $m$,
\begin{equation}
\begin{split}
\Gamma\left(-m-\frac{1}{2}\right)=&(-1)^{m+1}2^{m+1}\frac{\pi^{1/2}}{(2m+1)!!},\\
\Gamma\left(m+\frac{1}{2}\right)=&2^{-m}\pi^{1/2}(2m-1)!!.
\end{split}
\end{equation}
Assuming $|{\rm Im}\left[\tilde{\omega}\ell\right]|\ll1$, $D_2\left(\tilde{\omega},\kappa\right)$ takes a form
\begin{equation}
\label{AD2}
D_2\left(\tilde{\omega},\kappa\right)=\frac{2^{2l+1}\left(-1\right)^{l+1}}{\left(2l+1\right)!!\left(2l-1\right)!!}\left[\Sigma_{2,R}--i\frac{\Sigma_{2,I}}{2}{\rm Im}\left[\tilde{\omega}\ell\right]\right]+\mathcal{O}\left(\left({\rm Im}\left[\tilde{\omega}\ell\right]\right)^2\right),
\end{equation}
where
\begin{equation}
\begin{split}
\label{Sigma2R}
\Sigma_{2,R}=&\frac{\kappa \Gamma\left(1+\frac{1}{2}\sqrt{9+4\mu^2\ell^2}\right)}{\Gamma\left(\frac{{\rm Re}\left[\tilde{\omega}\ell\right]}{2}-\frac{l}{2}+\frac{1}{4}+\frac{1}{4}\sqrt{9+4\mu^2\ell^2}\right)\Gamma\left(-\frac{{\rm Re}\left[\tilde{\omega}\ell\right]}{2}-\frac{l}{2}+\frac{1}{4}+\frac{1}{4}\sqrt{9+4\mu^2\ell^2}\right)}\\
&+\frac{ \Gamma\left(1-\frac{1}{2}\sqrt{9+4\mu^2\ell^2}\right)}{\Gamma\left(\frac{{\rm Re}\left[\tilde{\omega}\ell\right]}{2}-\frac{l}{2}+\frac{1}{4}-\frac{1}{4}\sqrt{9+4\mu^2\ell^2}\right)\Gamma\left(-\frac{{\rm Re}\left[\tilde{\omega}\ell\right]}{2}-\frac{l}{2}+\frac{1}{4}-\frac{1}{4}\sqrt{9+4\mu^2\ell^2}\right)}
\end{split}
\end{equation}
and
\begin{equation}
\begin{split}
\label{Sigma2I}
\Sigma_{2,I}
=&\frac{\kappa \Gamma\left(1+\frac{1}{2}\sqrt{9+4\mu^2\ell^2}\right)}{\Gamma\left(\frac{{\rm Re}\left[\tilde{\omega}\ell\right]}{2}-\frac{l}{2}+\frac{1}{4}+\frac{1}{4}\sqrt{9+4\mu^2\ell^2}\right)\Gamma\left(-\frac{{\rm Re}\left[\tilde{\omega}\ell\right]}{2}-\frac{l}{2}+\frac{1}{4}+\frac{1}{4}\sqrt{9+4\mu^2\ell^2}\right)}\\
&\times\left(P\left(\frac{{\rm Re}\left[\tilde{\omega}\ell\right]}{2}-\frac{l}{2}+\frac{1}{4}+\frac{1}{4}\sqrt{9+4\mu^2\ell^2}\right)-P\left(-\frac{{\rm Re}\left[\tilde{\omega}\ell\right]}{2}-\frac{l}{2}+\frac{1}{4}+\frac{1}{4}\sqrt{9+4\mu^2\ell^2}\right)\right)\\
&+\frac{ \Gamma\left(1-\frac{1}{2}\sqrt{9+4\mu^2\ell^2}\right)}{\Gamma\left(\frac{{\rm Re}\left[\tilde{\omega}\ell\right]}{2}-\frac{l}{2}+\frac{1}{4}-\frac{1}{4}\sqrt{9+4\mu^2\ell^2}\right)\Gamma\left(-\frac{{\rm Re}\left[\tilde{\omega}\ell\right]}{2}-\frac{l}{2}+\frac{1}{4}-\frac{1}{4}\sqrt{9+4\mu^2\ell^2}\right)}\\
&\times\left(P\left(\frac{{\rm Re}\left[\tilde{\omega}\ell\right]}{2}-\frac{l}{2}+\frac{1}{4}-\frac{1}{4}\sqrt{9+4\mu^2\ell^2}\right)-P\left(-\frac{{\rm Re}\left[\tilde{\omega}\ell\right]}{2}-\frac{l}{2}+\frac{1}{4}-\frac{1}{4}\sqrt{9+4\mu^2\ell^2}\right)\right).
\end{split}
\end{equation}
Here, $P(z)$ is the digamma function. Eq.~\eqref{AD2} is the same as Eq.~\eqref{SD2}. 

\subsection{${\rm Re}[B_1/B2]$ and ${\rm Im}[B_1/B2]$}
The left-hand side of Eq.~\eqref{match} is explicitly
\begin{equation}
\begin{split}
\frac{B_1(\tilde{\omega},\sigma)}{B_2(\tilde{\omega},\sigma)}\ell^{-2l-1}=\frac{\Gamma\left(l+1\right)\Gamma\left(-2l-1\right)\Gamma\left(-2i\sigma+l+1\right)}{\Gamma\left(2l+1\right)\Gamma\left(-l\right)\Gamma\left(-2i\sigma-l\right)}\left(\frac{r_+}{\ell}-\frac{r_-}{\ell}\right)^{2l+1}.
\end{split}
\end{equation}
This is rewritten as
\begin{equation}
\begin{split}
\label{CB1B2}
\frac{B_1(\tilde{\omega},\sigma)}{B_2(\tilde{\omega},\sigma)}\ell^{-2l-1}=&2i\left(\tilde{\omega}-\frac{eQ}{r_+}\right)\frac
{{r_+}^2}{\ell}\left(\frac{r_+}{\ell}-\frac{r_-}{\ell}\right)^{2l}\frac{\left(l!\right)^2}{\left(2l\right)!\left(2l+1\right)!}\prod_{k=1}^{l}\left(k^2+4\sigma^2 \right),
\end{split}
\end{equation}
with the aid of properties and functional relations of the gamma functions,
\begin{align}
\Gamma(k+z)&=(k-1+z)(k-2+z)\cdots(1+z)\Gamma(1+z),\\
\frac{\Gamma\left(-2m-1\right)}{\Gamma\left(-m\right)}&=\left(-1\right)^{m+1}\frac{m!}{\left(2m+1\right)!},\\
\frac{\Gamma(-2 iz+m+1)}{\Gamma(-2i z-m)}&=2i z(-1)^{m+1}\prod_{k=1}^{m}\left(k^2+4z^2 \right),
\end{align}
for a complex number $z$, and non-negative integers $k,m$. Then, the real and imaginary parts of Eq.~\eqref{CB1B2} are respectively
\begin{equation}
\begin{split}
\label{RB1B2}
{\rm Re}\left[\frac{B_1}{B_2}\right]\ell^{-2l-1}=-2\left({\rm Im}\left[\tilde{\omega}\ell\right]\right)\left(\frac{{r_+}}{\ell}\right)^2\left(\frac{r_+}{\ell}-\frac{r_-}{\ell}\right)^{2l}\frac{\left(l!\right)^2}{\left(2l\right)!\left(2l+1\right)!}\prod_{k=1}^{l}\left(k^2+4\sigma^2 \right),
\end{split}
\end{equation}
and \begin{equation}
\begin{split}
\label{IB1B2}
{\rm Im}\left[\frac{B_1}{B_2}\right]\ell^{-2l-1}=&2\ell\left({\rm Re}\left[\tilde{\omega}\right]-\frac{eQ}{r_+}\right)\left(\frac{{r_+}}{\ell}\right)^2\left(\frac{r_+}{\ell}-\frac{r_-}{\ell}\right)^{2l}\frac{\left(l!\right)^2}{\left(2l\right)!\left(2l+1\right)!}\prod_{k=1}^{l}\left(k^2+4\sigma^2 \right).
\end{split}
\end{equation}
We notice here that ${\rm Re}[{B_1}/{B_2}]=\mathcal{O}({\rm Im}\left[\tilde{\omega}\ell\right],(r_+/\ell)^{2(l+1)})$ and ${\rm Im}[{B_1}/{B_2}]=\mathcal{O}(({\rm Im}\left[\tilde{\omega}\ell\right])^0,(r_+/\ell)^{2(l+1)})$.

\subsection{$\left(-1\right)^{l+1}\Sigma_{2,R}/\Sigma_{1,I}\to-\left(-1\right)^{l+1}\Sigma_{2,R}/\Sigma_{1,I}$ under ${\rm Re}[\tilde{\omega}]\to-{\rm Re}[\tilde{\omega}]$}
\label{proof}
We shall explicitly show that $\left(-1\right)^{l+1}\Sigma_{2,R}/\Sigma_{1,I}$ changes the sign under ${\rm Re}[\tilde{\omega}]\to-{\rm Re}[\tilde{\omega}]$. First, we express the boundary condition parameter $\kappa$ as a function of ${\rm Re}[\tilde{\omega}\ell],~\mu^2\ell^2$, and $l$ by using Eq.~\eqref{weakregularitycondition1}. Then, we can express the leading term of $\Sigma_{1,I}$ and $\Sigma_{2,R}$ in $r_+/\ell \ll1$ as a function of them:
\begin{equation}
\begin{split}
\label{explicitSigma1I}
\Sigma_{1,I}
=&\frac{ \Gamma\left(1-\frac{1}{2}\sqrt{9+4\mu^2\ell^2}\right)}{\Gamma\left(\frac{{\rm Re}\left[\tilde{\omega}\ell\right]}{2}+\frac{l}{2}+\frac{3}{4}-\frac{1}{4}\sqrt{9+4\mu^2\ell^2}\right)\Gamma\left(-\frac{{\rm Re}\left[\tilde{\omega}\ell\right]}{2}+\frac{l}{2}+\frac{3}{4}-\frac{1}{4}\sqrt{9+4\mu^2\ell^2}\right)}\\
&\times\left[P\left(\frac{{\rm Re}\left[\tilde{\omega}\ell\right]}{2}+\frac{l}{2}+\frac{3}{4}-\frac{1}{4}\sqrt{9+4\mu^2\ell^2}\right)-P\left(-\frac{{\rm Re}\left[\tilde{\omega}\ell\right]}{2}+\frac{l}{2}+\frac{3}{4}-\frac{1}{4}\sqrt{9+4\mu^2\ell^2}\right)\right.\\
&\left.-P\left(\frac{{\rm Re}\left[\tilde{\omega}\ell\right]}{2}+\frac{l}{2}+\frac{3}{4}+\frac{1}{4}\sqrt{9+4\mu^2\ell^2}\right)+P\left(-\frac{{\rm Re}\left[\tilde{\omega}\ell\right]}{2}+\frac{l}{2}+\frac{3}{4}+\frac{1}{4}\sqrt{9+4\mu^2\ell^2}\right)\right]\\
&+\mathcal{O}\left(\left(\frac{r_+}{\ell}\right)^{2(l+1)}\right),
\end{split}
\end{equation}
and
\begin{equation}
\begin{split}
\Sigma_{2,R}
=&\frac{ \Gamma\left(1-\frac{1}{2}\sqrt{9+4\mu^2\ell^2}\right)}{\Gamma\left(\frac{{\rm Re}\left[\tilde{\omega}\ell\right]}{2}+\frac{l}{2}+\frac{3}{4}-\frac{1}{4}\sqrt{9+4\mu^2\ell^2}\right)\Gamma\left(-\frac{{\rm Re}\left[\tilde{\omega}\ell\right]}{2}+\frac{l}{2}+\frac{3}{4}-\frac{1}{4}\sqrt{9+4\mu^2\ell^2}\right)}\\
&\times\left[\frac{\Gamma\left(\frac{{\rm Re}\left[\tilde{\omega}\ell\right]}{2}+\frac{l}{2}+\frac{3}{4}-\frac{1}{4}\sqrt{9+4\mu^2\ell^2}\right)\Gamma\left(-\frac{{\rm Re}\left[\tilde{\omega}\ell\right]}{2}+\frac{l}{2}+\frac{3}{4}-\frac{1}{4}\sqrt{9+4\mu^2\ell^2}\right)}{\Gamma\left(\frac{{\rm Re}\left[\tilde{\omega}\ell\right]}{2}-\frac{l}{2}+\frac{1}{4}-\frac{1}{4}\sqrt{9+4\mu^2\ell^2}\right)\Gamma\left(-\frac{{\rm Re}\left[\tilde{\omega}\ell\right]}{2}-\frac{l}{2}+\frac{1}{4}-\frac{1}{4}\sqrt{9+4\mu^2\ell^2}\right)}\right.\\
&\left.-\frac{\Gamma\left(\frac{{\rm Re}\left[\tilde{\omega}\ell\right]}{2}+\frac{l}{2}+\frac{3}{4}+\frac{1}{4}\sqrt{9+4\mu^2\ell^2}\right)\Gamma\left(-\frac{{\rm Re}\left[\tilde{\omega}\ell\right]}{2}+\frac{l}{2}+\frac{3}{4}+\frac{1}{4}\sqrt{9+4\mu^2\ell^2}\right)}{\Gamma\left(\frac{{\rm Re}\left[\tilde{\omega}\ell\right]}{2}-\frac{l}{2}+\frac{1}{4}+\frac{1}{4}\sqrt{9+4\mu^2\ell^2}\right)\Gamma\left(-\frac{{\rm Re}\left[\tilde{\omega}\ell\right]}{2}-\frac{l}{2}+\frac{1}{4}+\frac{1}{4}\sqrt{9+4\mu^2\ell^2}\right)}\right]\\
&+\mathcal{O}\left(\left(\frac{r_+}{\ell}\right)^{2(l+1)}\right).
\end{split}
\end{equation}
It follows from the above equations that $\Sigma_{1,I}\to -\Sigma_{1,I}$ under ${\rm Re}[\tilde{\omega}]\to-{\rm Re}[\tilde{\omega}]$, while $\Sigma_{2,R}\to\Sigma_{2,R}$. Hence, $\left(-1\right)^{l+1}\Sigma_{2,R}/\Sigma_{1,I}$ changes the sign for ${\rm Re}[\tilde{\omega}]\to-{\rm Re}[\tilde{\omega}]$. This is consistent with the results in Figures~\ref{SS1} and~\ref{SS2}.

\end{document}